\documentclass[12pt]{article}
\pdfoutput=1
\usepackage{amsmath}
\usepackage{amsfonts}
\usepackage{amssymb}
\usepackage{graphicx}
\usepackage{rotating}
\usepackage{color}
\usepackage{amsthm}
\usepackage{authblk}
\usepackage{natbib}
\usepackage{threeparttable}
\usepackage{pdflscape}
\usepackage{multirow}
\usepackage{accents}
\usepackage{longtable}
\usepackage{amsfonts,amsmath,bm,mathrsfs}
\usepackage{graphicx}
\usepackage{epsfig}
\usepackage[linesnumbered,ruled,vlined]{algorithm2e}
\usepackage{caption}
\usepackage{subcaption}
\usepackage{setspace}
\usepackage{nccmath}
\usepackage[table]{xcolor}
\usepackage{url}

\def\bfe{{\mathbf{e}}}

\def\bfx{{\mathbf{x}}}
\def\bfy{{\mathbf{y}}}
\def\bfz{{\mathbf{z}}}

\def\bfI{{\mathbf{I}}}

\def\bfX{{\mathbf{X}}}

\def\bfZ{{\mathbf{Z}}}
\def \pr {\mbox{Pr}}

\def \bfphi {\boldsymbol{\phi}}

\def \bfalpha {\boldsymbol{\alpha}}
\def \bfbeta {\boldsymbol{\beta}}
\def \bfeta {\boldsymbol{\eta}}
\def \bfdelta {\boldsymbol{\delta}}
\def \bfgamma {\boldsymbol{\gamma}}

\def \bftheta {\boldsymbol{\theta}}

\def \bfeta {\boldsymbol{\eta}}

\def\ie{\emph{i.e.}}

\allowdisplaybreaks[3]

\date{}
\begin{document}
\setlength{\textheight}{575pt}
\setlength{\baselineskip}{23pt}

\title{LPG: a four-groups probabilistic approach to leveraging pleiotropy in genome-wide association studies}

\author[1]{Yi Yang}
\affil[1]{School of Statistics and Management, The Shanghai University of Finance and Economics, Shanghai}
\author[2,4]{Mingwei Dai}
\affil[2]{Institute for Information and System Sciences, Xian Jiaotong University, Xian}
\author[3]{Jian Huang}
\affil[3]{Department of Applied Mathematics, Hong Kong Polytechnics University}
\author[5]{Xinyi Lin}
\author[4]{Can Yang}
\affil[4]{Department of Mathematics, Hong Kong University of Science and Technology}
\author[5,*]{Jin Liu}
\affil[5]{Centre for Quantitative Medicine, Duke-NUS Medical School}
\author[1,*]{Min Chen}
\maketitle

\begin{abstract}
To date, genome-wide association studies (GWAS) have successfully identified tens of thousands of genetic variants among a variety of traits/diseases, shedding a light on the genetic architecture of complex diseases. Polygenicity of complex diseases, which refers to the phenomenon that a vast number of risk variants collectively contribute to the heritability of complex diseases with modest individual effects, have been widely accepted. This imposes a major challenge towards fully characterizing the genetic bases of complex diseases. An immediate implication of polygenicity is that a much larger sample size is required to detect risk variants with weak/moderate effects. Meanwhile, accumulating evidence suggests that different complex diseases can share genetic risk variants, a phenomenon known as pleiotropy. %In principle, the pleiotropy is a phenomenon due to genetic correlations and thus the integrative analysis between genetic studies is an alternative to large sample recruitment.
In this study, we propose a statistical framework for \underline{L}everaging \underline{P}leiotropic effects in large-scale \underline{G}WAS data (LPG). % between genetically correlated traits. An efficient variational Bayes EM (VBEM) algorithm is developed to fit the model.
LPG utilizes a variational Bayesian expectation-maximization (VBEM) algorithm, making it computationally efficient and scalable for genome-wide scale analysis.
To demonstrate the advantage of LPG over existing methods that do not leverage pleiotropy, we conducted extensive simulation studies and also applied LPG to analyze three autoimmune disorders (Crohn's disease, rheumatoid arthritis and Type 1 diabetes). % from WTCCC data set.
The results indicate that LPG can improve the power of prioritization of risk variants and accuracy of risk prediction by leveraging pleiotropy. The software is available at https://github.com/Shufeyangyi2015310117/LPG.
\end{abstract}

\section{Introduction}
Genome-wide association studies (GWAS) have reported more than 51,000 single-nucleotide polymorphisms (SNPs) to be genome-wide significantly associated with complex human phenotypes, including quantitative traits and complex diseases (Accession of the GWAS Catalog database \url{https://www.ebi.ac.uk/gwas/} on October, 2017). Although discovery of these genetic risk variants has advanced our understanding of the genetic architecture of complex diseases/traits, these variants can explain only a small proportion of phenotypic variance~\citep{manolio2009finding}. For example, while the heritability of human height has been estimated to be about 70$\%$-80$\%$, the 697 genetic variants found from GWAS analysis of human height using 253,288 individuals could explain only 20$\%$ of the heritability for human height collectively. A more complete characterization of the genetic architecture of complex phenotypes remains a significant challenge.
%As of September 2017, more than 33,000 single-nucleotide polymorphisms (SNPs) have been discovered to be significantly ($p$-value $\leq 5.0 \times 10^{-8})$ associated with complex human phenotypes (including quantitative traits and complex diseases) in about 2,600 GWAS (see the GWAS Catalog database \url{https://www.ebi.ac.uk/gwas/}. The findings of significant risk variants have greatly deepened our understanding of the genetic architecture of human complex diseases/traits but these variants can often explain a small proportion of phenotypic variance~\citep{manolio2009finding}. For example, the heritability of human height is about 70$\%$-80$\%$. Wood \etal~\cite{wood2014defining} used
%genome-wide data of 253,288 individuals identified 697 variants ($p$-value$ \le 5\times 10^{-8}$) that can only explain 20$\%$ of heritability for human height collectively. A recent estimate suggests that there may be about 100,000 variants affecting human height with individually tiny effects~\citep{boyle2017expanded}.
%To advance our understanding of the genetic architecture of complex phenotypes, much larger sample sizes in GWAS are required but the sample recruitment is often very expensive and time consuming.
% The polygenicity of complex diseases is further supported by recent GWAS with larger sample sizes, in which more associated common variants with moderate effects have been identified(e.g. GWAS data from 34,8400 cases and 114,981 controls are analyzed to understand the genetic architecture of type-2 diabetes~\citep{morris2012large}). But large sample recruitment is often expensive and time-consuming.

To increase statistical power in a GWAS analysis, newer analytical methods leveraging pleiotropy have been developed. Pleiotropy, which refers to the phenomenon that a gene affects multiple phenotypes, was first proposed more than 100 years ago~\citep{wagner2011pleiotropic}. Since then, there have been an increasing number of human genetics studies reporting pleiotropic effects in various complex diseases such as autoimmune diseases~\citep{cotsapas2011pervasive}, metabolic disorders~\citep{kraja2014pleiotropic} and psychiatric disorders~\citep{wang2015pervasive}. Thus, the identification of genetic risk variants in GWAS can be significantly improved by incorporating pleiotropy into the statistical analysis. Existing statistical methods for incorporating pleiotropy in GWAS analysis can proceed by conducting joint GWAS analysis of multiple traits~\citep{zhou2014efficient, liu2016analyzing}. However, these methods assume that the GWAS data were collected from the same study individuals and cannot be applied when the GWAS data was collected from different study cohorts. Thus, there remains a methodology gap to leverage pleiotropy for joint GWAS analysis of multiple traits when the GWAS data for each trait was collected from different study cohorts. This article seeks to fill this gap.

In this article, we propose a novel statistical method for \underline{L}everaging \underline{P}leiotropic effects in large-scale \underline{G}WAS (LPG) data that is collected from different studies. LPG provides a statistical framework for the evaluation of local false discovery rate, prediction accuracy and formal test of pleiotropic effects between two traits. LPG utilizes a variational Bayesian expectation maximization (VBEM) algorithm, making it computationally efficient for genome-wide analysis. We conducted extensive simulation studies to evaluate the performance of LPG. We then applied it to conduct joint analysis of Crohn's disease, rheumatoid arthritis and type-I diabetes using data from the Welcome Trust Case Control Consortium (WTCCC)~\citep{burton2007genome}. The simulation studies and real data analysis suggest that LPG can steadily improve both the prediction accuracy and statistical power of risk variants identification over single-trait-based methods that do not leverage pleiotropy.

The remainder of this article is organized as follows. In section~\ref{method}, we introduce the statistical model and describe the VBEM algorithm used to estimate the parameters in the model. In section~\ref{stat_infer}, we describe the statistical inference procedure used to evaluate the local false discovery rate and the prediction accuracy of the identified genetic variants; we also describe a formal hypothesis test for pleiotropy. In sections~\ref{simstudies} and~\ref{realdata}, we evaluate the performance of LPG using simulations and real data analysis of WTCCC data, respectively. We conclude with discussions in section~\ref{discussion}.
\section{Methods}
\label{method}
\subsection{Model for quantitative traits}
\label{model_quant}
Suppose that we have a GWAS data set $\{\bfy,\bfX\}$ with $n$ independent samples, where  $\bfy\in\mathbb{R}^n$ is the vector of quantitative phenotype and $\bfX=[\bfx_1,\dots,\bfx_p]\in\mathbb{R}^{n\times p}$ is the genotype matrix for $n$ individuals and $p$ SNPs. Without loss of generality, we assume that both $\bfX$ and $\bfy$ have been centered. We assume the following standard linear model,
\begin{equation}
\bfy = \bfX\bfbeta + \bfe,
\end{equation}
where $\bfbeta = [\beta_1,\dots, \beta_p]^\top$ is a vector of effect sizes and $\bfe\sim\mathcal{N}(\bm{0},\sigma_{e}^2\bfI)$ is the random error. %corresponding to environmental and other unmeasured effects. %As a small proportion contributes to the quantitative trait $\bfy$, the vector of effect sizes $\bfbeta$ is sparse. Thus, inferring risk variants is identical to select non-zero entries in $\bfbeta$.
Let the vector of binary variables $\bfgamma=[\gamma_1,\dots,\gamma_p]^\top$ indicate the association status of all $p$ SNPs, where $\gamma_j=1$ indicates that the $j$-th SNP is associated with trait $\bfy$, and $\gamma_j=0$ otherwise. In this paper, we consider a spike-slab prior~\citep{kuo1998variable}%,zhou2009projections,soussen2011bernoulli}, %for each $\beta_j$,
\begin{equation}\label{BGM}
\bfy | \bfX,\bfbeta,\bfgamma,\sigma_e^2 \sim \mathcal{N}\left(\sum_j \gamma_j\beta_j\bfx_j,\sigma_e^2\right), \mbox{with }
\gamma_j \sim \mathrm{Ber}(\alpha),
\beta_j \sim \mathcal{N}(0,\sigma_{\beta}^2),
\end{equation}
where $\mathrm{Ber}(\alpha)$ is the Bernoulli distribution with probability $\pr(\gamma_j=1)=\alpha$ and $\mathcal{N}(m,\sigma^2)$ denotes the Gaussian distribution with mean $m$ and variance $\sigma^2$. The model~(\ref{BGM}) is known as a binary mask model as we can consider the indicator $\gamma_j$ as masking out the coefficient $\beta_j$. Then, the probabilistic model can be written as
\begin{equation}
\pr(\bfy,\bfbeta,\bfgamma|\bfX;\bftheta) = \pr(\bfy|\bfX,\bfbeta;\bftheta) \pr(\bfbeta|\bftheta) \pr(\bfgamma|\bftheta),
\end{equation}
where $\bftheta = \{\sigma_{\beta}^2,\sigma_e^2,\alpha\}$ is the collection of model parameters.
%and $\delta_0$ is the Dirac function centered at zero, and $\gamma_j$ is assumed to be drawn from the Bernoulli distribution\\
%\begin{equation}
%\gamma_j|\alpha = \alpha^{\gamma_j}(1-\alpha)^{1 - \gamma_j},
%\end{equation}
%where $\pr(\gamma_j=1)=\alpha$. To ease our derivation for variational inference, we introduce a new Gaussian random variable $\tilde\beta_j\sim\mathcal{N}(0,\sigma^2_\beta)$. It is trivia to show that the distribution for $\tilde\beta_j\gamma_j$ is identical to that of $\beta_j\gamma_j$ and the new parameterization is called, Bernoulli-Gaussian model in practice \citep{kuo1998variable,zhou2009projections,soussen2011bernoulli}. Hereafter, we abuse the notation $\beta_j$ for $\tilde\beta_j$.

Now we generalize the above two-groups model to leverage pleiotropy between two traits that are potentially genetically correlated. Suppose we have two GWAS datasets $\{\bfy_1, \bfX_1\}$, $\{\bfy_2, \bfX_2\}$ with $n_1$ and $n_2$ samples, respectively. Here, $\bfy_1 \in \mathbb{R}^{n_1}$ and $\bfy_2 \in \mathbb{R}^{n_2}$ are the vectors of phenotypic values, $\bfX_1=[\bfx_{11},\dots,\bfx_{1p}]\in\mathbb{R}^{n_1\times p}$ and $\bfX_2=[\bfx_{21},\dots,\bfx_{2p}]\in\mathbb{R}^{n_2\times p}$ are the corresponding genotype matrices for $p$ identical SNPs. Without loss of generality, we assume that both genotype data ($\bfX_1$ and $\bfX_2$) and phenotype data ($\bfy_1$ and $\bfy_2$) have been centered. Then we have
\begin{equation}
\label{4GPMP}
\bfy_k | \bfX_k,\bfbeta_k,\bfgamma_k,\sigma_{e_k}^2  \sim  \mathcal{N}\left(\sum_{j=1}^p \gamma_{kj}\beta_{kj}\bfx_{kj},\sigma_{e_k}^2\right), \mbox{with }
[\gamma_{1j},\gamma_{2j}] \sim \mathrm{Mu}_{l\in L}(\bfalpha),
\beta_{kj} \sim \mathcal{N}(0,\sigma_{\beta_k}^2),
\end{equation}
where $k(=1,2)$ refers to individual studies (hereafter, without further denotation, we take $k$ to be 1 and 2 by default), $\bfbeta_k = [\beta_{k1},\dots, \beta_{kp}]^\top$ is a vector of effect sizes for study $k$, $\sigma_{e_k}^2$ is the variance for the random error in study $k$, the vector of binary variables $\bfgamma_k=[\gamma_{k1},\dots,\gamma_{kp}]^\top$ indicates the association statuses in study $k$, $\bfgamma=[\bfgamma_1,\bfgamma_2] \in \mathbb{R}^{p \times 2}$ is a matrix for association statuses in two studies, $\bfalpha = (\alpha_{00},\alpha_{01},\alpha_{10},\alpha_{11})^\top$ is the vector of parameters in a multinomial distribution, and $\mathrm{Mu}_{l\in L}(\bfalpha)$ is the multinomial distribution with parameter $\bfalpha$ for each possible values $L = \{00,01,10,11\}$, \ie, $\alpha_{00} = \pr(\gamma_{1j}=0,\gamma_{2j}=0)$, $\alpha_{10} = \pr(\gamma_{1j}=1,\gamma_{2j}=0)$, $\alpha_{01} = \pr(\gamma_{1j}=0,\gamma_{2j}=1)$, and $\alpha_{11} = \pr(\gamma_{1j}=1,\gamma_{2j}=1)$.

Comparing model~(\ref{4GPMP}) with the basic model~(\ref{BGM}) for a single trait, the major difference lies in the joint sampling for hidden association statuses in the joint model of two studies. In the presence of pleiotropy, $\gamma_{1j}$ and $\gamma_{2j}$ are no longer independent. %, as indicated by their joint probability $\bfalpha$.
We will demonstrate in the supplementary document, that all the parameters in our model can be adaptively estimated from the data, without any ad-hoc tuning. Let $\bftheta (= \{\sigma_{\beta_1}^2,\sigma_{\beta_2}^2,\sigma_{e_1}^2,\sigma_{e_2}^2,\bfalpha\})$ be the collection of model parameters. The joint probabilistic model can be written as
\begin{equation}
\label{joint_4GPMP}
\pr(\bfy_1,\bfy_2,\bfbeta_1,\bfbeta_2,\bfgamma_1,\bfgamma_2| \bfX_1,\bfX_2;\bftheta) = \prod_{k=1}^2 \bigg(\pr(\bfy_k|\bfX_k,\bfbeta_k,\bfgamma_k;\bftheta) \pr(\bfbeta_k|\bftheta)\bigg) \pr(\bfgamma|\bftheta).
\end{equation}
Marginalizing over latent variables ($\bfbeta_1,\bfbeta_2,\bfgamma_1,\bfgamma_2$), the probabilistic model of observed data becomes
\begin{equation}
\label{marg_4GPMP}
\pr(\bfy_1,\bfy_2| \bfX_1,\bfX_2;\bftheta) = \sum_{\bfbeta_1,\bfbeta_2, \bfgamma_1,\bfgamma_2}  \pr(\bfy_1,\bfy_2,\bfbeta_1,\bfbeta_2,\bfgamma_1,\bfgamma_2| \bfX_1,\bfX_2;\bftheta),
\end{equation}
where we have abused the operation $\sum$ to represent integration of continuous variables.
Then, by Bayes rule, the posterior probability distribution for the variables of interest can be calculated as
\begin{equation}
	\label{post}
 \pr(\bfbeta_1,\bfbeta_2,\bfgamma_1,\bfgamma_2 |\bfy_1,\bfy_2, \bfX_1,\bfX_2;\bftheta) = \frac{\pr(\bfy_1,\bfy_2,\bfbeta_1,\bfbeta_2,\bfgamma_1,\bfgamma_2| \bfX_1,\bfX_2;\bftheta) }{\pr(\bfy_1,\bfy_2| \bfX_1,\bfX_2;\bftheta)}.
\end{equation}
Computing the posterior distribution~(\ref{(post}) is difficult as it requires evaluation of marginal likelihood (\ref{marg_4GPMP}), which is computationally intractable.
%The difficulty of the above strategy lies in the evaluation of marginal likelihood (\ref{marg_4GPMP}), which is computationally intractable.

\subsection{Algorithm for quantitative trait model}
\label{alg_qtm}
To overcome the intractability of the marginal likelihood (\ref{marg_4GPMP}), we derive an efficient algorithm based on variational inference, which makes our model scalable to genome-wide data analysis. The key idea is that we make use of Jensen's inequality to iteratively obtain an adjustable lower bound on the marginal log likelihood \citep{jordan1999introduction}. First, we have a lower bound of the logarithm of the marginal likelihood (\ref{marg_4GPMP}),
%\begin{eqnarray}
%\label{lob}
%& &\log \pr(\bfy_1,\bfy_2| \bfX_1,\bfX_2;\bftheta) \notag\\ &\geq&\mathbb{E}_q[\log\pr(\bfy_1,\bfy_2,\bfbeta_1,\bfbeta_2,\bfgamma_1,\bfgamma_2| \bfX_1,\bfX_2;\bftheta)] - \mathbb{E}_q[\log q(\bfbeta_1,\bfbeta_2,\bfgamma_1,\bfgamma_2)]
%:=\mathcal{L}(q),
%\end{eqnarray}
\begin{eqnarray}
\label{lob}
& &\log \pr(\bfy_1,\bfy_2| \bfX_1,\bfX_2;\bftheta) = \mathcal{L}(q,\bftheta) + \mathbb{KL}(q||p)
\notag \\
&\ge & \mathbb{E}_q[\log\pr(\bfy_1,\bfy_2,\bfbeta_1,\bfbeta_2,\bfgamma_1,\bfgamma_2| \bfX_1,\bfX_2;\bftheta)] - \mathbb{E}_q[\log q(\bfbeta_1,\bfbeta_2,\bfgamma_1,\bfgamma_2)],
\end{eqnarray}
where we define
\begin{eqnarray}
\label{lb1}
  \mathcal{L}(q,\bftheta) &= &\sum_{\bfbeta_1,\bfbeta_2,\bfgamma_1,\bfgamma_2} q(\bfbeta_1,\bfbeta_2,\bfgamma_1,\bfgamma_2)\log \frac{p(\bfy_1,\bfy_2, \bfbeta_1,\bfbeta_2,\bfgamma_1,\bfgamma_2|\bfX_1,\bfX_2;\bftheta) }{q(\bfbeta_1,\bfbeta_2,\bfgamma_1,\bfgamma_2)}, \notag \\
  \mathbb{KL}(q||p) &=& \sum_{\bfbeta_1,\bfbeta_2,\bfgamma_1,\bfgamma_2} q(\bfbeta_1,\bfbeta_2,\bfgamma_1,\bfgamma_2)\log \frac{q(\bfbeta_1,\bfbeta_2,\bfgamma_1,\bfgamma_2) }{p(\bfbeta_1,\bfbeta_2,\bfgamma_1,\bfgamma_2 | \bfy_1, \bfy_2, \bfX_1, \bfX_2; \bftheta)} .
\end{eqnarray}
%where the inequality follows Jensen's inequality and the equality holds if, and only if, $q(\bfbeta_1,\bfbeta_2,\bfgamma_1,\bfgamma_2)$ is the true posterior $p^*=\pr(\bfbeta_1,\bfbeta_2,\bfgamma_1,\bfgamma_2 | \bfy_1, \bfy_2, \bfX_1, \bfX_2; \bftheta)$. One can easily verify that the difference between left-hand side and right-hand side of the lower bound (\ref{lob}) is the Kullback-Leibler (KL) divergence between the variational posterior probability ($q$) and the true posterior probability ($p^*$),
Note that Kullback-Leibler (KL) divergence satisfies $\mathbb{KL}(q||p)\ge 0$, with equality holds if, and only if, that variational posterior probability ($q$) and the true posterior probability ($p$) are equal. Similar to expectation-maximization (EM) algorithm, we can maximize the lower bound $\mathcal{L}(q,\bftheta)$ with respect to the variational distribution $q$, which is equivalent to minimizing the KL divergence~\citep{bishop2006pattern}. To make it computationally efficient to evaluate the lower bound, we use mean-field theory~\citep{opper2001advanced} and assume that $q(\bfbeta_1,\bfbeta_2,\bfgamma_1,\bfgamma_2)$ can be factorized as %Thus, in each iteration, it is equivalent to minimize the KL divergence with $q$ subject to some factorization. To be specific, we use mean-field method \citep{parisi1988statistical,opper2001advanced}, assuming that $q(\bfbeta_1,\bfbeta_2,\bfgamma_1,\bfgamma_2)$ can be factorized as
\begin{equation}
\label{factorization}
q(\bfbeta_1,\bfbeta_2,\bfgamma_1,\bfgamma_2) =
\prod_{j=1}^p q_j(\bfbeta_{1j},\bfbeta_{2j},\bfgamma_{1j},\bfgamma_{2j}).
\end{equation}
No additional assumptions on the posterior distribution are required. This factorization~(\ref{factorization}) is used as a surrogate for the posterior distribution $\pr(\bfbeta_1,\bfbeta_2,\bfgamma_1,\bfgamma_2| \bfy_1,\bfy_2, \bfX_1,\bfX_2;\bftheta)$.
Using the properties of factorized distributions in variational inference \citep{bishop2006pattern}, we can obtain the optimal approximation using the following formula:
\begin{equation}
\log q_j(\beta_{1j},\beta_{2j},\gamma_{1j},\gamma_{2j}) =
\mathbb{E}_{j'\neq j}[\log \pr(\bfy_1,\bfy_2,\bfbeta_1,\bfbeta_2,\bfgamma_1,\bfgamma_2|
\bfX_1,\bfX_2,\bftheta)]
 + \mbox{const}
\end{equation}
where the expectation is taken with respect to all of the other factors
$\{q_{j'}(\beta_{1j'},\beta_{2j'},\gamma_{1j'},\gamma_{2j'})\}$ for $j'\neq j$. After some derivations (details are given in the supplementary document), we have
\begin{equation}
\label{variational_q}
q(\beta_{1j},\beta_{2j},\gamma_{1j},\gamma_{2j}) = f_{1j}(\beta_{1j})^{\gamma_{1j}}f_0(\beta_{1j})^{1-\gamma_{1j}} f_{2j}(\beta_{2j})^{\gamma_{2j}}f_0(\beta_{2j})^{1-\gamma_{2j}} \prod_{l}\alpha_{lj}^{\textbf{1}{(\gamma_{1k}=l_1,\gamma_{2k}=l_2)}}, \\
%&=&
%[\mathcal{N}(\mu_{1j},s_{1j}^2)]^{\gamma_{1j}}[\mathcal{N}(0,\sigma_{\beta_1}^2)]^{1 - \gamma_{1j}}
%[\mathcal{N}(\mu_{2j},s_{2j}^2)]^{\gamma_{2j}}[\mathcal{N}(0,\sigma_{\beta_2}^2)]^{1 - \gamma_{2j}}
%\prod_{l}\phi_{lj}^{\textbf{1}_{(\gamma_{1k}=l_1,\gamma_{2k}=l_2)}}
\end{equation}
where $\alpha_{lj}$ is the posterior probability of $[\gamma_{1j},\gamma_{2j}] = l$, $f_0(\beta_{kj})$ is the posterior distribution of $\beta_{kj}$ when $\gamma_{kj}=0$, $f_{kj}(\beta_{kj})$ is the posterior distribution of $\beta_{kj}$ under $\gamma_{kj}=1$. With some algebra, it is easy to show that $f_0(\beta_{kj})$ and $f_{kj}(\beta_{kj})$ are the density functions of Gaussian distributions $\mathcal{N}(0,\sigma_{\beta_k}^2)$ and $\mathcal{N}(\mu_{kj},s_{kj}^2)$. %with
The derivation details on the updating equations and corresponding VBEM algorithm (Algorithm S1) can be found in the supplementary document. The VBEM algorithm performs similarly to coordinate descent algorithm, which comes from the factorization of variational distribution (\ref{variational_q}). Hence, VBEM algorithm developed here is scalable to ultra-high dimensions.

\subsection{Accommodating case-control data}
Suppose that we have  two GWAS case-control datasets $\{\bfy_1,\bfX_1, \bfZ_1\}$ and $\{\bfy_2,\bfX_2, \bfZ_2\}$ with $n_1$ and $n_2$ samples, respectively. All other definitions are identical to those introduced in Section \ref{model_quant} except that $\bfy_k \in\mathbb{R}^{n_k\times 1}$ is the vector for disease status taking values -1 and 1 for controls and cases, respectively, and $\bfZ_k=[\bfz_{k1},\dots,\bfz_{kp_0}]\in\mathbb{R}^{n_k\times p_0}$ is a matrix for $p_0$ covariates in study $k$. Note that the first column of $\bfZ_k$ is a vector of ones corresponding to an intercept. Then conditional on observed genotype $\bfX_k$, hidden status $\bfgamma$, and effects $\bfbeta_k$, we have
\begin{equation}
  \bfy_k|\bfX_k, \bfZ_k, \bfbeta_k, \bfgamma_k,\bfphi_k  \sim \mathrm{Ber}(\bfdelta_k),
\end{equation}
where $\bfdelta_k = [\delta_{k1},\dots,\delta_{kn_k}]^\top$, $\delta_{ki} \left(= \pr(y_{ki}=1|\bfX, \bfbeta_k, \bfgamma_k) = \frac{1}{1+e^{-y_{ki}\eta_{ki}}}\right)$ is the sigmoid function of linear predictor $\eta_{ki}$, $i$ is the index for individuals, $\bfeta_k (= [\eta_{k1},\dots,\eta_{kn_k}]^\top\in\mathbb{R}^{n_k\times 1})$ is the linear predictor of the all individuals in study $k$ such that $\bfeta_{k} = \sum_{j=1}^{p_0}\bfz_{kj}\phi_{kj} + \sum_{j=1}^p \gamma_{kj}\beta_{kj}\bfx_{kj}$. %Note that $\bfphi_k$ is a vector of fixed effects including intercept in the model.
Here, we include fixed-effect covariates in the binary studies to adjust potential population stratification and confounders in samples. %The rationale for including covariates in the model for binary studies is that most times, we need to adjust the first few principal components from data to correct the population stratification in samples \citep{reed2015guide}.
$\bfbeta$ and $\bfgamma$ are effect sizes and indicator variables as defined in Section \ref{model_quant}. Let $\bftheta = \{\sigma_{\beta_1}^2, \sigma_{\beta_2}^2,\bfphi_1, \bfphi_2, \bfalpha\}$ be the collection of model parameters. The probabilistic model can be written as
\begin{equation}
\label{joint_4GPMP_binary}
 \pr(\bfy_1,\bfy_2,\bfbeta_1,\bfbeta_2,\bfgamma_1,\bfgamma_2| \bfX_1,\bfZ_1, \bfX_2, \bfZ_2 ;\bftheta) = \prod_{k=1}^2 \bigg(\pr(\bfy_k|\bfX_k,\bfZ_k, \bfbeta_k,\bfgamma_k;\bftheta) \pr(\bfbeta_k|\bftheta)\bigg) \pr(\bfgamma|\bftheta).
\end{equation}
Note that we take coefficients for covariates ($\bfZ_1$ and $\bfZ_2$) as fixed effects, which are included in parameter space $\bftheta$. Marginalizing over latent variables ($\bfbeta_1,\bfbeta_2,\bfgamma_1,\bfgamma_2$), we can get the marginal likelihood similar to expression (\ref{marg_4GPMP}). The primary difficulty for the binary model (\ref{joint_4GPMP_binary}) comes from the evaluation of sigmoid function $\delta_{ki}$. %Similar to the quantitative case, it is tempting make model (\ref{joint_4GPMP_binary}) full Bayes by adding hyper prior on $\bftheta$. In fact, this cannot be done easily as there is no convenient conjugate prior for logistic regression.
%As variational approaches iteratively finds a sequence of increasing lower bounds, it is essential to evaluate lower bounds involving sigmoid functions, which is not trivia at all.
As there is no convenient conjugate prior for sigmoid function, it is not analytically feasible to compute the full posterior over the parameter space. To overcome this limitation,, we use the Bohning bound~\citep{bohning1992multinomial}. %many quadratic lower bounds of sigmoid functions can be applied, \eg,, the Bohning bound~\citep{bohning1992multinomial} and the JJ bound~\citep{jaakkola1997variational}. Here, we adopt the Bohning bound. %because it is computationally efficient than the JJ bound which requires iteratively re-evaluation of the variational parameter for each individual.
Here, we first derive a lower bound of the complete-data likelihood as follows
\begin{equation}
\label{lb_bohn}
\begin{aligned}
&\text{Pr}(\textbf{y}_1,\textbf{y}_2,{\bm{\beta}}_1,{\bm{\beta}}_2,\bm{\gamma}_1,\bm{\gamma}_2|
\textbf{X}_1,\textbf{X}_2,\textbf{Z}_1,\textbf{Z}_2,\bm{\theta})\\
%%%%%%%%%%%%%%%%%%%%%%%%%%%%%%%%%%%%%%%%%%%%%%%%%%%%%%%%%%%%%%%%%%%%%%%%%%%%%%%%%%%%%%%%%%%%%%%%%%%%%%%%%%%%%%%%%%%
\geq& \left( \prod_{k=1}^2 B(\textbf{y}_k|\textbf{X}_k,\textbf{Z}_k,\bm{\beta}_k,\bm{\gamma}_k;\bm{\theta})
\text{Pr}(\bm{\beta}_k;\bm{\theta})\right)
\text{Pr}(\bm{\gamma};\bm{\theta})\\
=&h(\textbf{y}_1,\textbf{y}_2,{\bm{\beta}}_1,{\bm{\beta}}_2,\bm{\gamma}_1,\bm{\gamma}_2
|\textbf{X}_1,\textbf{X}_2,\textbf{Z}_1,\textbf{Z}_2;\tilde{\bm{\theta}}),
\end{aligned}
\end{equation}
where $B(\textbf{y}_k|\textbf{X}_k,\textbf{Z}_k,\bm{\beta}_k,\bm{\gamma}_k;\tilde{\bm{\theta}}) \left(=
\prod_{i=1}^{n_k}\exp(-\frac{1}{2}a\eta_{ki}^2y_{ki}^2+(1+b_{ki})\eta_{ki}y_{ki} - c_{ki})\right)$ denotes the product of lower bound of sigmoid functions with $a=1/4$, $b_{kn} = a\psi_{kn} - (1 + e^{-\psi_{kn}})^{-1}$ and $c_{kn} = \frac{1}{2}a\psi_{kn}^2 - (1 + e^{-\psi_{kn}})^{-1}\psi_{kn} + \log(1 + e^{\psi_{kn}})$, and $\tilde{\bm{\theta}} =\{\sigma_{\beta_1}^2,\sigma_{\beta_2}^2,\bfphi_1,\bfphi_2,\bm{\alpha},\bm{\psi}_1,\bm{\psi}_2\}$ is the new parameter which combines the model parameters $\bm{\theta}$ with variational parameters $\bm{\psi}_1$,$\bm{\psi}_2$ (details are provided in the supplementary document). Using Jensen's inequality and the lower bound of complete-data likelihood (\ref{lb_bohn}), we have the following lower bound
\begin{equation}
\begin{aligned}
\label{def_lowerbound_logis}
  &\log \text{Pr}(\textbf{y}_1,\textbf{y}_2| \textbf{X}_1,\textbf{X}_2,\textbf{Z}_1,\textbf{Z}_2,\bm{\theta})\\
=&\log \sum_{\bfbeta_1,\bfbeta_2, \bfgamma_1,\bfgamma_2}  \text{Pr}(\textbf{y}_1,\textbf{y}_2,\bfbeta_1,\bfbeta_2,\bfgamma_1,\bfgamma_2| \textbf{X}_1,\textbf{X}_2,\textbf{Z}_1,\textbf{Z}_2,\bm{\theta})  \\
\geq&\log \sum_{\bfbeta_1,\bfbeta_2, \bfgamma_1,\bfgamma_2}  h(\textbf{y}_1,\textbf{y}_2,\bfbeta_1,\bfbeta_2,\bfgamma_1,\bfgamma_2| \textbf{X}_1,\textbf{X}_2,\textbf{Z}_1,\textbf{Z}_2,\bm{\theta}) \\
\geq&\mathbb{E}_q[\log h(\textbf{y}_1,\textbf{y}_2, \bm{\beta}_1,\bm{\beta}_2,\bm{\gamma}_1,\bm{\gamma}_2
|\textbf{X}_1,\textbf{X}_2,\textbf{Z}_1,\textbf{Z}_2;\tilde{\bm{\theta}})] - \mathbb{E}_q[\log q(\bm{\beta}_1,\bm{\beta}_2,\bm{\gamma}_1,\bm{\gamma}_2)] \\
:=&\mathcal{L}(q),
\end{aligned}ß
\end{equation}
where the first inequality is based on Bohning bound and the second one follows from Jensen's inequality as in lower bound~(\ref{lob}).
By maximizing the lower bound (\ref{def_lowerbound_logis}) with respect to $\mu_{kj}$ and $s_{kj}^2$, we can again obtain the variational distribution in the same fashion as expression (\ref{variational_q}). Details of updating equations and corresponding VBEM algorithm (Algorithm S2) are given in supplementary document.% With some algebra, we have
\section{Statistical inference}
\label{stat_infer}
\subsection{Evaluation of local false discovery rate (lfdr)}
After fitting an LPG model with all the parameters estimated, SNPs can be prioritized based on their local false discovery rates (lfdr)~\citep{efron2012large}. As discussed in~\cite{efron2008microarrays}, although false discovery rate (FDR) methods were  developed in a strict frequentist framework, they also have a convincing Bayesian rationale. Since $\sum_{l\in L_k} \alpha_{lj}$ is a good approximation to the true posterior $\pr(\gamma_{kj} = 1|\textbf{y}_1,\textbf{y}_2,\textbf{X}_1,\textbf{X}_2;\bm{\theta})$, $\mathrm{lfdr}_{kj} (= 1 - \sum_{l\in L_k} \alpha_{lj})$ can be used as lfdr of SNP $j$ in the $k$-th trait, where $k=1$ or 2, $L_1=\{10,11\}$ and $L_2=\{01,11\}$. Namely, the smaller the lfdr, the more we are confident in prioritizing a SNP. Then we use the {\sl direct posterior probability approach}~\citep{newton2004detecting} to control global false discovery rate to select a list of SNPs to be as large as possible while bounding the rate of false discoveries by a pre-specified threshold $\tau$. In details, with data and fitted model in hand, we rank the SNPs according to their local false discovery rate in the ascending order. We can increase the threshold for lfdr $\zeta$ from zero to one and find the largest $\zeta$ that satisfies
\begin{equation}
\widehat{\mathrm{FDR}}(\tau) =\frac{\sum_{j=1}^{p} \widehat{\mathrm{lfdr}}_{kj} \mathbb{I}\left[\widehat{\mathrm{lfdr}}_{kj}  \le \zeta\right]}{\sum_{j=1}^{p} \mathbb{I}\left[\widehat{\mathrm{lfdr}}_{kj}\le \zeta\right]}\le\tau,
\end{equation}
where $\tau$ is a pre-specified bound of global FDR, $\mathbb{I}(\cdot)$ is the indicator function which returns 1 if the argument is true, 0 otherwise. By doing so, it is convenient for users to control FDR either in terms of global FDR or lfdr.  %Finally, we can determine the list of SNPs such that $\left[\widehat{\mathrm{lfdr}}_k \le \zeta\right]$, which are SNPs found to be associated with the trait at a given false discovery rate $\tau$. As demonstrated in simulation studies, we showed that our FGP and its two-group counterpart, \eg BVSR, can control global FDR at a pre-specified value.

\subsection{Evaluation of prediction performance}
In addition to the identification of risk variants, we can also use the LPG approach to conduct risk prediction. In the LPG model, the effect size of SNP $j$ in the $k$-th study is given as $\mathbb{E}(\gamma_{kj}{\beta}_{kj}) = \sum_{l\in L_k}\alpha_{lj}\mu_{kj}$. Given the genotype vector of an individual
${\textbf{x}} = [{x}_1 ,\dots, {x}_p]^\top$, the predicted phenotypic value
is $\hat{y}_k = c_{k0} + \sum_j\left(({x}_{kj} - c_{kj})\sum_{l\in L_k} \alpha_{lj} \mu_{kj}\right)$ , where $c_{k0}$ and $c_{k1}, \dots, c_{kp}$ are the
mean of the phenotype and each SNP before centering for the $k$-th study, respectively. We measured Pearson's correlation between the observed phenotypic values and predicted phenotypic values in the testing set for quantitative trait. For case-control studies, predicted linear predictor is $\hat\eta_k = \bfz_k \bfphi_k + \sum_j\left(({x}_{kj} - c_{kj})\sum_{l\in L_k} \alpha_{lj} \mu_{kj}\right)$ and odds of being a case for such an individual can be found by logit transformation. For the predicted odds from the testing set, we can evaluate the area under the receiver operating characteristic (ROC) curve (AUC)~\citep{fogarty2005case}.  %$\hat{y}$ should only be interpreted as the relative risk score rather than the absolute risk score. % [Chatterjee et al., 2016]

\subsection{Hypothesis testing of pleiotropy}
%After estimating the parameters $\alpha$ about pleiotropy, we want to test whether there exists pleiotropy between two disease/traits. So we provide a hypothesis as following.
It is of great interest to quantify the significance of pleiotropy between two traits. Based on the definition of pleiotropy, the presence of pleiotropy means that the null and non-null groups in two traits are not distributed independently. Formally, we can set up a likelihood ratio test (LRT) as follows:
\begin{equation}
H_0: \alpha_{11} = \alpha_{1*}\alpha_{*1}, \qquad \mbox{vs.} \qquad H_a: \alpha_{11} \ne \alpha_{1*}\alpha_{*1}
\end{equation}
where $\alpha_{1*} = \alpha_{10} + \alpha_{11}$ and  $\alpha_{*1} = \alpha_{01} + \alpha_{11}$ are marginal probability. Clearly, a LRT statistic is
\begin{equation}
\lambda = 2\left(\log \pr(\bfy_1,\bfy_2|\bfX_1,\bfX_2;\widehat{\boldsymbol{\Theta}}) - \log \pr(\bfy_1,\bfy_2|\bfX_1,\bfX_2;\widehat{\boldsymbol{\Theta}}_0)\right),% \frac{\text{Pr}(\bfy_1,\bfy_2|\bfX_1,\bfX_2;\widehat{\boldsymbol{\Theta}}_0
%)}{\text{Pr}(\bfy_1,\bfy_2|\bfX_1,\bfX_2;\widehat{\boldsymbol{\Theta}})}
\end{equation}
where $\widehat{\boldsymbol{\Theta}}_0$ and $\widehat{\boldsymbol{\Theta}}$ denote the parameters estimated under the null and alternative hypotheses respectively. Due to the intractability of marginal distribution (\ref{marg_4GPMP}), we use the lower bound as a surrogate to approximate the marginal likelihood. Under the null hypothesis, the test statistic $\lambda$ approximately follows a $\chi^2$  distribution with $\mathrm{df}=1$. %As the probabilistic BG model is approximated by the lower bound, we adopt lower bound as the substitude likelihood here. According to the property of likelihood ratio test, $-2\log\lambda$ is asymptotically distributed as Chi-square distribution with degree of freedom 1.

\section{Simulation studies}
\label{simstudies}
In this section, we evaluated the performance of LPG approach described in the Section~\ref{method} using simulation studies. We examined its performance in risk variants identification as measured by AUC, statistical power and FDR and risk prediction as measured by Pearson's correlation and AUC for quantitative traits and binary traits, respectively. We compared its performance with two other methods that do not leverage pleiotropy, including the two-groups model (BVSR~\citep{guan2011bayesian}) and Lasso~\citep{friedman2010regularization}. %Then we compared the accuracy on risk prediction of LPG with that of BVSR and Lasso.
The number of replicates in simulation studies was 100 for all settings.
\subsection{Simulation settings}
The simulation data sets were generated as follows. The genotype matrices $\bfX_k$ ($k=1,2$) were first simulated from normal distribution, where autoregressive correlation (AR) $\rho^{|j-j'|}$ was set to mimic the linkage disequilibrium(LD) between variants $j$ and $j'$ with $\rho$ = 0.2, 0.5 and 0.7. Next, the entries of both $\bfX_k$ ($k=1,2$) were discretized to numerically coded genotypes $\{0,1,2\}$ according to the Hardy-Weinberg equilibrium based on the minor allele frequencies, which were drawn from  a uniform distribution on $[0.05,0.5]$. In all scenarios, the sample sizes for each study were set to $n_k = 3,000$ ($k=1,2$) and the number of variants were set to $p = 20,000$. To evaluate prediction performance, we generated additional $n_\mathrm{test}=500$ samples for each study under the same model. Denote the proportions of the null and non-null SNPs of both GWAS as $\alpha_0$ and $\alpha_1$, respectively. Then the hidden association status in first study ($\bfgamma_1$) could be sampled randomly with the number of nonzero entries -- $p\alpha_1$. $\alpha_1$ is set to 0.005 for the quantitative traits and 0.0025 for the binary traits. To account pleiotropy between two GWAS, we controlled the number of SNPs with pleiotropic effects as $p\alpha_1(\alpha_1+g\alpha_0)$. Clearly, $g=0$ and $g=1$ refer to two extreme cases -- no pleiotropy and full pleiotropy, respectively. We considered $g = 0$ to 1 equally spaced by 0.2. Next, effect sizes $\bfbeta$ were simulated from $\mathcal{N}(0,1)$. For the quantitative trait, as heritability of each study is defined as $h_k^2 = \frac{\mathrm{Var}(\bfX_k\bfbeta_k \bfgamma_k)}{\mathrm{Var}(\bfX_k\bfbeta_k \bfgamma_k) + \sigma_{e_k}^2}$, the noise level was chosen to control heritability at 0.3, 0.4 and 0.5. For the binary trait, we set population prevalence to be 0.1 and case-control ratio to be 0.5 while controlling heritability at 0.3, 0.4 and 0.5 using liability model~\citep{lee2011estimating}.

\subsection{Simulation results}
For both the quantitative and binary traits, we analyzed the simulated data using the proposed LPG jointly on two traits in comparison with other alternative methods, including BVSR and Lasso on each separate trait. For probabilistic approaches, LPG and BVSR, we evaluated their performance of risk variants identification using the area under the receiver operating characteristic (ROC) curve (AUC), statistical power, and false discovery rate (FDR). Note that for all settings, we evaluated statistical power to identify risk variants under the global FDR controlled at 0.2. As Lasso is a deterministic approach and FDR is not controllable, there are no points to evaluate their statistical power. The tuning parameter in Lasso was chosen by using 5-fold cross-validation~\citep{friedman2001elements}. We evaluated the performance of risk prediction measured by Pearson's correlation between the observed phenotypic values and the predicted values in testing sets for the quantitative trait while AUC was used to measure the performance of classification accuracy for case-control studies.

\begin{figure}
  \centering
  \includegraphics[width=\textwidth]{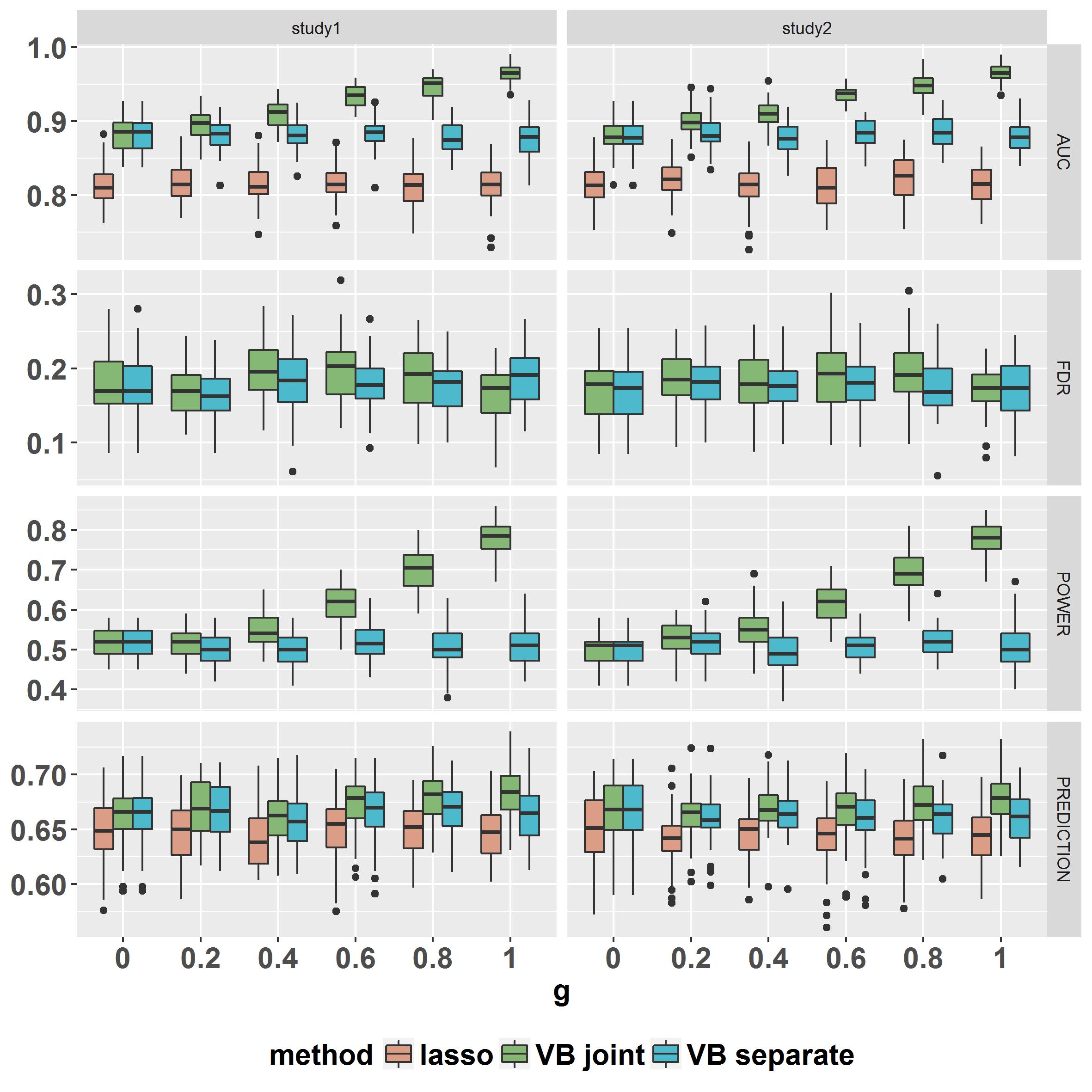}\\
  \caption{The comparison of LPG (VB joint) with its alternative methods, BVSR (VB separate) and Lasso, for quantitative traits demonstrated the increased power in the ascending order of pleiotropy $g$ while FDR of both LPG and BVSR were controlled at 0.2. Panels from top to bottom are AUC, FDR, Power and Prediction, respectively. Choices of $g$ range from 0 to 1. The parameter setting of the model is : $p$ = 20,000, $n_1=n_2$ = 3000, $h^2$ = 0.5, $\rho = 0.5$, $\alpha_1$ = 0.005.}
  \label{linrho5}
\end{figure}

\begin{figure}
  \centering
  \includegraphics[width=\textwidth]{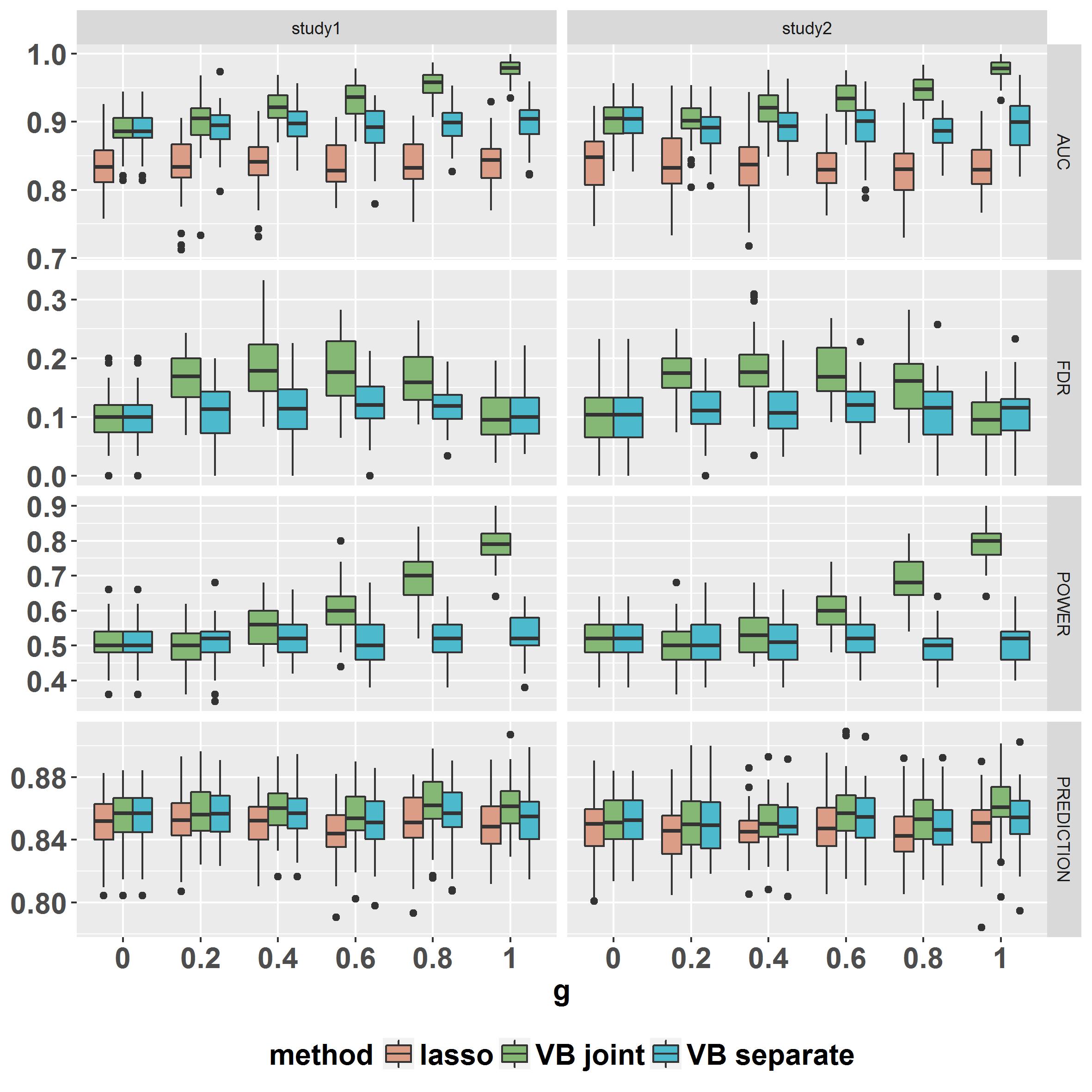}\\
  \caption{The comparison of LPG (VB joint) with its alternative methods, BVSR (VB separate) and Lasso, for binary traits demonstrated the increased power in the ascending order of pleiotropy $g$ while FDR of both LPG and BVSR were controlled at 0.2. Panels from top to bottom are AUC, FDR, Power and Prediction, respectively. Choices of $g$ range from 0 to 1. The parameter setting of the model is : $p$ = 20,000, $n_1=n_2$ = 3000, $h^2$ = 0.5, $\rho = 0.5$, $\alpha_1$ = 0.0025.}
  \label{logisrho5}
\end{figure}

For the quantitative traits, Figure \ref{linrho5} shows the performance of risk variants identification and prediction under $\rho=0.5$ and $h^2 = 0.5$.
%First, in terms of the performance of risk SNP identification, Figure \ref{linrho5} shows the result for autoregressive correlation $\rho= 0.5$ and heritability $h^2 = 0.5$.
It demonstrates that LPG incorporating the pleiotropy between two traits can actually improve the risk SNP identification over two-groups model (BVSR) in general. In particular, when there is no pleiotropy ($g$=0), the performance of LPG is the same as two-groups model (BVSR) suggesting that LPG can explore available pleiotropic information automatically. Another observation is that the performance of risk SNP identification (AUC and statistical power) of LPG improves in the ascending trend with the proportion of shared risk SNPs. Additionally, probablistic approaches (LPG and BVSR) outperform the Lasso in terms of risk SNP identification in the presence of pleiotropy or not as Lasso does not leverage pleiotropy between two traits and its performance depends on the extend of sparsity and strong signals. The FDR of both probabilistic models (LPG and BVSR) is well controlled by our target, 0.2. In terms of prediction performance, as pleiotropy becomes stronger, the Pearson's correlation coefficients between the observed and predicted phenotypic values in LPG increase slightly over the BVSR.
For the binary traits, the results with similar patterns are shown in Figure \ref{logisrho5} under $\rho=0.5$ and $h^2 = 0.5$. First, the improved AUC and statistical power of LPG are in the ascending order of the strength of pleiotropy while global FDR of LPG and BVSR is well under control. The prediction performance of LPG slightly improves over the BVSR when pleiotropy is strong. In our simulation studies, we found the performance of Lasso is worse than its probabilistic counter part, BVSR. The similar pattern can be found in~\citep{dai2017igess}. More simulation results under different settings (Tables S1 and S2) are shown in Figures S1 - S18 in the supplementary document. %The similar pattern can be found there.

We evaluated the type 1 error and power of the hypothesis test for pleiotropy. As shown in Figure S19 in the supplementary document, the power increases with the ascending trend of pleiotropy ($g$) and the similar performance among different $\rho$ for both the quantitative and binary traits can be found there. Type 1 error for different choice of $\rho$ is shown in Figure S20 in the supplementary document. One can find that type 1 error is a little conservative for both the quantitative and binary traits.

% As for the power to identify the true risk SNPs, the simulation result in
% Figure \ref{linrho5} shows that the FGP can actually improve the statistical power over two-group model (BVSR) if there exists pleiotropic information, which is also true for binary trait model simulation.
%As shown in Figure\ref{logisrho5}, the improved statistical power of FGP is in the ascending order of the strength of pleiotropy.
%We have to note that the quantitative trait model have more power to identify risk SNPs than binary trait model in the same model setting because its phenotype have more information.
%We also calculated the actual FDR and it is indeed controlled at 0.2 in both FGP and two-group model (BVSR). Binary trait model simulation has the similar result in Figure \ref{logisrho5}.
%For the risk SNPs prediction, incorporating the pleiotropy between two traits can slightly improve the risk prediction in both quantitative and binary traits model in Figure \ref{linrho5} and \ref{logisrho5}. Lasso show lower risk SNPs prediction accuracy compared with FGP.

\section{Real data analysis}
\label{realdata}
Crohn's disease (CD), rheumatoid arthritis (RA) and type-I diabetes (T1D) are autoimmune diseases and previous work suggests they can share common genetic risk variants~\citep{solovieff2013pleiotropy}.
We applied LPG for the analysis of CD, RA and T1D uisng data reported by the Welcome Trust Case Control Consortium (WTCCC)~\citep{burton2007genome}. The dataset consists of approximately 2,000 cases for all CD, RA and T1D and 3,000 shared controls, with genotypes at 500,568 SNPs. We performed strict quality control data using plink~\citep{purcell2007plink}. First, we removed individuals with missing genotypes call-rates greater than 2$\%$. For cases from each disease and samples from each control data set, we removed SNPs with minor allele frequencies smaller than 5$\%$ and SNPs with missing rate greater than 1$\%$. We further excluded SNPs with $p$-value $<0.001$ in Hardy-Weinberg equilibrium test for samples in control groups. In addition, pairs of subjects with estimated relatedness exceeding 0.025$\%$ were identified and one individual from each pair was removed at random by GCTA~\citep{yang2011gcta}.

\subsection{RA and T1D}
\begin{figure}
    \centering
    \begin{subfigure}{0.24\textwidth}
        \centering
        \includegraphics[width=\textwidth]{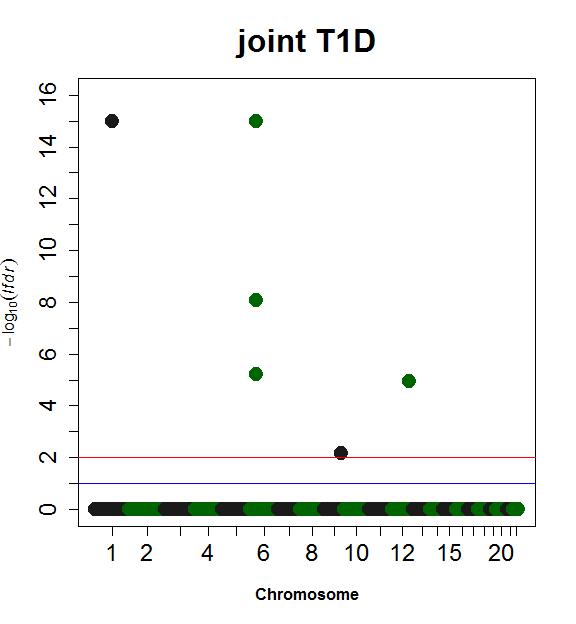}
    \end{subfigure}
    \begin{subfigure}{0.24\textwidth}
        \centering
        \includegraphics[width=\textwidth]{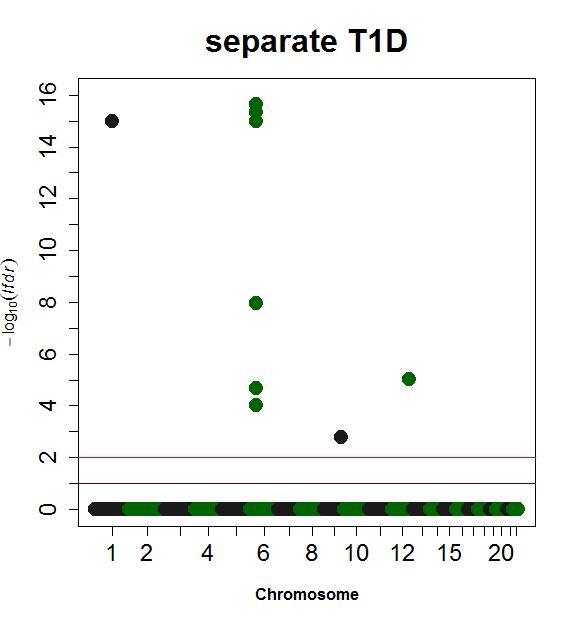}
    \end{subfigure}
    \begin{subfigure}{0.24\textwidth}
        \centering
        \includegraphics[width=\textwidth]{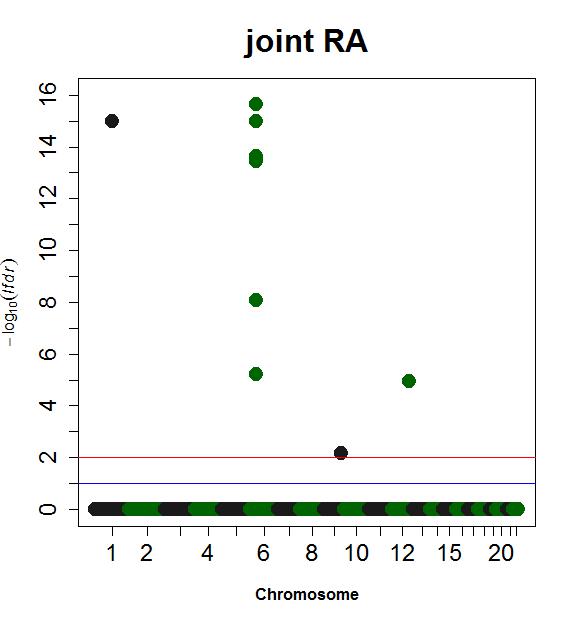}
    \end{subfigure}
    \begin{subfigure}{0.24\textwidth}
        \centering
        \includegraphics[width=\textwidth]{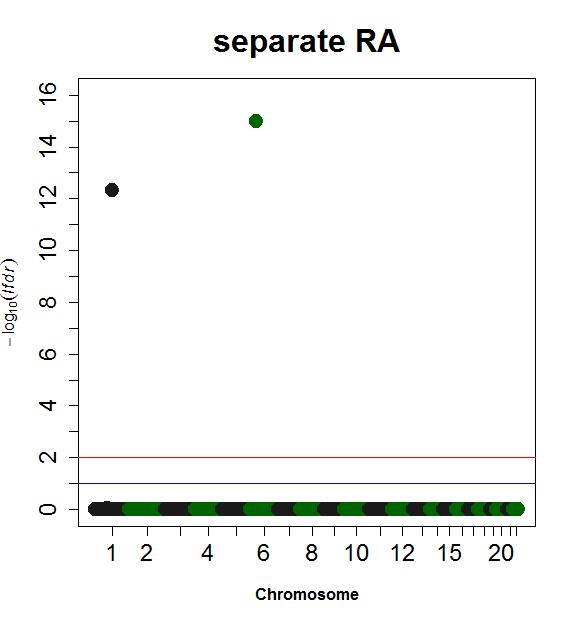}
    \end{subfigure}
   \caption{For the data consisting of 58C controls with RA and UKBS controls
with T1D, manhattan plots of separate analysis using BVSR and joint analysis using
LPG.}
\label{fig1}
\end{figure}

Since WTCCC used shared controls among seven diseases and samples in the control group were from two cohorts (the 1958 British Birth Cohort (58C) and UK Blood Services (UKBS)), we used one control cohort for RA and the other for T1D. After quality control filtering, 240,101 SNPs in 1,812 cases from RA, 1,932 cases from T1D, and 2,897 controls from two data sets were retained for the following analysis. First, we conducted the analysis for 58C controls with RA and UKBS controls with T1D using LPG and BVSR. The prioritization results are shown in Figure~\ref{fig1} in addition to a complete list of findings in Table~\ref{tab:02}, in which the cut off point is 0.2 in terms of lfdr. As shown in Table~\ref{tab:01}, LPG identified slightly more SNPs than these from BVSR. The estimated proportions were $\hat{\alpha}_{00} = 0.9999$, $\hat{\alpha}_{01} = 1.488 \times 10^{-10}$, $\hat{\alpha}_{10} = 1.499 \times 10^{-18}$, $\hat{\alpha}_{11} = 5.209 \times 10^{-05}$ and $p$-value for the pleiotropy test was $1.68\times 10^{-17}$ suggesting the existence of pleiotropy between RA and T1D. In summary, leveraging the pleiotropic effects enabled LPG to identify more risk SNPs over the single-trait analysis (BVSR). For example, rs6679677 within {\it PTPN22} gene and rs9272346 within MHC region exhibited higher significance using the joint analysis of LPG, which was previously reported~\citep{burton2007genome}. In addition, T1D-associated SNP rs9272346 was also found to be associated with RA in Pakistani patients~\citep{kiani2015genetic}, rs10484565 within {\it TAP2} gene was shown to confer risk enriched in the RA group~\citep{lee2008several}, and rs241427 near {\it TAP1/TAP2} was shown to have relatively strong associations in the collective interaction analysis of RA and T1D~\citep{woo2017collective}. We also evaluated the prediction performance using RA and T1D. Specifically, we quantitatively assessed the performance of risk prediction using 10-fold cross validation. The prediction accuracy of both LPG and BVSR is shown in Table~\ref{tab:01}. Clearly, it shows that the joint analysis of RA and T1D consistently outperformed the separate analysis of each study in terms of prediction accuracy, which improved from 62.8$\%$ to 64.4$\%$ for RA and from 76.7$\%$ to 78.3$\%$ for T1D as shown in Table~\ref{tab:01}. It took around 7 minutes to finish the joint analysis as shown in Table S3 in the supplementary document. %Figure~\ref{linrho5}{\color{red}{?}}.
To demonstrate the robustness of our LPG, we switched the control cohorts for RA and T1D and conducted the same analysis again and the results were similar. %To avoid the effect of MHC region harboring SNPs(chr6:25Mb-34Mb), we removed the SNPs in MHC region, leaving a total of 239,641 SNPs, and performed the same analysis above. In total, there are four comparisons, two include MHC region, two exclude MHC region. The results of rest three comparisons are detailed in the supplementary document.

\subsection{CD and T1D}
After the basic quality control filtering as described above, 240,393 SNPs remained in 1,675 cases from CD, 1,932 cases from T1D, and 2,895 controls from two data sets were used for analysis. After excluding the MHC region SNPs, leaving us a total of 239,931 SNPs, we performed the same four comparisons. Here we are presenting the one with 58C controls for CD and UKBS controls for T1D by excluding MHC region. The Manhattan plots are shown in Figure S33 and all SNP findings are shown in Table S17 in the supplementary document, where the threshold for lfdr is set to 0.2. Obviously, the LPG identified slightly more risk SNPs than these from two-groups model (BVSR). The estimated proportions were $\hat{\alpha}_{00} = 0.9999$, $\hat{\alpha}_{01} = 4.477 \times 10^{-5}$, $\hat{\alpha}_{10} = 2.058 \times 10^{-9}$ and $\hat{\alpha}_{11} = 3.898 \times 10^{-05}$, and $p$-values for the pleiotropy test was $2.96\times 10^{-2}$ suggesting the existence of pleiotropy between CD and T1D.
For example, rs11805303 in the {\it IL23R} gene was identified to be strongly associated with CD, which was previously reported~\citep{buzdugan2016assessing}. Another intergenic SNP rs2542151 identified by LPG was also reported to be significantly associated with CD~\citep{parkes2007sequence}. Besides, rs17234657 on chromosome 5 was identified to be associated with CD by both LPG and BVSR, which was also previously reported in~\citep{burton2007genome}. The prediction performance of both LPG and BVSR is shown in Table S16 in the supplementary document. The results demonstrate that the prediction of the jointly analysis of CD and T1D was slightly better than the separate analysis of each study, which improved from 58.1$\%$ to 58.7$\%$ for CD and from 60.1$\%$ to 60.3$\%$ for T1D. The rest results of other comparisons are detailed in the supplementary document and they were similar to the one we presented here.

\section{Conclusion}
\label{discussion}
In this article, we proposed a novel statistical framework for leveraging pleiotropy in GWAS data. %As a genetic phenomenon reported more than 100 years ago, pleiotropy plays an essential role to make an efficient use of available GWAS data.
Compared with a single-trait-based analysis that does not leverage pleiotropy, LPG offers improved statistical power and prediction accuracy in the identification of risk variants. %in the presence of pleiotropy. Although the four-groups models have been developed to analyze the summary statistics from GWAS~\citep{chung2014gpa, liu2016eps}, its usage on individual-level data is still missing, which is mainly due to the computational inscalability.
In this article, an efficient algorithm based on VBEM was developed, which not only enables us to evaluate the posterior quantities of interest, but also makes it scalable to ultra-high dimensions. These advantages make LPG a powerful tool to analyze GWAS data exhibiting pleiotropic effects. In this article, we analyzed two pairs of traits from WTCCC, namely, RA vs T1D and CD vs T1D. The findings are consistent with the recent advances in biology and genetics.

Despite these advantages, modelling pleiotropic effects in a combinatorial fashion limits the usage of LPG to no more than two traits as the number of hidden association status increases exponentially with the number of traits. This feature limits the use of LPG for analysis of two traits. LPG was designed to leverage pleiotropy when GWAS data for multiple traits is collected in different study individuals and complements the earlier methods proposed for incorporating pleiotropy when GWAS data is collected in the same study individuals~\citep{zhou2014efficient, liu2016analyzing}. However, to date no method has been proposed for leveraging pleiotropy when the GWAS data for multiple traits is collected from partially shared study samples, and is an avenue for future work. %Although LPG is capable of analyzing multiple traits from different cohorts, there are many cases that a single cohort have multiple measurement on traits. It remains elusive to conduct the joint analysis of mixture of shared cohorts and independent cohorts.
%Analyzing multiple traits using individual-level GWAS data will be left for future work.
\section{Acknowledgement}
This work was supported in part by grant NO. 61501389 from National Science Funding of China, grants No.22302815, NO. 12316116 and NO. 12301417 from the Hong Kong Research Grant Council, and grant R-913-200-098-263 from Duke-NUS Graduate Medical School, and AcRF Tier 2 (MOE2016-T2-2-029) from Ministry of Eduction, Singapore.

\clearpage

\clearpage

%\begin{figure}
%  \centering
%    \includegraphics[width=0.5\textwidth]{Figure/logispreT1D-RA}
%  \caption{For the data consisting of 58C controls with RA and UKBS controls
%with T1D, prediction performance of separate analysis using BVSR and joint analysis using FGP.}
%\label{fig:03}
%\end{figure}

\begin{table}
\centering
\begin{tabular}{rlrrllll}
  \hline
 & snp & chr & position & sep T1D(fdr) & sep RA(fdr) & joi T1D(fdr) & joi RA(fdr) \\
  \hline
1 & rs6679677 &   1 & 114303808 & 0e+00* & 4.66e-13* & 0e+00* & 0e+00* \\
  2 & rs13200022 &   6 & 31098957 & 2.22e-16* & 1e+00 & 0e+00* & 2.29e-14* \\
  3 & rs550513 &   6 & 31920687 & 2.14e-05* & 9.96e-01 & 8.76e-09* & 8.76e-09* \\
  4 & rs3130287 &   6 & 32050544 & 0e+00* & 1e+00 & 0e+00* & 3.29e-14* \\
  5 & rs17421624 &   6 & 32066177 & 1.1e-08* & 0e+00* & 0e+00* & 0e+00* \\
  6 & rs9272346 &   6 & 32604372 & 0e+00* & 1e+00 & 0e+00* & 3.73e-14* \\
  7 & rs2070121 &   6 & 32781554 & 4.44e-16* & 1e+00 & 0e+00* & 2.22e-16* \\
  8 & rs10484565 &   6 & 32795032 & 0e+00* & 1e+00 & 0e+00* & 0e+00* \\
  9 & rs241427 &   6 & 32804414 & 1e-04* & 9.98e-01 & 6.1e-06* & 6.1e-06* \\
  10 & rs10759987 &   9 & 121364134 & 1.66e-03* & 1e+00 & 6.76e-03* & 6.76e-03* \\
  11 & rs17696736 &  12 & 112486818 & 9.86e-06* & 1e+00 & 1.18e-05* & 1.18e-05* \\
   \hline
\end{tabular}
\caption{For the data consisting of 58C controls with RA and UKBS controls
with T1D, list of SNPs of two modes: separate analysis and joint analysis. * denotes the local
fdr \textless 0.2, sep means separate, joi means joint.}
\label{tab:02}
\end{table}

\begin{table}
\centering
\begin{tabular}{rlll}
  \hline
 & Data & number of hits & prediction accuracy(AUC) \\
  \hline
1 & Type-I diabetes(T1D)joint & 11 & 78.3\%(2.9\%) \\
  2 & Rheumatoid arthritis(RA)joint & 11 & 64.4\%(1.8\%) \\
  3 & Type-I diabetes(T1D)separate & 11 & 76.7\%(2.9\%) \\
  4 & Rheumatoid arthritis(RA)separate & 2 & 62.8\%(2.4\%) \\
   \hline
\end{tabular}
\caption{For the data consisting of 58C controls with RA and UKBS controls
with T1D, summary of separate and joint analysis of T1D and RA}
\label{tab:01}
\end{table}
%
%\clearpage
%\begin{figure}
%    \centering
%    \begin{subfigure}{0.4\textwidth}
%        \centering
%        \includegraphics[width=\textwidth]{Figure/linpower}
%    \end{subfigure}
%    \begin{subfigure}{0.4\textwidth}
%        \centering
%        \includegraphics[width=\textwidth]{Figure/logpower}
%    \end{subfigure}
%   \caption{Power of pleiotropy test for both quantitative (left panel) and binary (right panel) trait. $\rho$ is chosen to be {0.2, 0.5, 0.7}, $h_2$ is 0.5 and the pleiotropy parameter $g$ is controlled at {0.1, 0.2, 0.3}. The number of replicates is 100.}
%\label{linpleiotropy}
%\end{figure}
%
%
%\begin{figure}
%    \centering
%    \begin{subfigure}{0.4\textwidth}
%        \centering
%        \includegraphics[width=\textwidth]{Figure/lintype1error}
%    \end{subfigure}
%    \begin{subfigure}{0.4\textwidth}
%        \centering
%        \includegraphics[width=\textwidth]{Figure/logtype1error}
%    \end{subfigure}
%  \caption{Type I error of pleiotropy test for quantitative (left panel) and binary (right panel) trait. $\rho$ is chosen to be {0.2, 0.5, 0.7} and $h_2$ is 0.5. The number of replicates is 500.}
%\label{logpleiotropy}
%\end{figure}

\rowcolors{2}{gray!25}{white}
\begin{table}
\centering
\begin{tabular}{rlllll}
  \hline
 &  $\widehat{\alpha}_{01}$ & $\widehat{\alpha}_{10}$ & $\widehat{\alpha}_{11}$ & LRT & p-value \\
  \hline
  %\rowcolor{gray!50}
  CD-T1D-inMHC &  5.49e-05 & 6.17e-05 & 4.71e-08 &  2.27e-05 & 1.00e+00 \\
  RA-T1D-inMHC &  7.98e-13 & 5.68e-11 & 5.19e-05 &  1.03e+02 & 2.75e-24 \\
  T1D-CD-inMHC &  4.97e-05 & 4.40e-05 & 1.83e-05 & -8.87e-02 & 1.00e+00 \\
  T1D-RA-inMHC &  1.49e-10 & 1.50e-18 & 5.21e-05 &  7.25e+01 & 1.68e-17 \\
  CD-T1D-exMHC &  1.10e-11 & 4.41e-05 & 3.93e-05 &  8.22e+00 & 4.13e-03 \\
  RA-T1D-exMHC &  6.16e-06 & 2.64e-14 & 1.47e-05 &  2.33e+01 & 1.38e-06 \\
  T1D-CD-exMHC &  4.48e-05 & 2.06e-09 & 3.90e-05 &  4.73e+00 & 2.96e-02 \\
  T1D-RA-exMHC &  5.99e-17 & 2.01e-05 & 1.09e-05 &  2.07e+01 & 5.29e-06 \\
   \hline
\end{tabular}
\caption{Pleiotropy estimated and inference, inMHC means include MHC region and exMHC means exclude MHC region}
\label{pleiotropytesttable}
\end{table}
\bibliographystyle{plainnat}
%\bibliographystyle{abbrv}

%\bibliographystyle{abbrv}
%
%\bibliography{Document}

\bibliography{4group_ref}

\end{document}

\setlength{\textheight}{575pt}
\setlength{\baselineskip}{23pt}

\title{Supplementary information for ``FGP: a four-group probabilistic model to account for pleiotropic effects for multiple GWAS" }
\author[1]{Yi Yang}
\affil[1]{School of Statistics and Management, The Shanghai University of Finance and Economics, Shanghai}
\author[2,4]{Mingwei Dai}
\affil[2]{Institute for Information and System Sciences, Xian Jiaotong University, Xian}
\author[3]{Jian Huang}
\affil[3]{Department of Applied Mathematics, Hong Kong Polytechnics University}
\author[5]{Xinyi Lin}
\author[4]{Can Yang}
\affil[4]{Department of Mathematics, Hong Kong University of Science and Technology}
\author[5,*]{Jin Liu}
\affil[5]{Centre for Quantitative Medicine, Duke-NUS Medical School}
\author[1,*]{Min Chen}

\maketitle
\clearpage
\tableofcontents
\clearpage
\section{Simulation studies: Quantitative trait}
\begin{figure}[H]
    \centering
        \includegraphics[width=\textwidth]{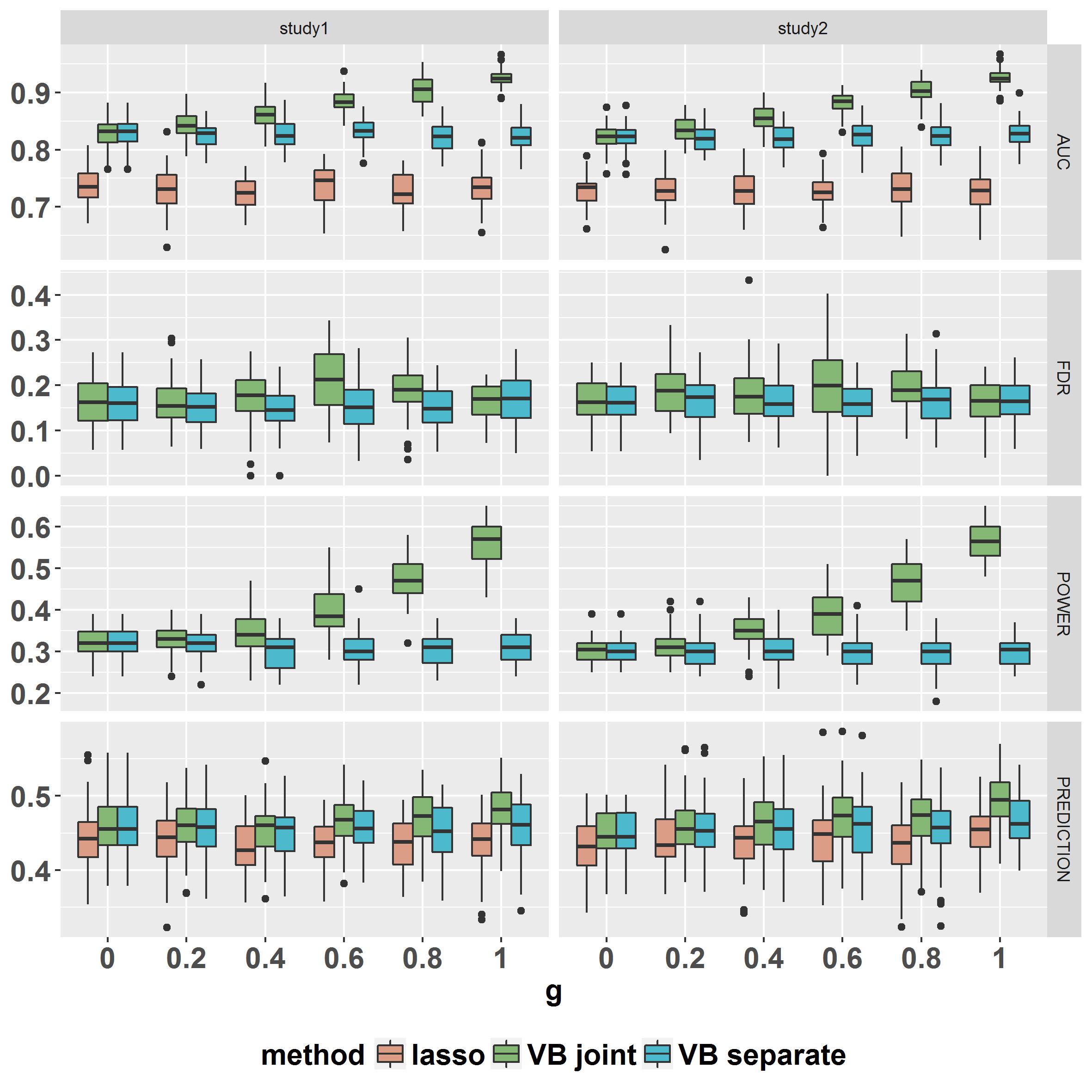}
        \caption{Comparison of FGP (VB joint), BVSR (VB separate), and Lasso for different choice $g$ ranging from 0 to 1 for quantitative trait. Panels from top to bottom are AUC, FDR, Power and Prediction, respectively. The parameter setting of the model is : $p$ = 20,000, $n_1=n_2$ = 3000, $h^2$ = 0.3, $\rho = 0.2$, $\alpha_1$ = 0.005.}
        \label{begin_quan_simulation}
\end{figure}

\begin{figure}[H]
    \centering
        \includegraphics[width=\textwidth]{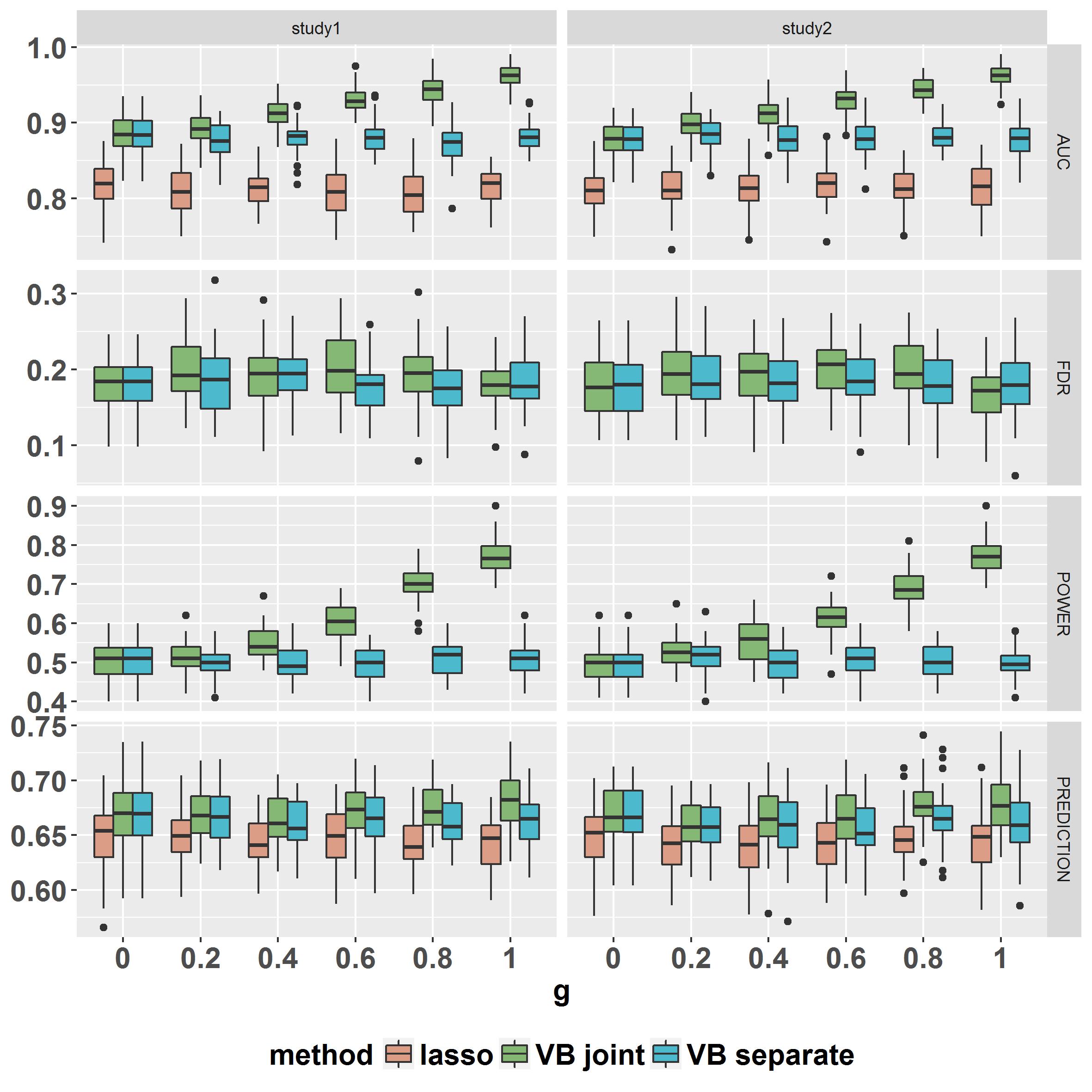}
        \caption{Comparison of FGP (VB joint), BVSR (VB separate), and Lasso for different choice $g$ ranging from 0 to 1 for quantitative trait. Panels from top to bottom are AUC, FDR, Power and Prediction, respectively. The parameter setting of the model is : $p$ = 20,000, $n_1=n_2$ = 3000, $h^2$ = 0.5, $\rho = 0.7$, $\alpha_1$ = 0.005.}
        \label{end_quan_simulation}
        %\label{}
\end{figure}

For the quantitative trait, we conducted simulation studies to evaluate the performance of FGP with its alternative, \ie BVSR and Lasso. We first evaluated their performance of the identification of risk variants using AUC, statistical power, and FDR. Note that we can only evaluate AUC for Lasso and FDR was controlled at 0.2. We then evaluated the prediction performance by measuring the Pearson's correlation between observed phenotypic values and estimated phenotypic values in a testing set for all three methods. The number of replicates in simulation studies was 50 for all settings. The results are provided in Figure \ref{begin_quan_simulation}-\ref{end_quan_simulation} and they demonstrate that the performance of FGP gradually gets better as the pleiotropic effects become stronger.

\begin{table}
\center
\begin{tabular}{lcccccc}
\toprule
 & $h^2$=0.3 & $h^2$=0.4 & $h^2$=0.5  \\
\midrule
$\rho$=0.2   & \multicolumn{1}{@{\hspace{\tabcolsep}Figure S}l}{$1$} & \multicolumn{1}{@{\hspace{\tabcolsep}Figure S}l}{$4$} & \multicolumn{1}{@{\hspace{\tabcolsep}Figure S}l}{$7$} & \\
$\rho$=0.5   & \multicolumn{1}{@{\hspace{\tabcolsep}Figure S}l}{$2$} & \multicolumn{1}{@{\hspace{\tabcolsep}Figure S}l}{$5$} & \multicolumn{1}{@{\hspace{\tabcolsep}Figure S}l}{$8$} & \\
$\rho$=0.7   & \multicolumn{1}{@{\hspace{\tabcolsep}Figure S}l}{$3$} & \multicolumn{1}{@{\hspace{\tabcolsep}Figure S}l}{$6$} & \multicolumn{1}{@{\hspace{\tabcolsep}Figure S}l}{$9$} & \\
\bottomrule
\end{tabular}
\caption{A table of simulation settings for quantitative trait scenario, Figure S means Supplementary Figure}
\end{table}
\vspace{0.5cm}

\begin{table}
\center
\begin{tabular}{lcccc}
\toprule

  & $h^2$=0.3 & $h^2$=0.4 & $h^2$=0.5  \\
\midrule
$\rho$=0.2  & \multicolumn{1}{@{\hspace{\tabcolsep}Figure S}c}{$10$} & \multicolumn{1}{@{\hspace{\tabcolsep}Figure S}c}{$13$} & \multicolumn{1}{@{\hspace{\tabcolsep}Figure S}c}{$16$}  \\
$\rho$=0.5  & \multicolumn{1}{@{\hspace{\tabcolsep}Figure S}c}{$11$} & \multicolumn{1}{@{\hspace{\tabcolsep}Figure S}c}{$14$} & \multicolumn{1}{@{\hspace{\tabcolsep}Figure S}c}{$17$}  \\
$\rho$=0.7  & \multicolumn{1}{@{\hspace{\tabcolsep}Figure S}c}{$12$} & \multicolumn{1}{@{\hspace{\tabcolsep}Figure S}c}{$15$} & \multicolumn{1}{@{\hspace{\tabcolsep}Figure S}c}{$18$} \\
\bottomrule
\end{tabular}
\caption{A table of simulation settings for binary trait scenario, Figure S means Supplementary Figure}
\end{table}

\clearpage

%%%%%%%%%%%%%%%%%%%%%%%%%%%%%    h^2 = 0.3       %%%%%%%%%%%%%%%%%%%%%%%%%%%%%%%%%%%%%%%%%%%%%%%%%%%%%%%%%%%%%%%
\begin{figure}[H]
    \centering
        \includegraphics[width=\textwidth]{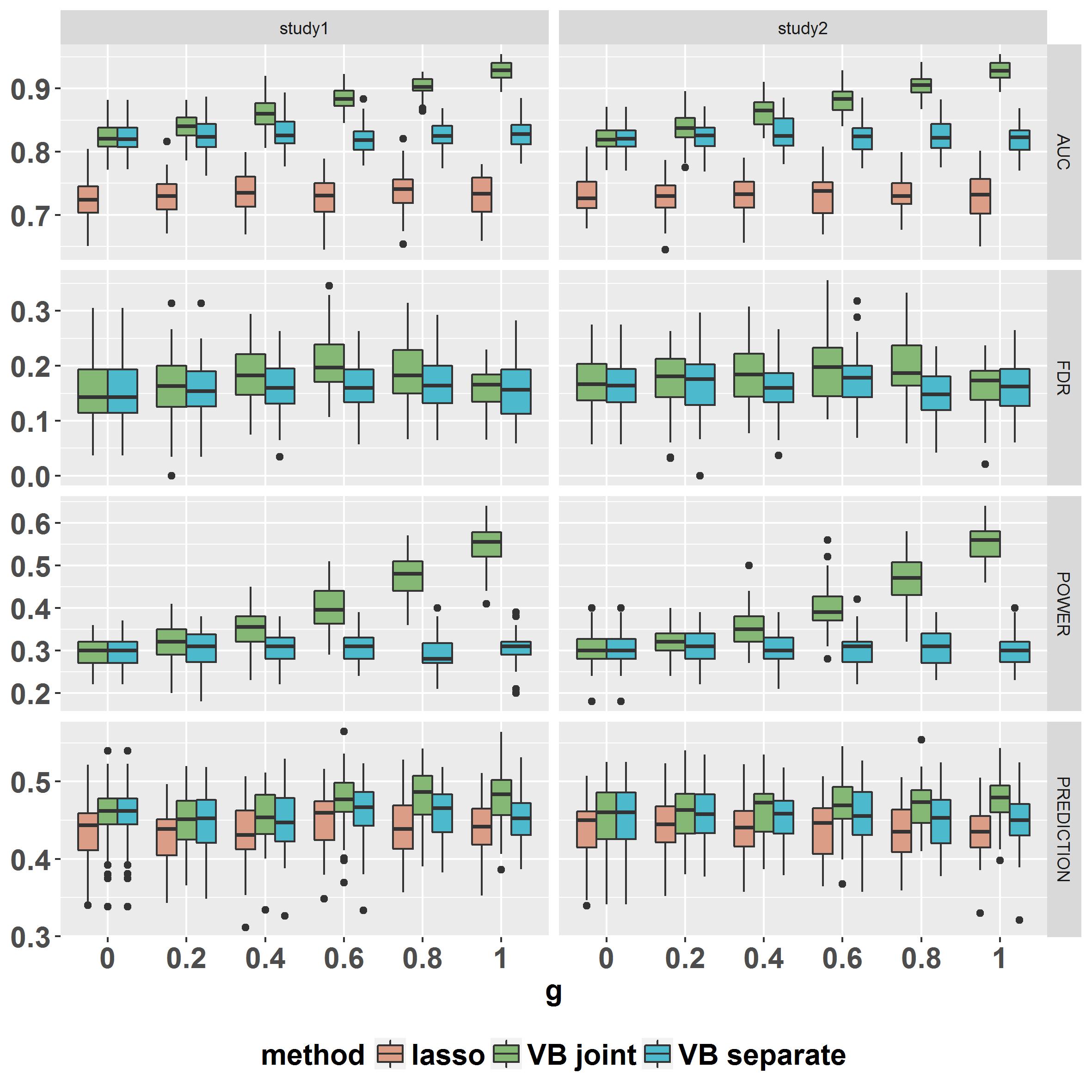}
        \caption{Comparison of FGP (VB joint), BVSR (VB separate), and Lasso for different choice $g$ ranging from 0 to 1 for quantitative trait. Panels from top to bottom are AUC, FDR, Power and Prediction, respectively. The parameter setting of the model is : $p$ = 20,000, $n_1=n_2$ = 3000, $h^2$ = 0.3, $\rho = 0.5$, $\alpha_1$ = 0.005.}
        %\label{}
\end{figure}
\begin{figure}[H]
    \centering
        \includegraphics[width=\textwidth]{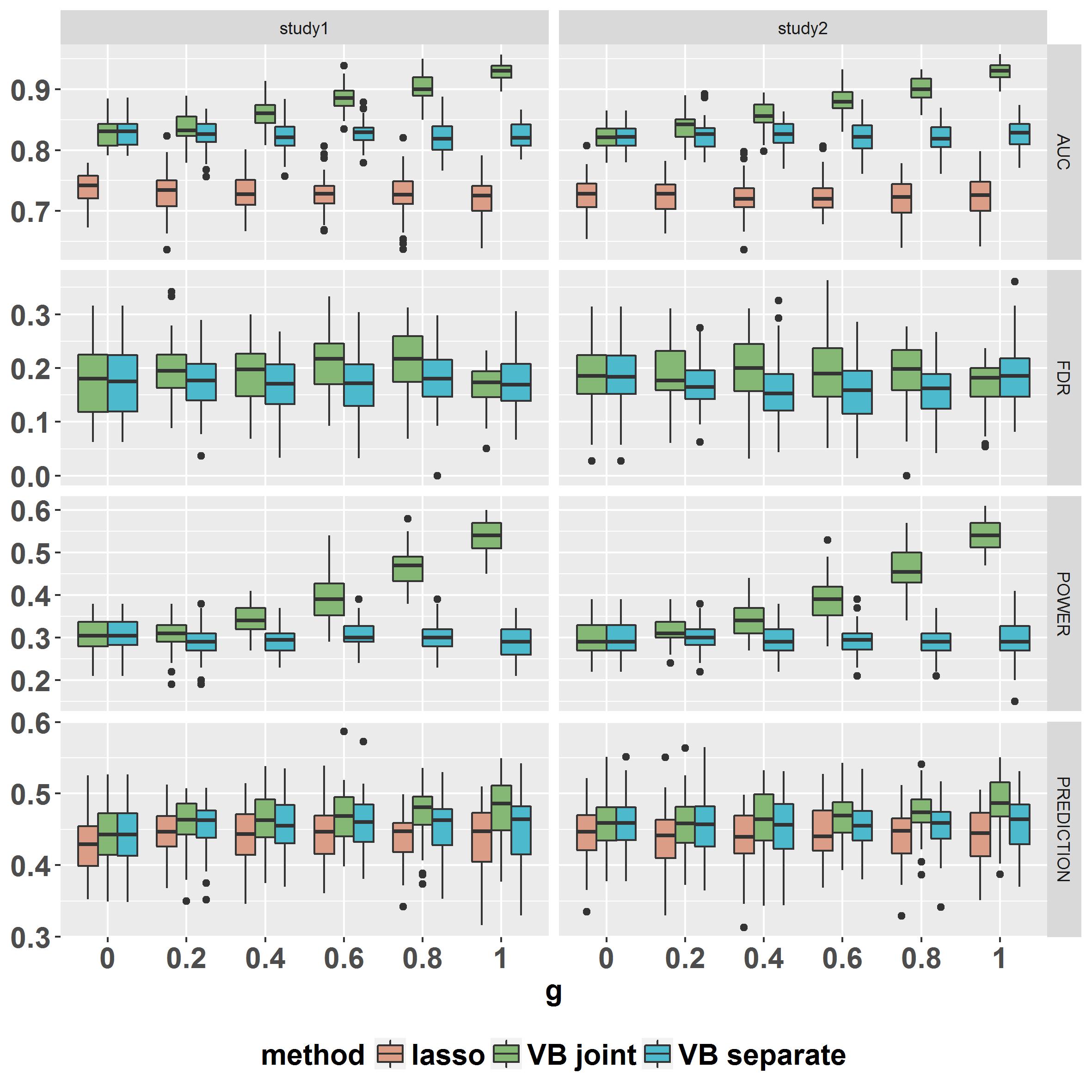}
        \caption{Comparison of FGP (VB joint), BVSR (VB separate), and Lasso for different choice $g$ ranging from 0 to 1 for quantitative trait. Panels from top to bottom are AUC, FDR, Power and Prediction, respectively. The parameter setting of the model is : $p$ = 20,000, $n_1=n_2$ = 3000, $h^2$ = 0.3, $\rho = 0.7$, $\alpha_1$ = 0.005. }
        %\label{}
\end{figure}

%%%%%%%%%%%%%%%%%%%%%%%%%%%%%%%%    h^2=0.4  %%%%%%%%%%%%%%%%%%%%%%%%%%%%%%%%%%%%%%%%%%%%%%%%%%%%%%%%%%%%%%%%
\begin{figure}[H]
    \centering
        \includegraphics[width=\textwidth]{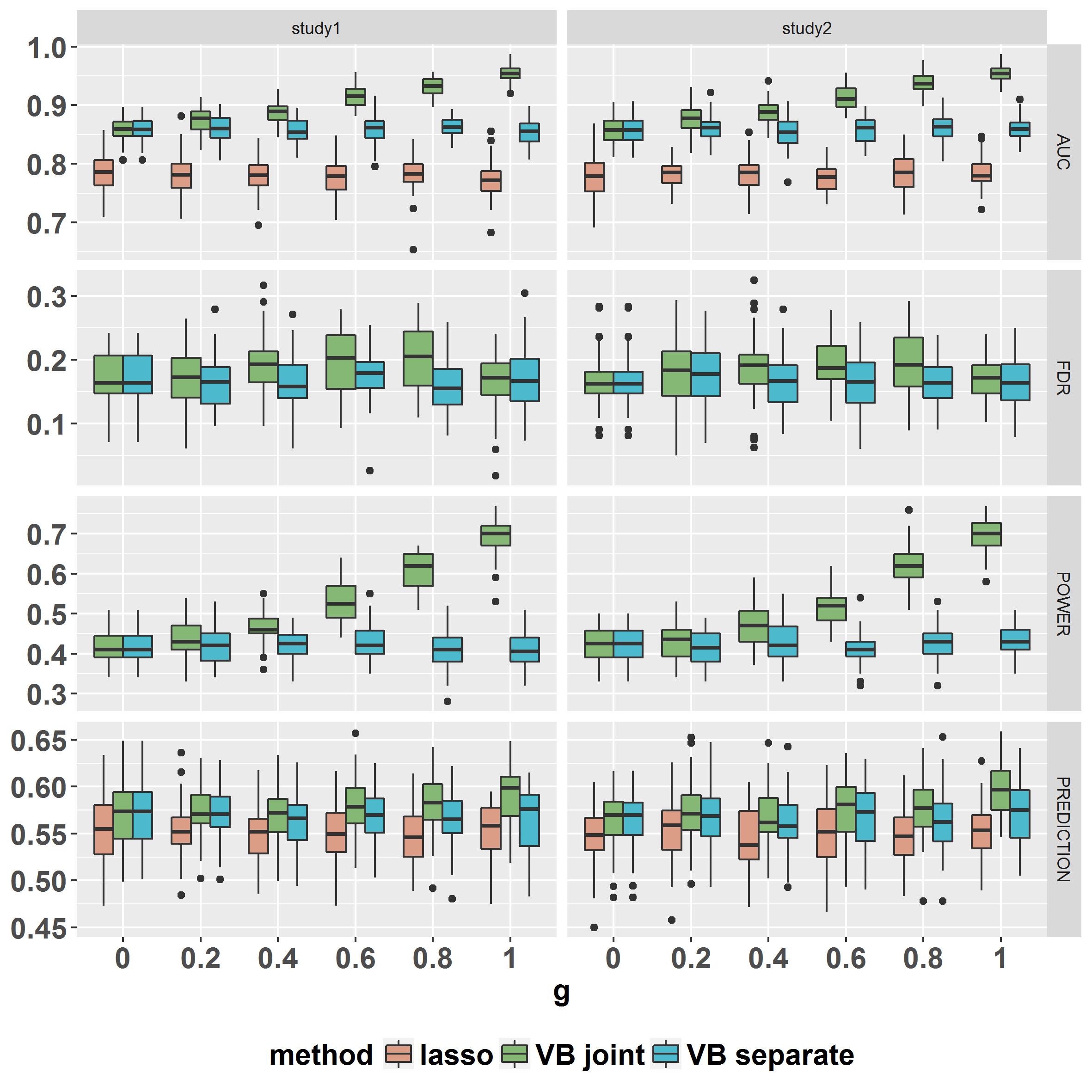}
        \caption{Comparison of FGP (VB joint), BVSR (VB separate), and Lasso for different choice $g$ ranging from 0 to 1 for quantitative trait. Panels from top to bottom are AUC, FDR, Power and Prediction, respectively. The parameter setting of the model is : $p$ = 20,000, $n_1=n_2$ = 3000, $h^2$ = 0.4, $\rho = 0.2$, $\alpha_1$ = 0.005.}
\end{figure}
\begin{figure}[H]
    \centering
        \includegraphics[width=\textwidth]{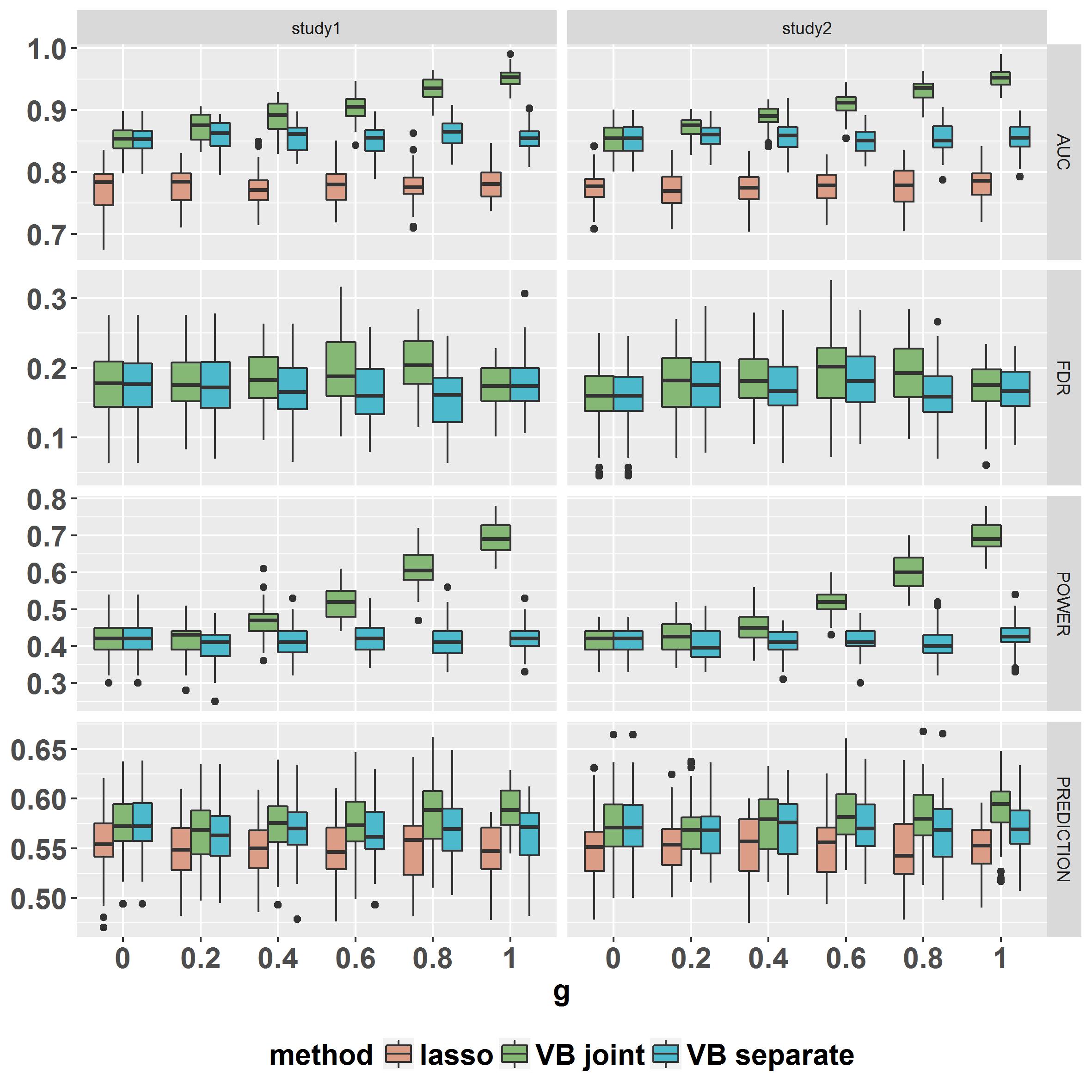}
        \caption{Comparison of FGP (VB joint), BVSR (VB separate), and Lasso for different choice $g$ ranging from 0 to 1 for quantitative trait. Panels from top to bottom are AUC, FDR, Power and Prediction, respectively. The parameter setting of the model is : $p$ = 20, 000, $n_1=n_2$ = 3000, $h^2$ = 0.4, $\rho = 0.5$, $\alpha_1$ = 0.005.}
        %\label{}
\end{figure}
\begin{figure}[H]
    \centering
        \includegraphics[width=\textwidth]{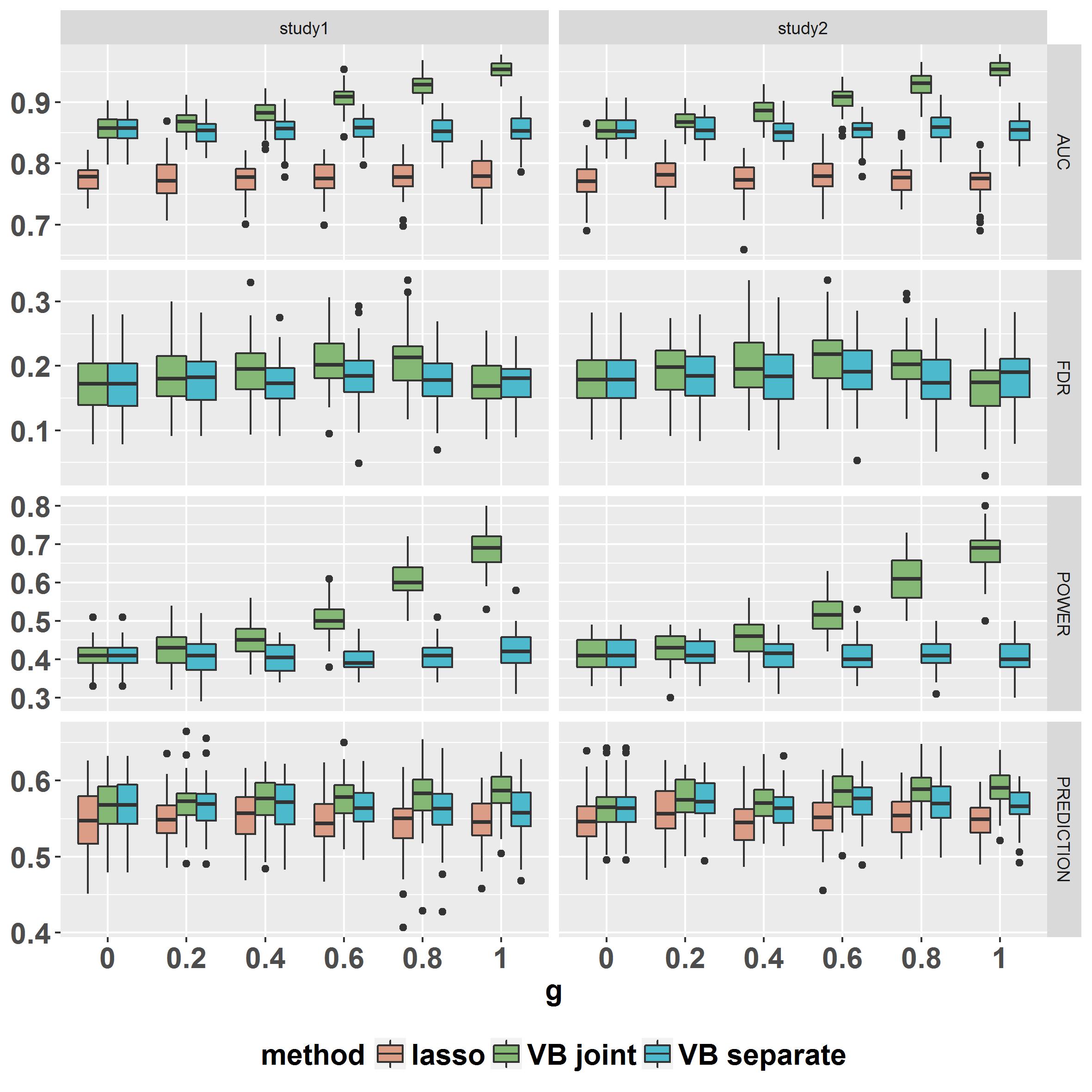}
        \caption{Comparison of FGP (VB joint), BVSR (VB separate), and Lasso for different choice $g$ ranging from 0 to 1 for quantitative trait. Panels from top to bottom are AUC, FDR, Power and Prediction, respectively. The parameter setting of the model is : $p$ = 20,000, $n_1=n_2$ = 3000, $h^2$ = 0.4, $\rho = 0.7$, $\alpha_1$ = 0.005.}
        %\label{}
\end{figure}

%%%%%%%%%%%%%%%%%%%%%%%%%%%%%%%    h^2=0.5  %%%%%%%%%%%%%%%%%%%%%%%%%%%%%%%%%%%%%%%%%%%%%%%%%%%%%%%%%%%%%%%%
\begin{figure}[H]
    \centering
        \includegraphics[width=\textwidth]{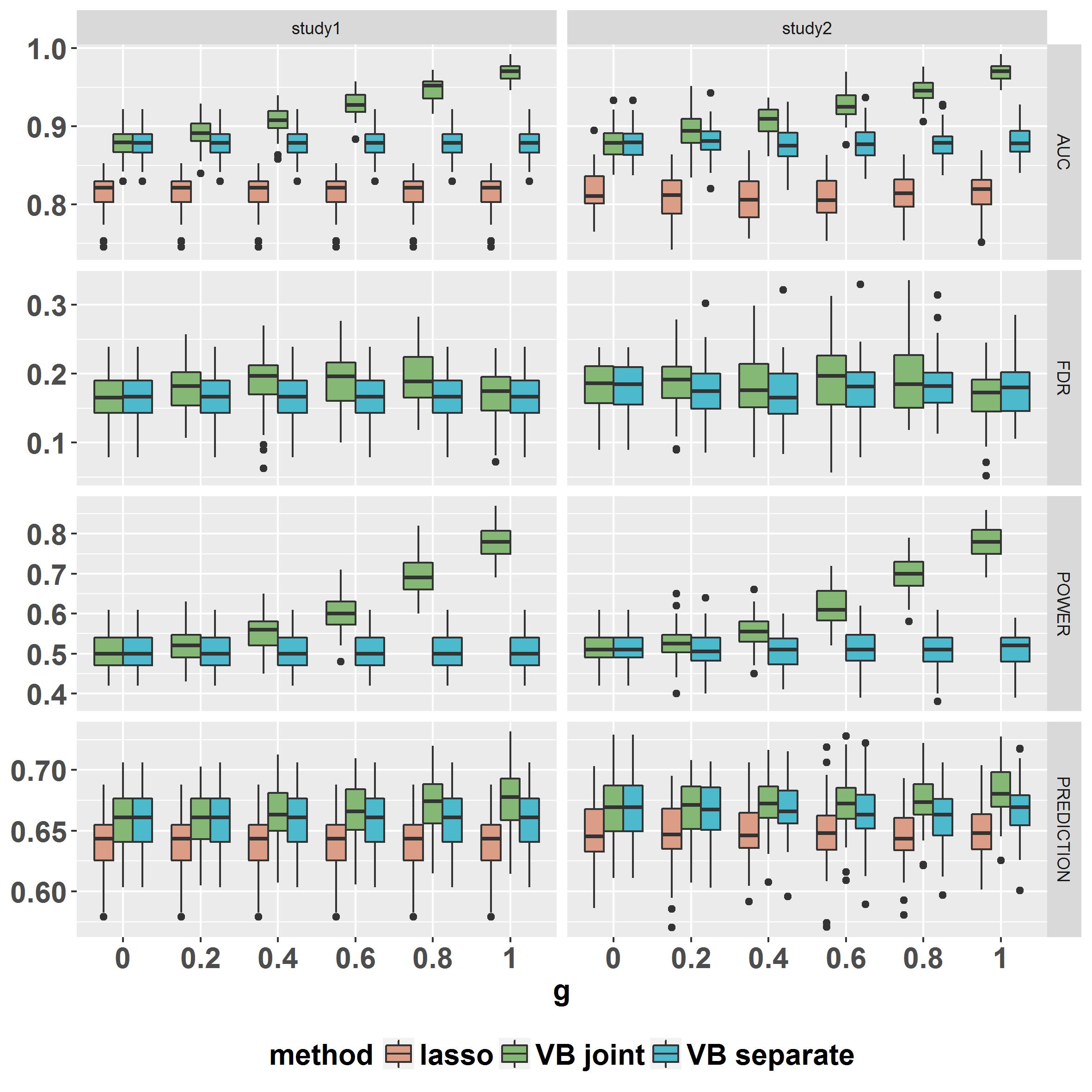}
        \caption{Comparison of FGP (VB joint), BVSR (VB separate), and Lasso for different choice $g$ ranging from 0 to 1 for quantitative trait. Panels from top to bottom are AUC, FDR, Power and Prediction, respectively. The parameter setting of the model is : $p$ = 20,000, $n_1=n_2$ = 3000, $h^2$ = 0.5, $\rho = 0.2$, $\alpha_1$ = 0.005.  }
        %\label{}
\end{figure}
\begin{figure}[H]
    \centering
        \includegraphics[width=\textwidth]{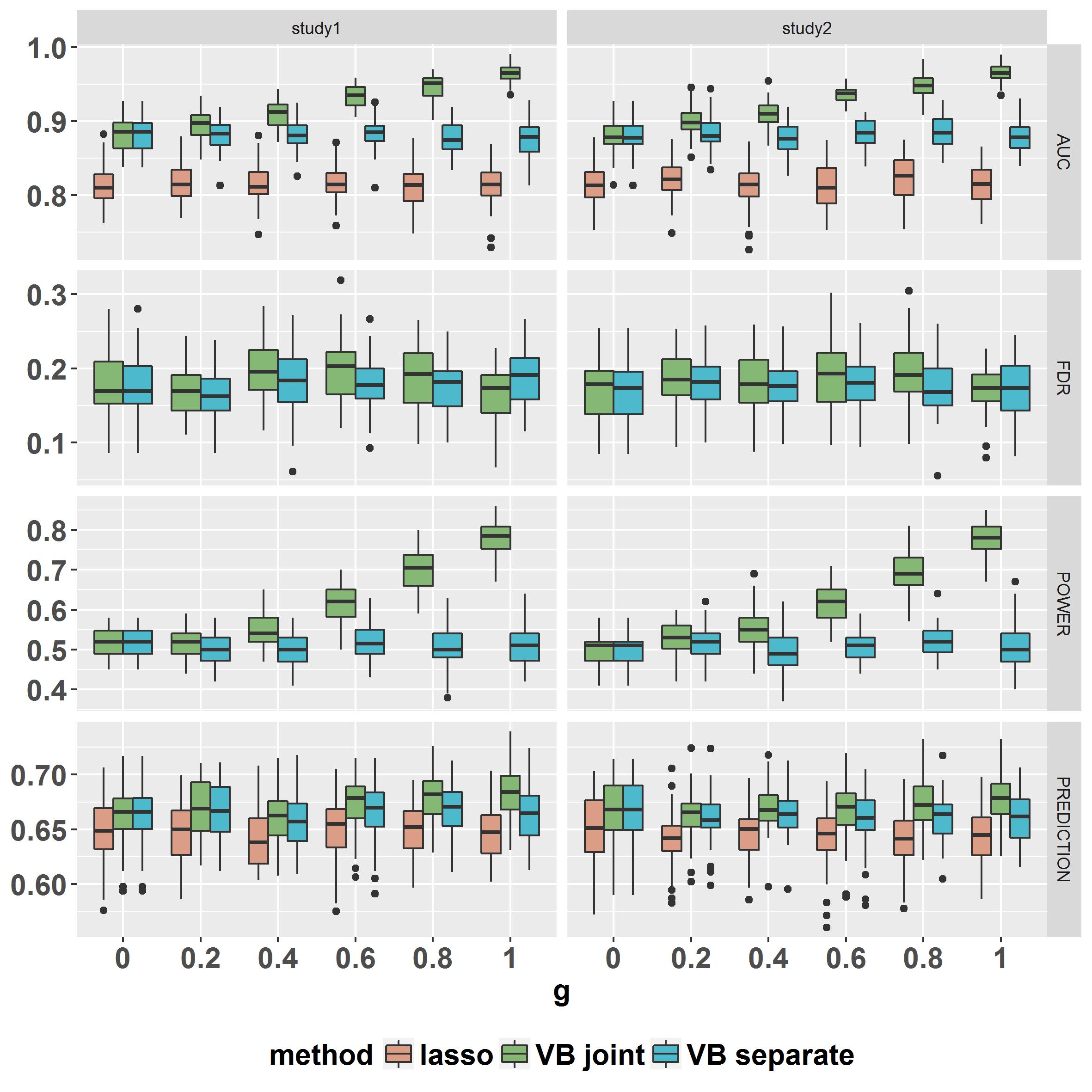}
        \caption{Comparison of FGP (VB joint), BVSR (VB separate), and Lasso for different choice $g$ ranging from 0 to 1 for quantitative trait. Panels from top to bottom are AUC, FDR, Power and Prediction, respectively. The parameter setting of the model is : $p$ = 20,000, $n_1=n_2$ = 3000, $h^2$ = 0.5, $\rho = 0.5$, $\alpha_1$ = 0.005.}
        %\label{}
\end{figure}

\clearpage
\section{Simulation studies: Binary trait}
For the binary trait in case-control studies,we conducted simulation studies to evaluate the performance of FGP with its alternative, \ie BVSR and Lasso. We first evaluated their performance of the identification of risk variants using AUC, statistical power, and FDR. Note that we can only evaluate AUC for Lasso and FDR was controlled at 0.2. Thirdly, we measure the proportion of true risk SNPs discovered by Power. We then evaluated the prediction performance in a testing set using AUC as described in Section 3.2 for all three methods The results are provided in Supplementary Figure \ref{begin_binary_simulation}-\ref{end_binary_simualtion} and they demonstrate that the performance of FGP gradually gets better as the pleiotropic
effects become stronger.
%%%%%%%%%%%%%%%%%%%%%%%%%%%%%% h^2 = 0.3  %%%%%%%%%%%%%%%%%%%%%%%%%%%%%%%%%%%%%%%%%%%%%%%%%%%%%%%%%%%%%%%%%%%%%%
\clearpage
\begin{figure}[H]
  \centering
  \includegraphics[width=\textwidth]{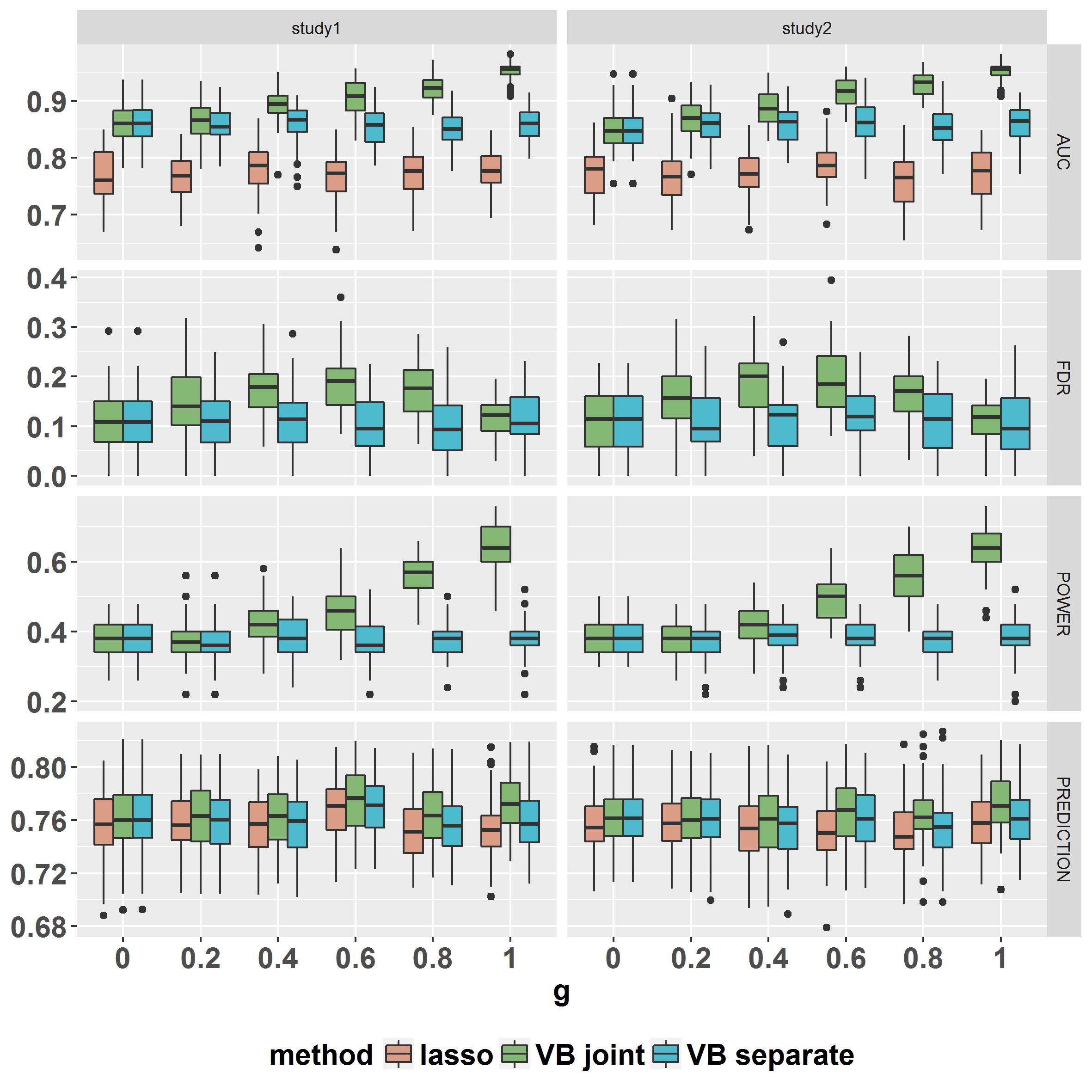}\\
  \caption{Comparison of FGP (VB joint), BVSR (VB separate), and Lasso with different $g$ ranging from 0 to 1 for binary trait. Panels from top to bottom are AUC, FDR, Power and Prediction, respectively.  The parameter setting of the model is : $p$ = 20,000, $n_1=n_2$ = 3000, $h^2$ = 0.3, $\rho = 0.2$, $\alpha_1$ = 0.0025.}
  \label{begin_binary_simulation}
\end{figure}
\begin{figure}[H]
  \centering
  \includegraphics[width=\textwidth]{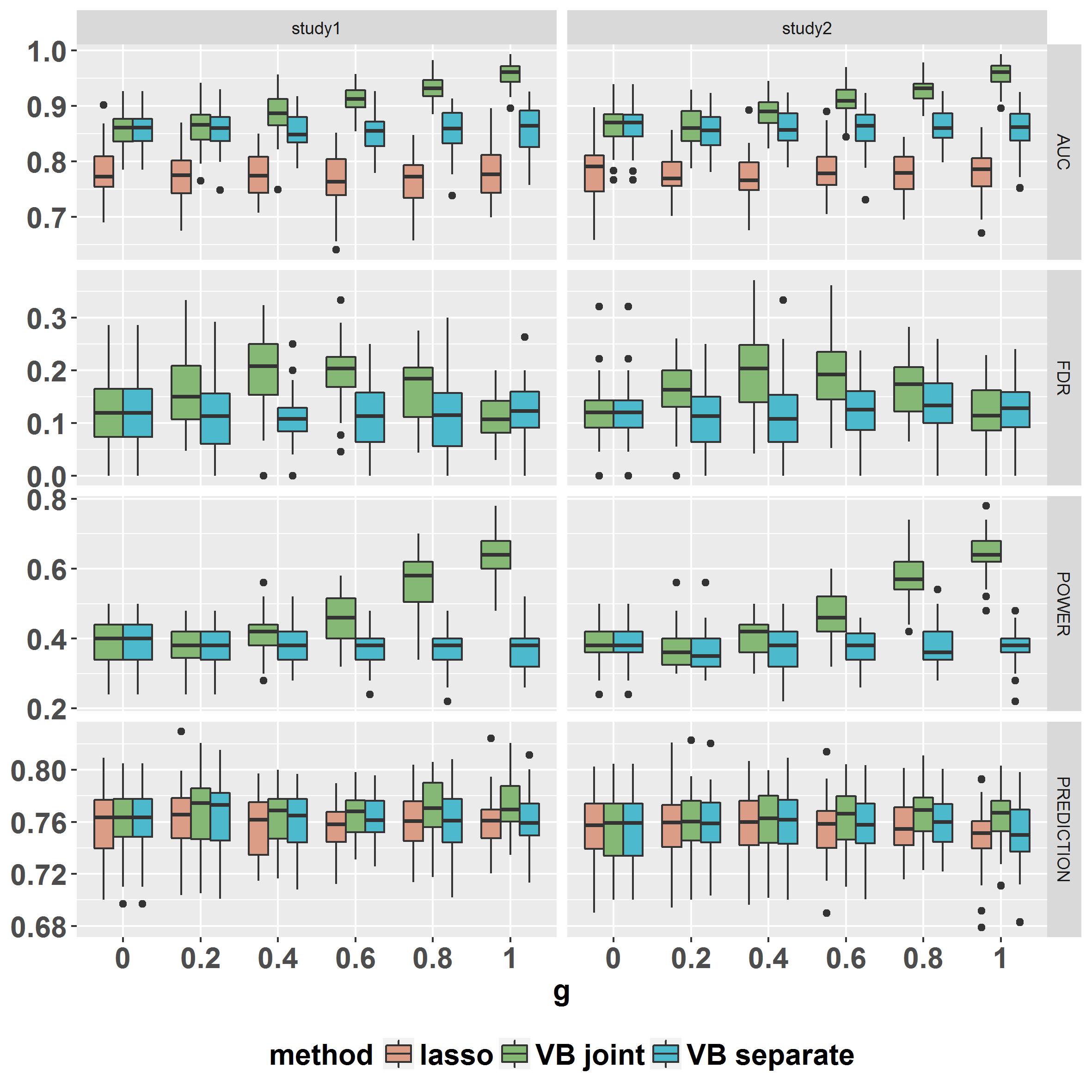}\\
  \caption{Comparison of FGP (VB joint), BVSR (VB separate), and Lasso with different $g$ ranging from 0 to 1 for binary trait. Panels from top to bottom are AUC, FDR, Power and Prediction, respectively.  The parameter setting of the model is : $p$ = 20,000, $n_1=n_2$ = 3000, $h^2$ = 0.3, $\rho = 0.5$, $\alpha_1$ = 0.0025.}
\end{figure}
\begin{figure}[H]
  \centering
  \includegraphics[width=\textwidth]{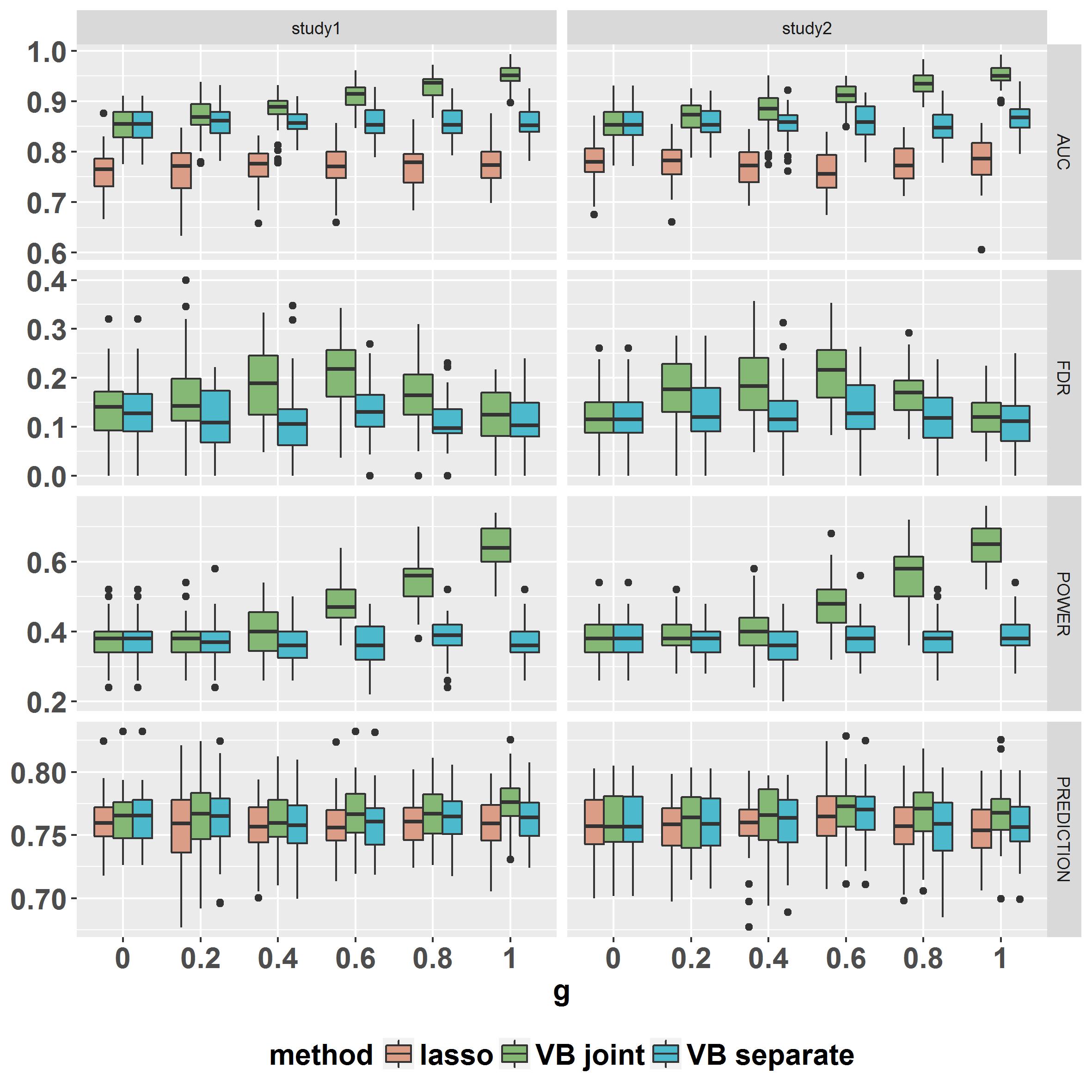}\\
  \caption{Comparison of FGP (VB joint), BVSR (VB separate), and Lasso with different $g$ ranging from 0 to 1 for binary trait. Panels from top to bottom are AUC, FDR, Power and Prediction, respectively. The parameter setting of the model is : $p$ = 20,000, $n_1=n_2$ = 3000, $h^2$ = 0.3, $\rho = 0.7$, $\alpha_1$ = 0.0025.}
\end{figure}

%%%%%%%%%%%%%%%%%%%%%%%%%%%%%% h^2 = 0.4  %%%%%%%%%%%%%%%%%%%%%%%%%%%%%%%%%%%%%%%%%%%%%%%%%%%%%%%%%%%%%%%%%%%%%%
\clearpage
\begin{figure}[H]
  \centering
  \includegraphics[width=\textwidth]{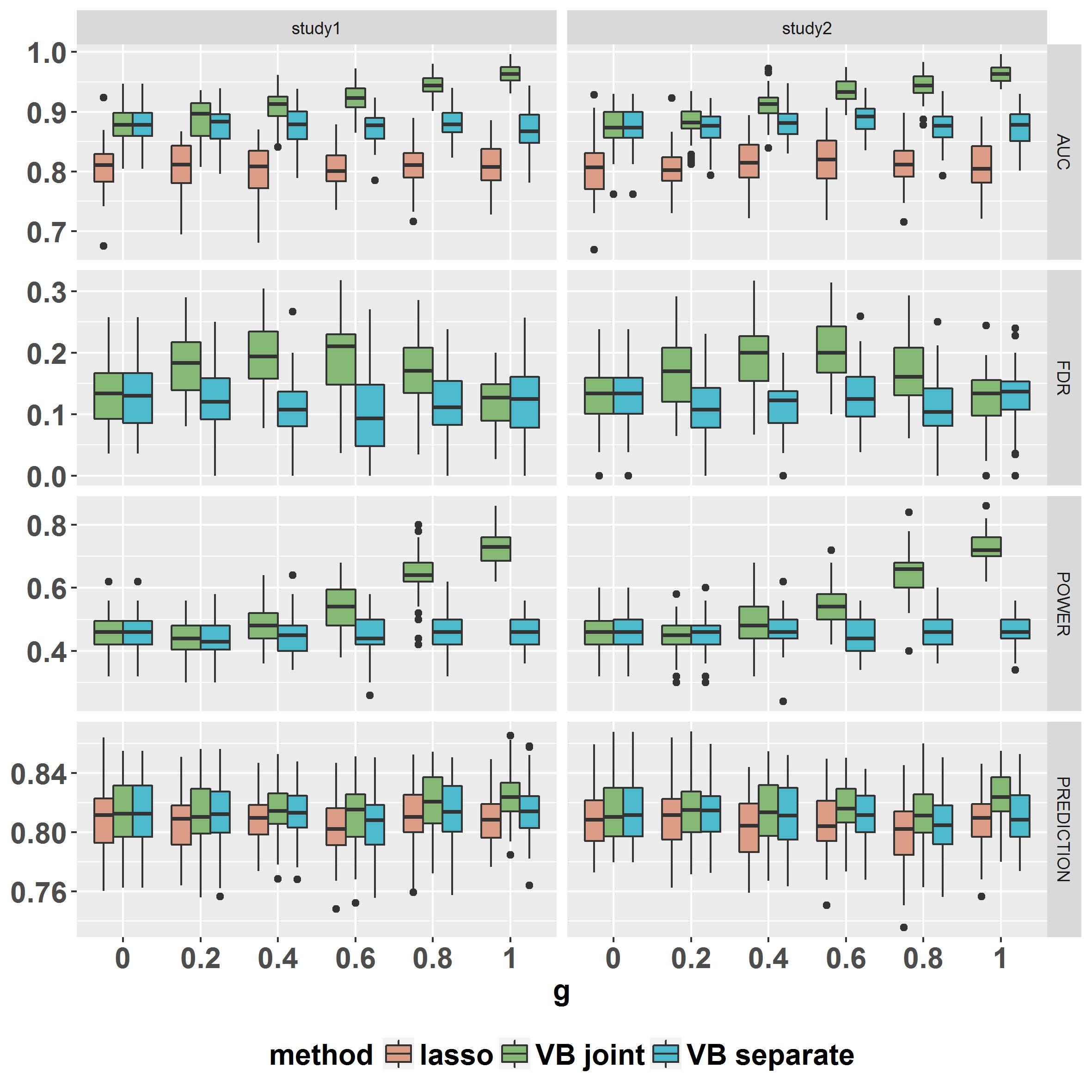}\\
  \caption{Comparison of FGP (VB joint), BVSR (VB separate), and Lasso with different $g$ ranging from 0 to 1 for binary trait. Panels from top to bottom are AUC, FDR, Power and Prediction, respectively.  The parameter setting of the model is : $p$ = 20,000, $n_1=n_2$ = 3000, $h^2$ = 0.4, $\rho = 0.2$, $\alpha_1$ = 0.0025.}
\end{figure}
\begin{figure}[H]
  \centering
  \includegraphics[width=\textwidth]{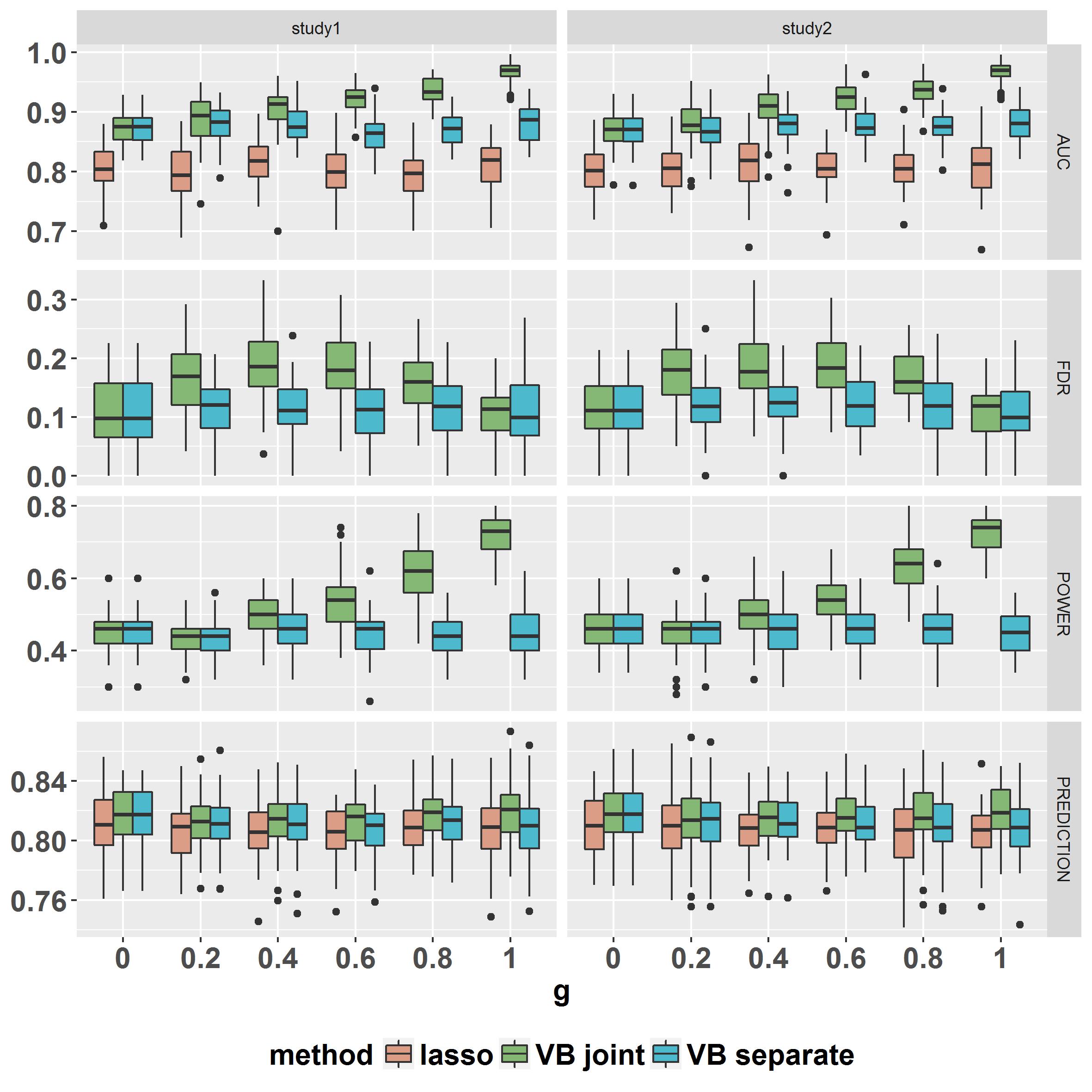}\\
  \caption{Comparison of FGP (VB joint), BVSR (VB separate), and Lasso with different $g$ ranging from 0 to 1 for binary trait. Panels from top to bottom are AUC, FDR, Power and Prediction, respectively.  The parameter setting of the model is : $p$ = 20,000, $n_1=n_2$ = 3000, $h^2$ = 0.4, $\rho = 0.5$, $\alpha_1$ = 0.0025.}
\end{figure}
\begin{figure}[H]
  \centering
  \includegraphics[width=\textwidth]{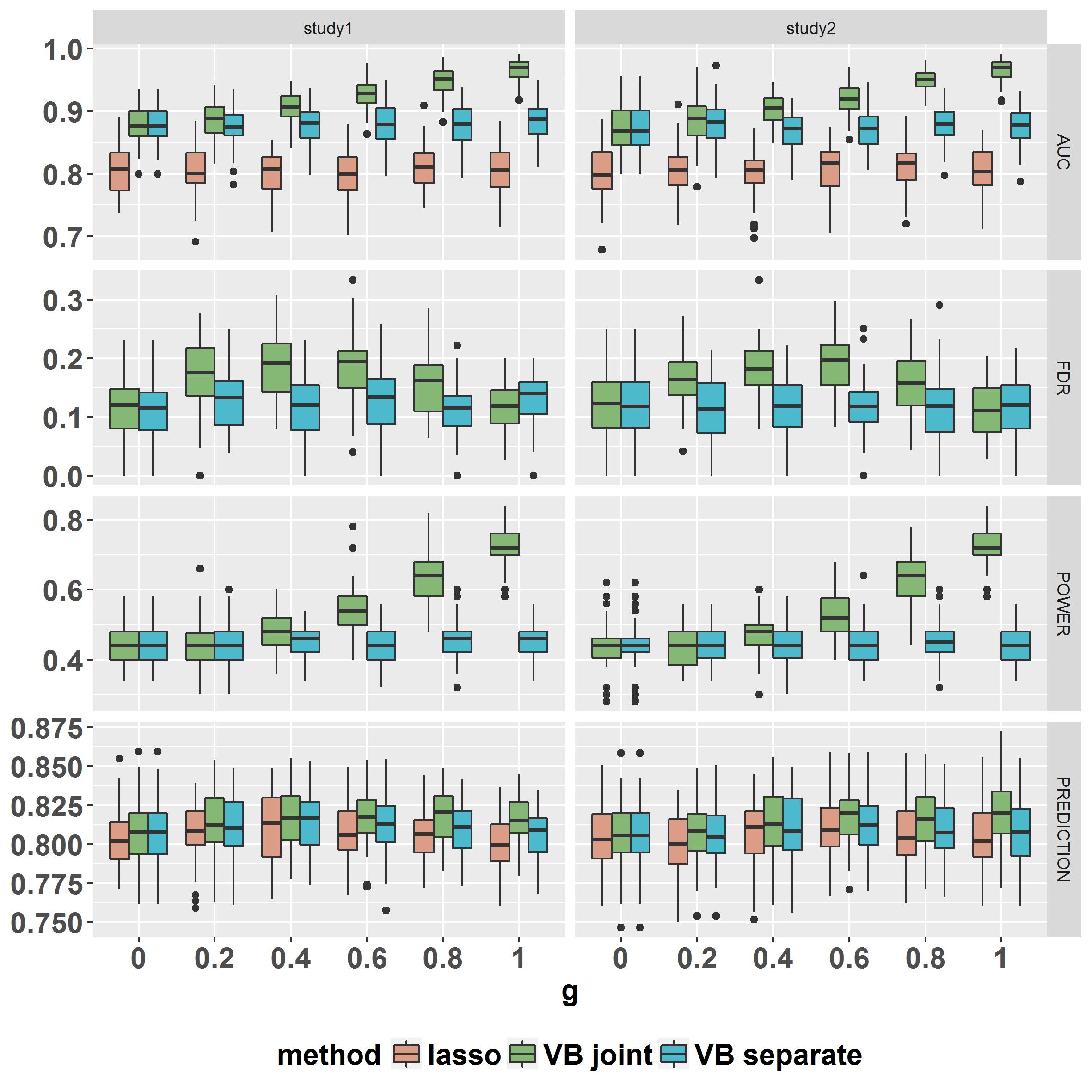}\\
  \caption{Comparison of FGP (VB joint), BVSR (VB separate), and Lasso with different $g$ ranging from 0 to 1 for binary trait. Panels from top to bottom are AUC, FDR, Power and Prediction, respectively.  The parameter setting of the model is : $p$ = 20,000, $n_1=n_2$ = 3000, $h^2$ = 0.4, $\rho = 0.7$, $\alpha_1$ = 0.0025.}
\end{figure}

%%%%%%%%%%%%%%%%%%%%%%%%%%%%%%  h^2 = 0.5  %%%%%%%%%%%%%%%%%%%%%%%%%%%%%%%%%%%%%%%%%%%%%%%%%%%%%%%%%%%%%%%%%%%%%%
\begin{figure}[H]
  \centering
  \includegraphics[width=\textwidth]{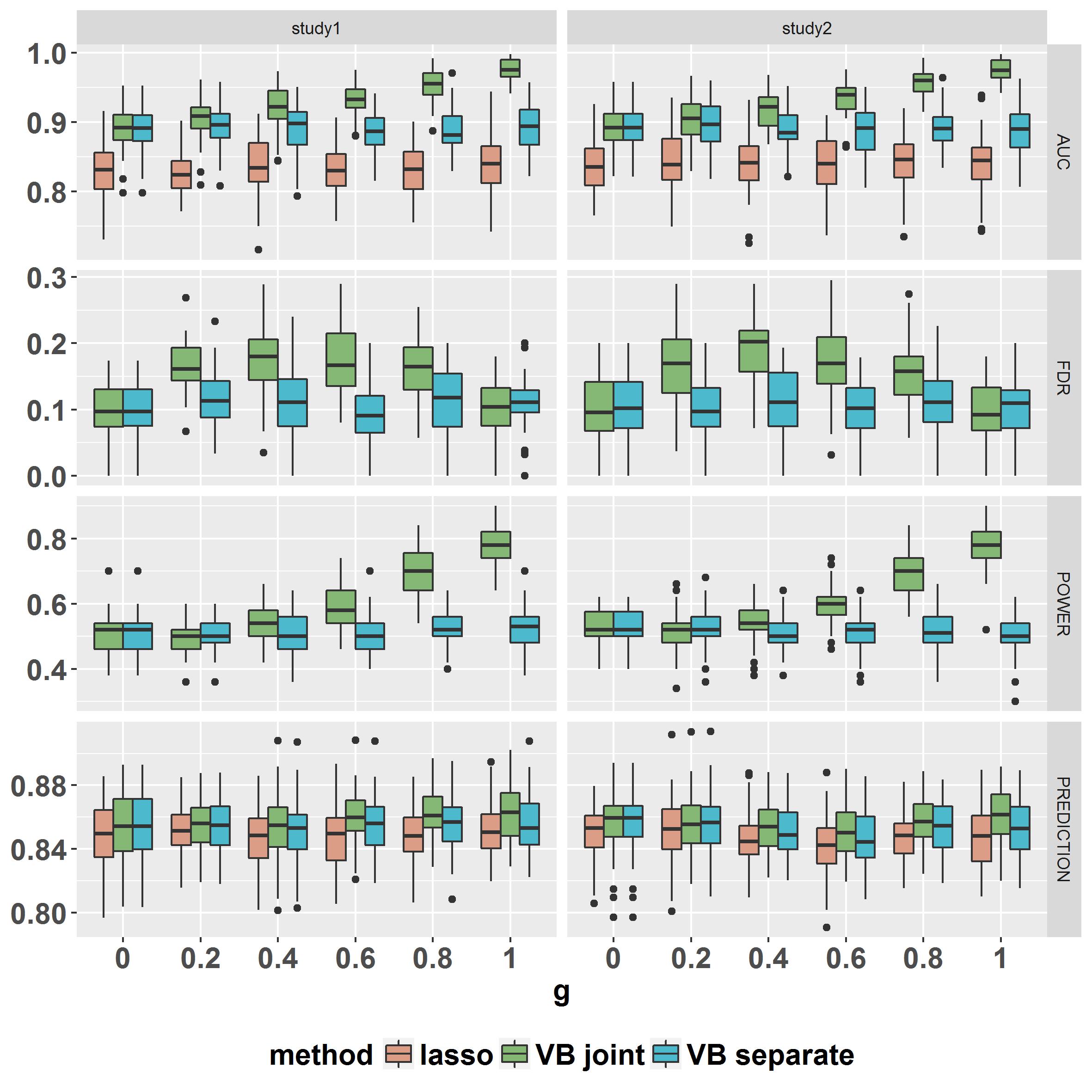}\\
  \caption{Comparison of FGP (VB joint), BVSR (VB separate), and Lasso with different $g$ ranging from 0 to 1 for binary trait. Panels from top to bottom are AUC, FDR, Power and Prediction, respectively. The parameter setting of the model is : $p$ = 20,000, $n_1=n_2$ = 3000, $h^2$ = 0.5, $\rho = 0.2$, $\alpha_1$ = 0.0025.}
\end{figure}
\begin{figure}[H]
  \centering
  \includegraphics[width=\textwidth]{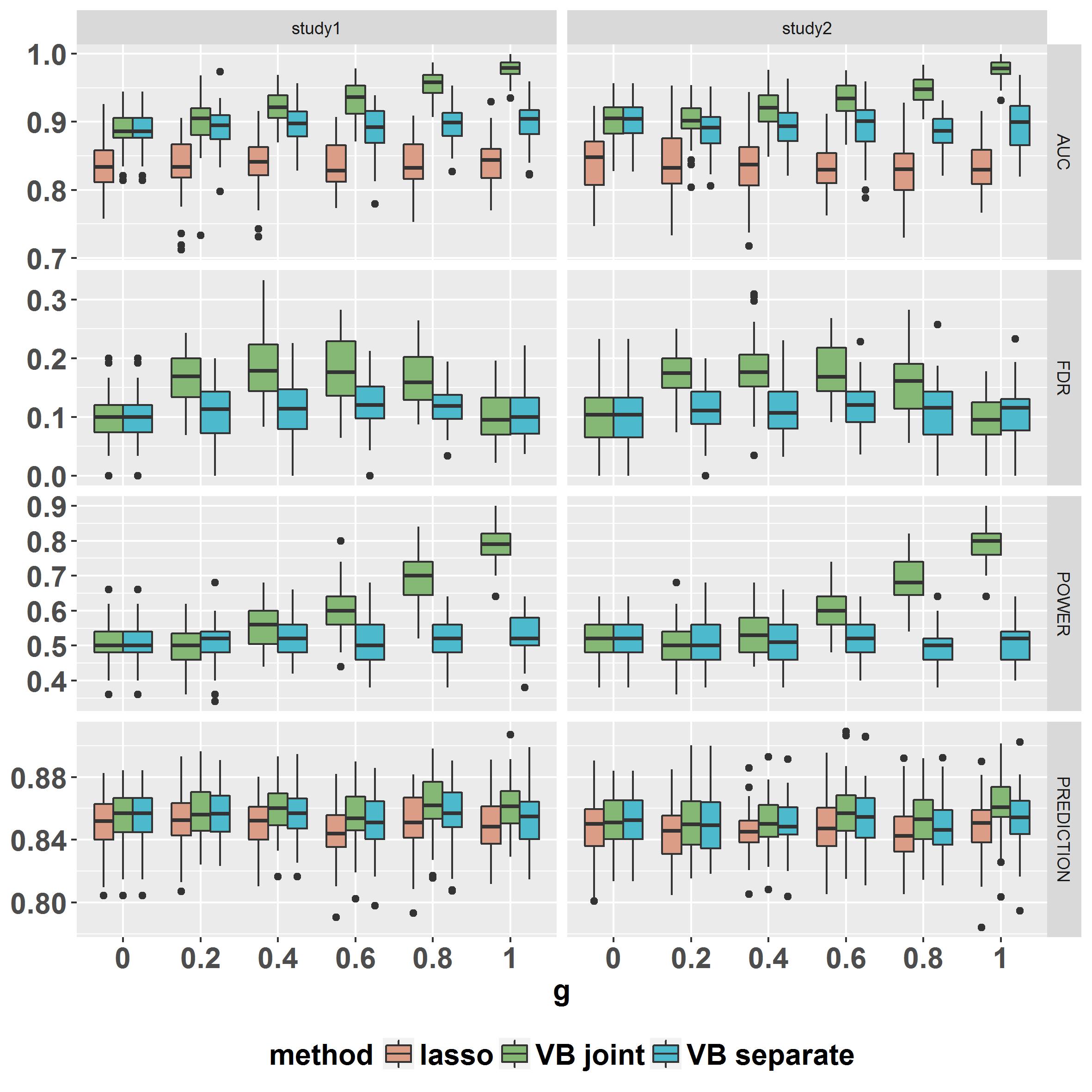}\\
  \caption{Comparison of FGP (VB joint), BVSR (VB separate), and Lasso with different $g$ ranging from 0 to 1 for binary trait. Panels from top to bottom are AUC, FDR, Power and Prediction, respectively.  The parameter setting of the model is : $p$ = 20,000, $n_1=n_2$ = 3000, $h^2$ = 0.5, $\rho = 0.5$, $\alpha_1$ = 0.0025.}
\end{figure}
\begin{figure}[H]
  \centering
  \includegraphics[width=\textwidth]{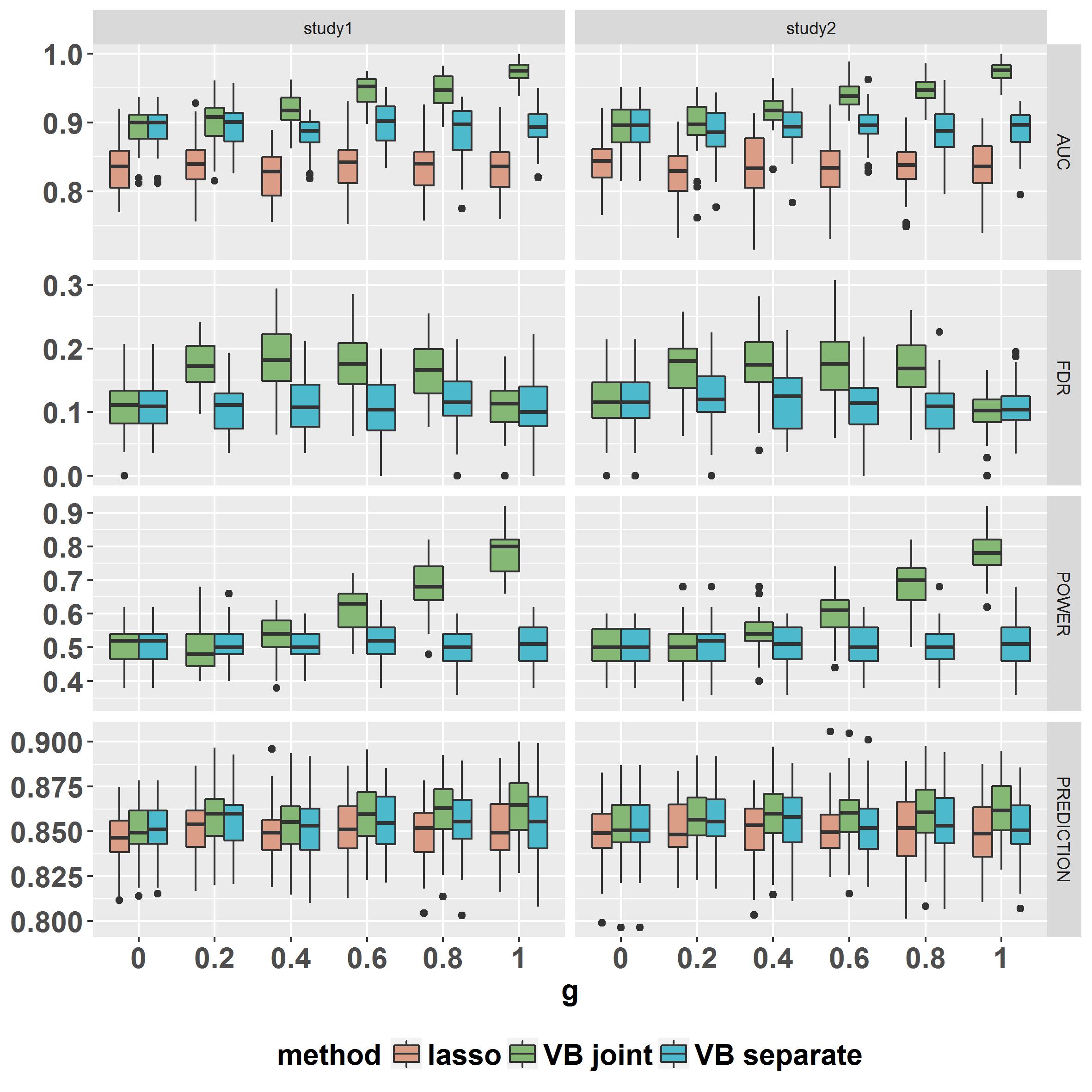}\\
  \caption{Comparison of FGP (VB joint), BVSR (VB separate), and Lasso with different $g$ ranging from 0 to 1 for binary trait. Panels from top to bottom are AUC, FDR, Power and Prediction, respectively.  The parameter setting of the model is : $p$ = 20,000, $n_1=n_2$ = 3000, $h^2$ = 0.5, $\rho = 0.7$, $\alpha_1$ = 0.0025.}
  \label{end_binary_simualtion}
\end{figure}

%% type 1 error and power
\clearpage
\begin{figure}
    \centering
    \begin{subfigure}{0.7\textwidth}
        \centering
        \includegraphics[width=\textwidth]{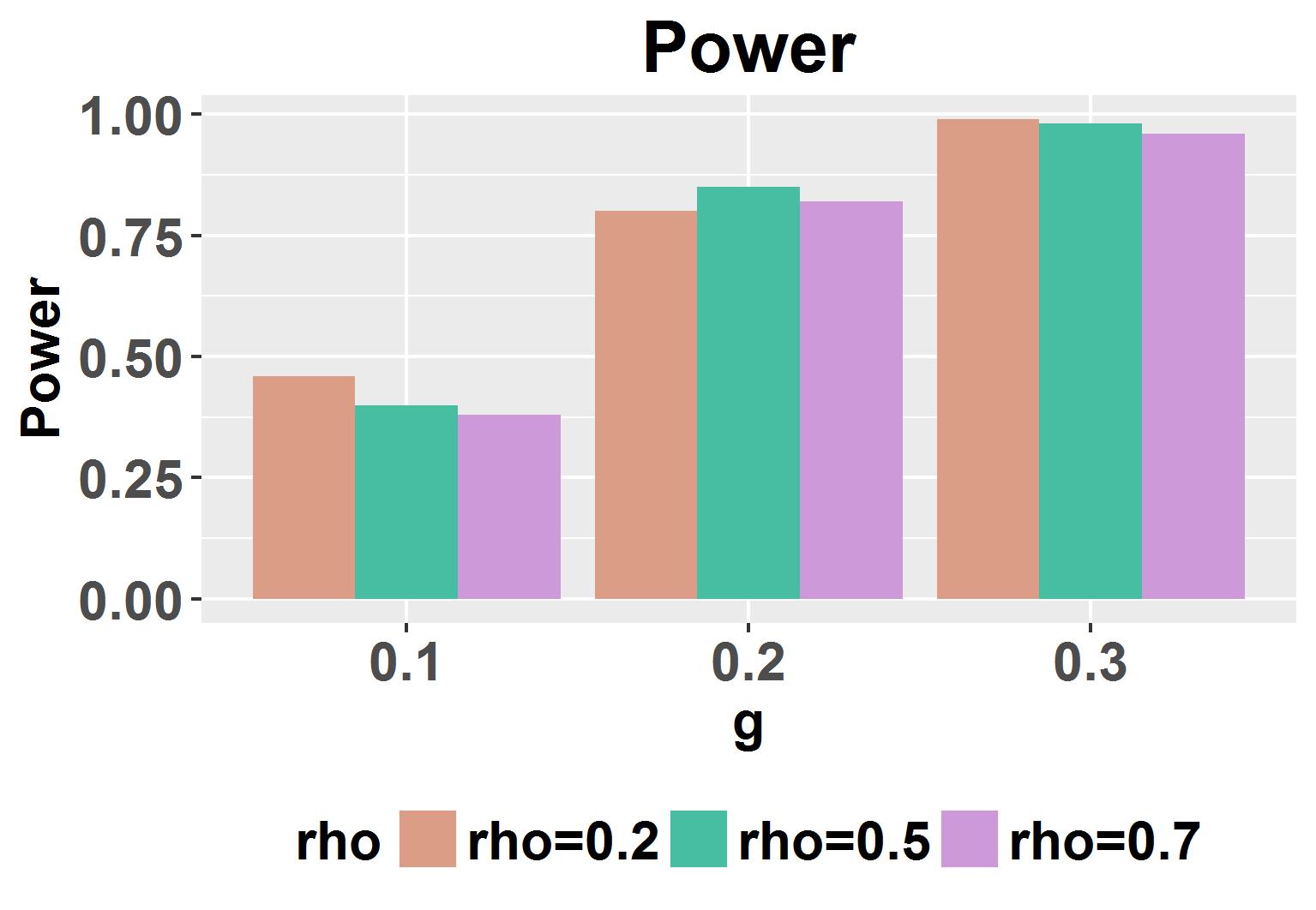}
    \end{subfigure}
    \begin{subfigure}{0.7\textwidth}
        \centering
        \includegraphics[width=\textwidth]{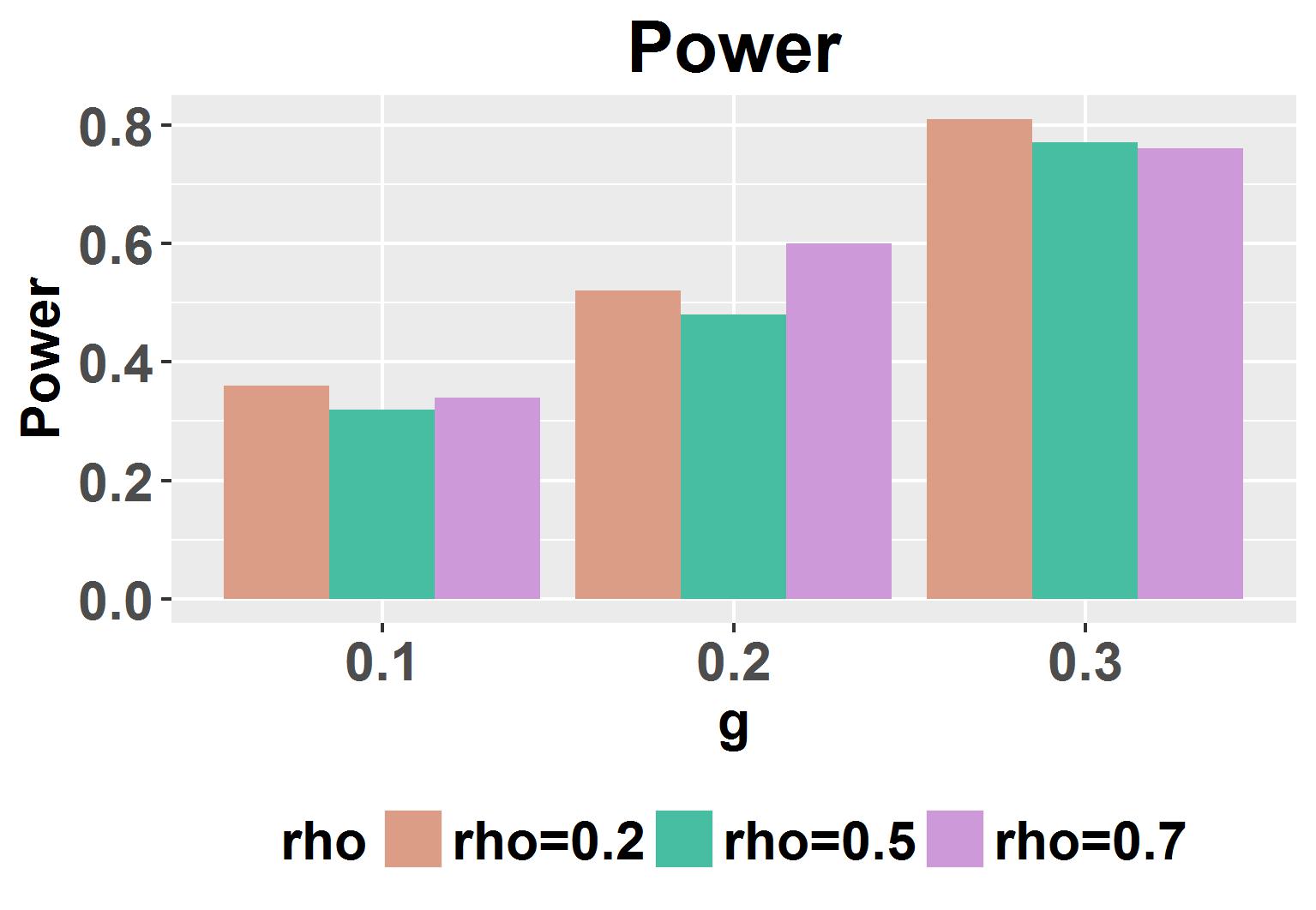}
    \end{subfigure}
   \caption{Power of pleiotropy test for both quantitative (left panel) and binary (right panel) trait. $\rho$ is chosen to be {0.2, 0.5, 0.7}, $h_2$ is 0.5 and the pleiotropy parameter $g$ is controlled at {0.1, 0.2, 0.3}. The number of replicates is 100.}
\label{linpleiotropy}
\end{figure}

\begin{figure}
    \centering
    \begin{subfigure}{0.4\textwidth}
        \centering
        \includegraphics[width=\textwidth]{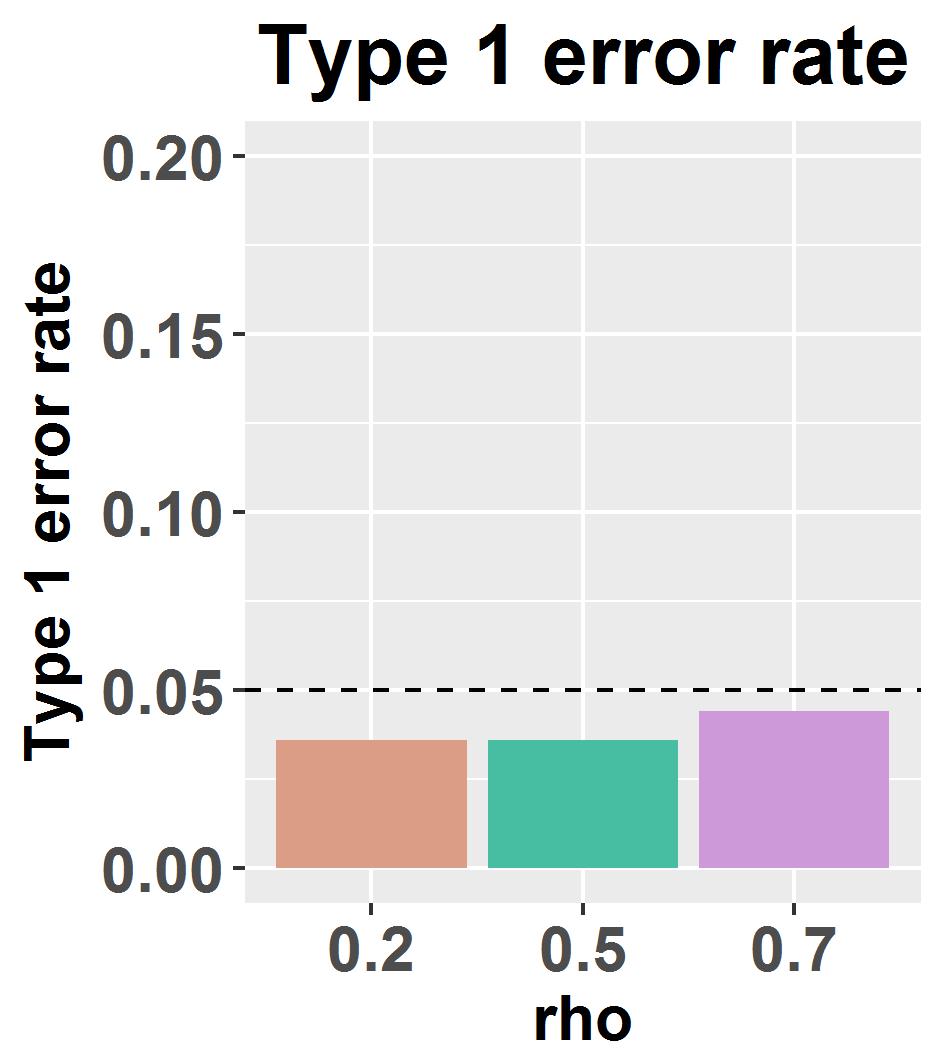}
    \end{subfigure}
    \begin{subfigure}{0.4\textwidth}
        \centering
        \includegraphics[width=\textwidth]{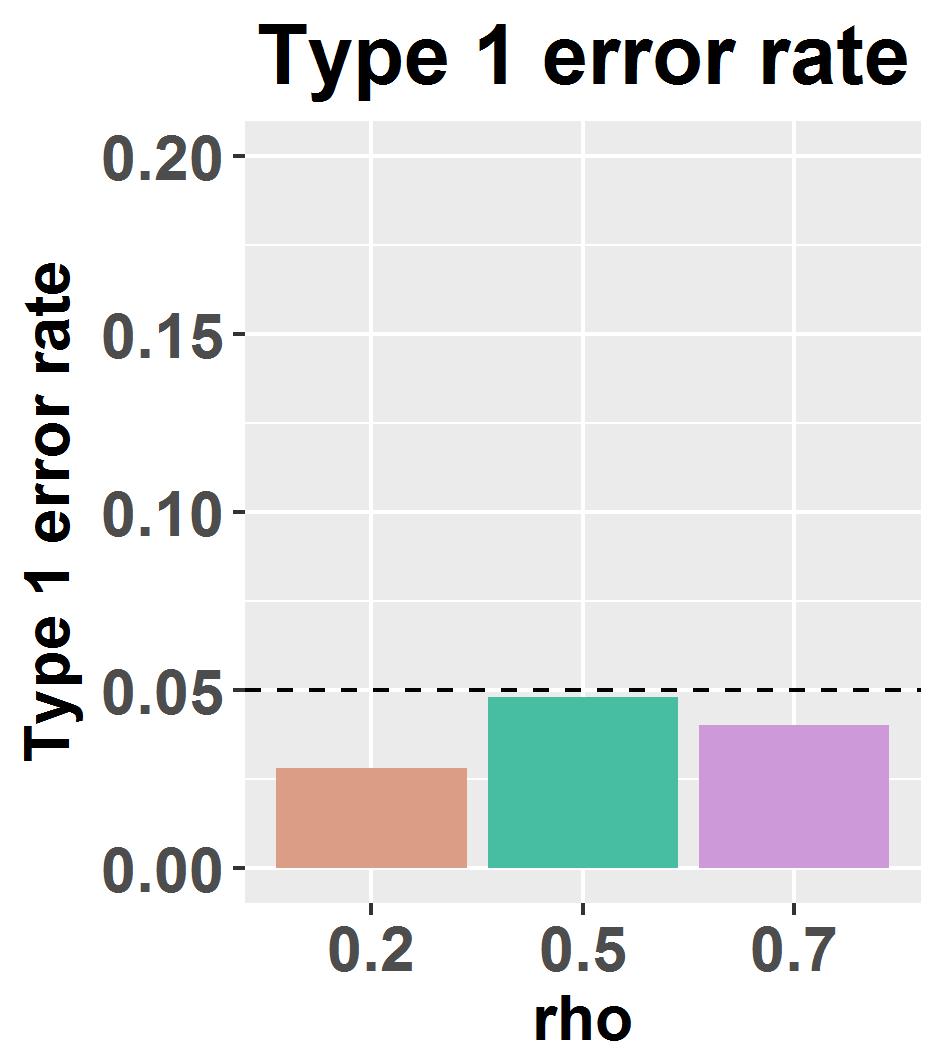}
    \end{subfigure}
  \caption{Type I error of pleiotropy test for quantitative (left panel) and binary (right panel) trait. $\rho$ is chosen to be {0.2, 0.5, 0.7} and $h_2$ is 0.5. The number of replicates is 500.}
\label{logpleiotropy}
\end{figure}

%%%%%%%%%%%%%%%%%%%%%%%%%%%%%% alpha_1 = 0.005 %%%%%%%%%%%%%%%%%%%%%%%%%%%%%%%%%%%%%%%%%%%%%%%%%%%%%%%%%%%%%%%%%%
%\clearpage
%\begin{figure}[H]
%  \centering
%  % Requires \usepackage{graphicx}
%  \includegraphics[width=\textwidth]{Figure/rho2snr11_pi0005}\\
%  \caption{Comparison of FGP (VB joint), BVSR (VB separate), and Lasso with different $g$ ranging from 0 to 1 for binary trait. Panels from top to bottom are AUC, FDR, Power and Prediction, respectively.  The parameter setting of the model is : $p$ = 20,000, $n_1=n_2$ = 3000, $h^2$ = 0.5, $\rho = 0.2$, $\alpha_1$ = 0.005.}
%\end{figure}
%\begin{figure}[H]
%  \centering
%  % Requires \usepackage{graphicx}
%  \includegraphics[width=\textwidth]{Figure/rho5snr11_pi0005}\\
%  \caption{Comparison of FGP (VB joint), BVSR (VB separate), and Lasso with different $g$ ranging from 0 to 1 for binary trait. Panels from top to bottom are AUC, FDR, Power and Prediction, respectively.  The parameter setting of the model is : $p$ = 20,000, $n_1=n_2$ = 3000, $h^2$ = 0.5, $\rho = 0.5$, $\alpha_1$ = 0.005.}
%\end{figure}
%\begin{figure}[H]
%  \centering
%  % Requires \usepackage{graphicx}
%  \includegraphics[width=\textwidth]{Figure/rho7snr11_pi0005}\\
%  \caption{Comparison of FGP (VB joint), BVSR (VB separate), and Lasso with different $g$ ranging from 0 to 1 for binary trait. Panels from top to bottom are AUC, FDR, Power and Prediction, respectively.  The parameter setting of the model is : $p$ = 20,000, $n_1=n_2$ = 3000, $h^2$ = 0.5, $\rho = 0.7$, $\alpha_1$ = 0.005.}
%  \label{end_binary_simualtion}
%\end{figure}
\clearpage
\section{Real data analysis}
\begin{table}[H]
\centering
\begin{tabular}{rrrrrrr}
  \hline
 & iter-sep1 & time-sep1(s) & iter-sep2 & time-sep2(s) & iter-joint & time-joint(s) \\
  \hline
  RA-T1D-inMHC & 69 & 271.70 & 27 & 139.88 & 69 & 713.53 \\
  T1D-RA-inMHC & 29 & 72.98 & 66 & 167.67 & 75 & 451.86 \\
  RA-T1D-exMHC & 160 & 540.37 & 100 & 403.49 & 175 & 1736.89 \\
  T1D-RA-exMHC & 106 & 374.78 & 134 & 522.26 & 178 & 1441.98 \\
  CD-T1D-inMHC & 177 & 427.43 & 27 & 73.75 & 305 & 1778.91 \\
  T1D-CD-inMHC & 27 & 128.62 & 183 & 583.80 & 237 & 2187.62 \\
  CD-T1D-exMHC & 177 & 372.41 & 101 & 263.57 & 217 & 1170.94 \\
  T1D-CD-exMHC & 107 & 319.80 & 183 & 422.41 & 220 & 1366.07 \\
   \hline
\end{tabular}
\caption{Summary of real data analysis, inMHC and exMHC mean SNPs include and exclude MHC region, respectively, iter-sep1 and iter-sep2 mean iterations of separate analysis using BVSR for first and second study, respectively, iter-joint means iterations of joint analysis using FGP, time-sep1 and time-sep2 mean time of separate analysis using BVSR for first and second study, respectively, time-joint means time of joint analysis}
\end{table}
\subsection{Comparison of FGP and BVSR for the data consisting of 58C controls with T1D and UKBS
controls with RA}\begin{figure}[H]
    \centering
    \begin{subfigure}[b]{0.24\textwidth}
        \centering
        \includegraphics[width=\textwidth]{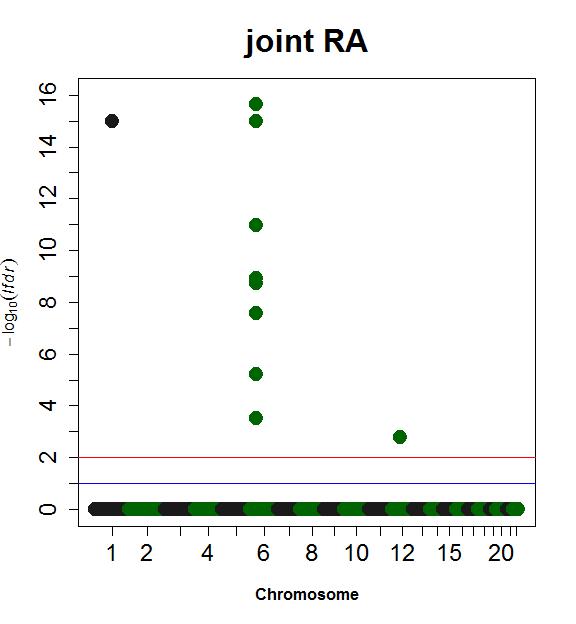}
    \end{subfigure}
    \begin{subfigure}[b]{0.24\textwidth}
        \centering
        \includegraphics[width=\textwidth]{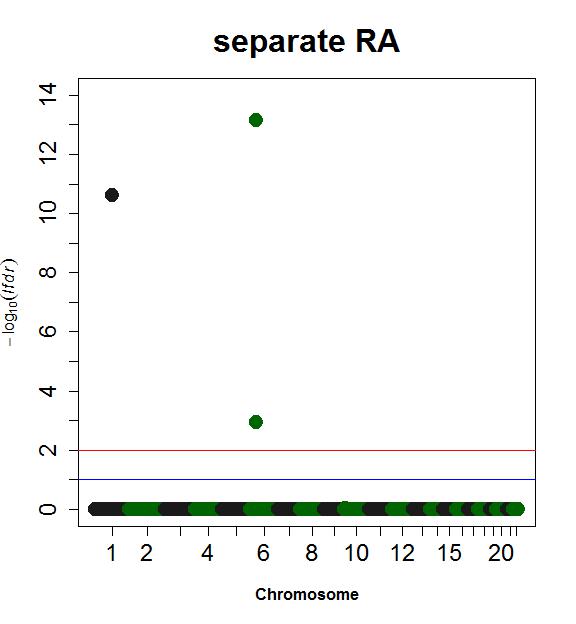}
    \end{subfigure}
    \begin{subfigure}[b]{0.24\textwidth}
        \centering
        \includegraphics[width=\textwidth]{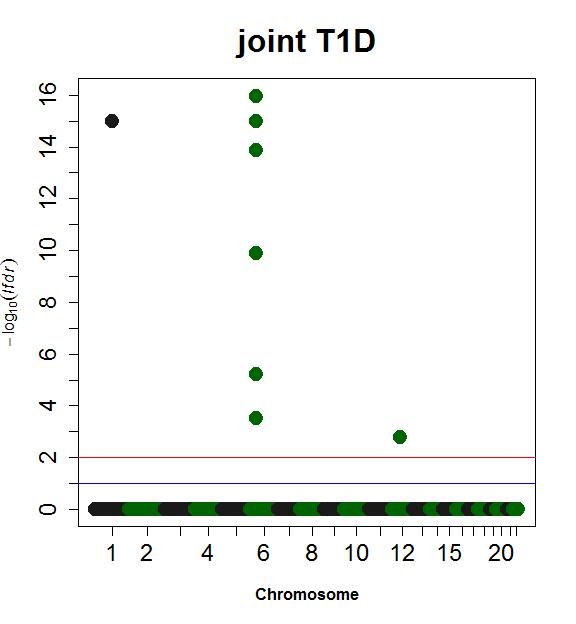}
    \end{subfigure}
    \begin{subfigure}[b]{0.24\textwidth}
        \centering
        \includegraphics[width=\textwidth]{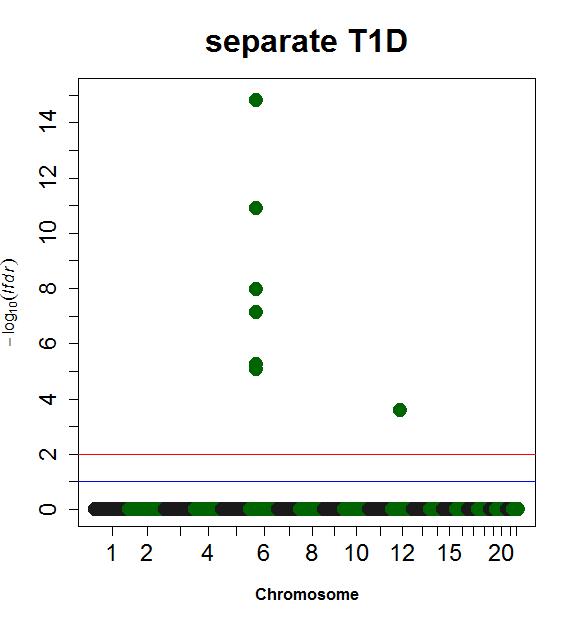}
    \end{subfigure}
   \caption{For the data consisting of 58C controls with T1D and UKBS controls
with RA, manhattan plots of separate analysis using BVSR and joint analysis using
FGP.}
\end{figure}

\begin{figure}[H]
  \centering
    \includegraphics[width=0.5\textwidth]{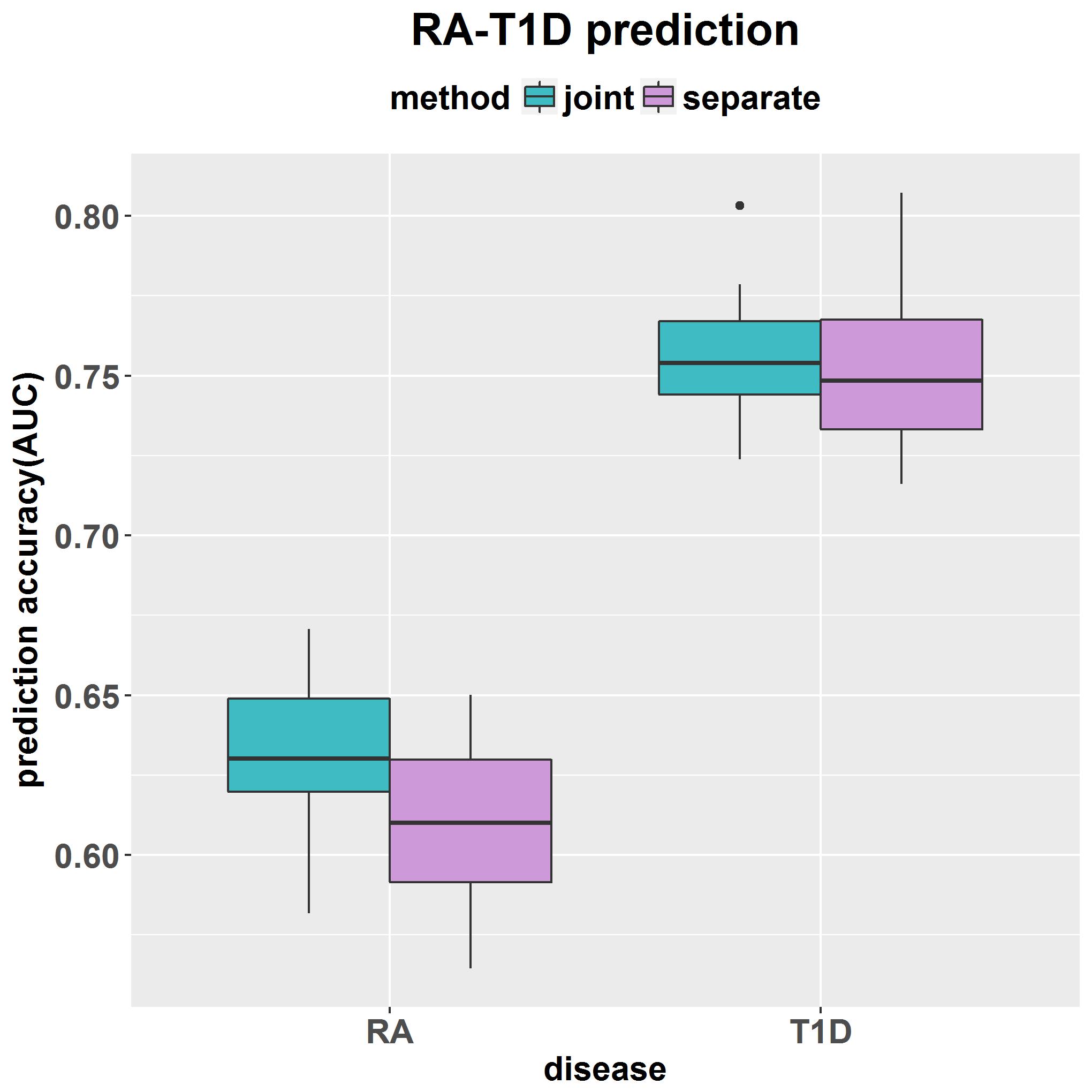}
  \caption{For the data consisting of 58C controls with T1D and UKBS controls
with RA, prediction performance of separate analysis using BVSR and joint analysis using
FGP.}
\end{figure}

% latex table generated in R 3.4.1 by xtable 1.8-2 package
% Sun Aug 27 15:14:49 2017
\begin{table}[H]
\centering
\begin{tabular}{rlll}
  \hline
 & Data & number of hits & prediction accuracy(AUC) \\
  \hline
1 & Rheumatoid arthritis(RA)joint & 11 & 63.2\%(2.5\%) \\
  2 & Type 1 diabetes(T1D)joint & 11 & 75.7\%(2.2\%) \\
  3 & Rheumatoid arthritis(RA)separate & 3 & 60.9\%(2.8\%) \\
  4 & Type 1 diabetes(T1D)separate & 10 & 75.2\%(2.8\%) \\
   \hline
\end{tabular}
\caption{For the data consisting of 58C controls with T1D and UKBS controls
with RA, summary of separate and joint analysis of RA and T1D}
\end{table}

% latex table generated in R 3.4.1 by xtable 1.8-2 package
% Sun Sep 10 14:56:27 2017
\begin{table}[H]
\centering
\begin{tabular}{rlrrllll}
  \hline
 & snp & chr & position & sep RA(fdr) & sep T1D(fdr) & joi RA(fdr) & joi T1D(fdr) \\
  \hline
1 & rs6679677 &   1 & 114303808 & 2.5e-11* & 0e+00* & 0e+00* & 0e+00* \\
  2 & rs13200022 &   6 & 31098957 & 1e+00 & 5.77e-06* & 1.82e-09* & 1.37e-14* \\
  3 & rs3130484 &   6 & 31715882 & 1e+00 & 8.35e-06* & 1e+00 & 1e+00 \\
  4 & rs2075800 &   6 & 31777946 & 6.91e-14* & 1e+00 & 1e+00 & 1e+00 \\
  5 & rs550513 &   6 & 31920687 & 9.88e-01 & 1e+00 & 6.3e-06* & 6.3e-06* \\
  6 & rs3130287 &   6 & 32050544 & 1e+00 & 0e+00* & 2.79e-08* & 1.11e-16* \\
  7 & rs17421624 &   6 & 32066177 & 1e+00 & 1e+00 & 2.22e-16* & 1.11e-16* \\
  8 & rs9272346 &   6 & 32604372 & 1.14e-03* & 0e+00* & 1.03e-11* & 0e+00* \\
  9 & rs2070121 &   6 & 32781554 & 1e+00 & 1.22e-11* & 2.22e-16* & 0e+00* \\
  10 & rs10484565 &   6 & 32795032 & 9.97e-01 & 1.55e-15* & 0e+00* & 0e+00* \\
  11 & rs241427 &   6 & 32804414 & 1e+00 & 1.1e-08* & 1.24e-09* & 1.28e-10* \\
  12 & rs12529313 &   6 & 32817130 & 1e+00 & 7.3e-08* & 3.05e-04* & 3.05e-04* \\
  13 & rs11171739 &  12 & 56470625 & 1e+00 & 2.61e-04* & 1.63e-03* & 1.63e-03* \\
   \hline
\end{tabular}
\caption{For the data consisting of 58C controls with T1D and UKBS controls
with RA, list of SNPs of two modes: separate analysis and joint analysis. * denotes the local
fdr \textless 0.2, sep means separate, joi means joint.}
\end{table}
\clearpage
\subsection{Comparison of FGP and BVSR for the data consisting of 58C controls with T1D and UKBS
controls with RA exclude MHC region}\begin{figure}[H]
    \centering
    \begin{subfigure}[b]{0.24\textwidth}
        \centering
        \includegraphics[width=\textwidth]{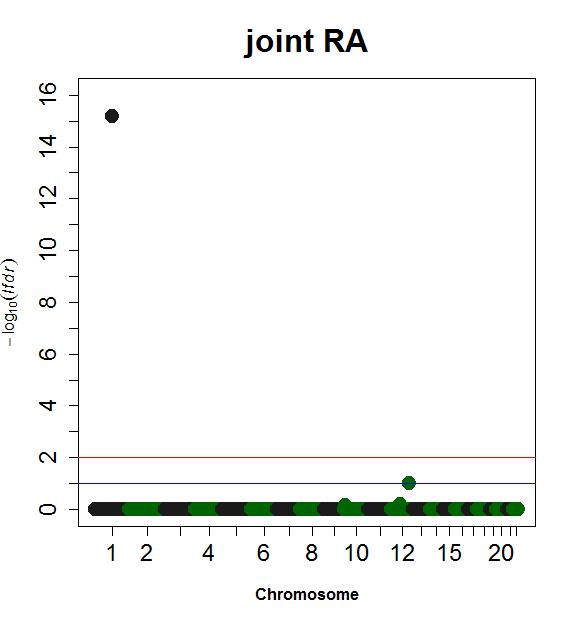}
    \end{subfigure}
    \begin{subfigure}[b]{0.24\textwidth}
        \centering
        \includegraphics[width=\textwidth]{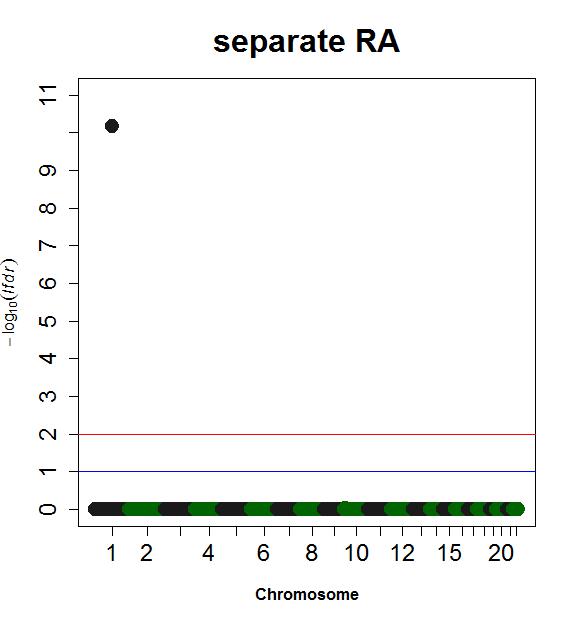}
    \end{subfigure}
    \begin{subfigure}[b]{0.24\textwidth}
        \centering
        \includegraphics[width=\textwidth]{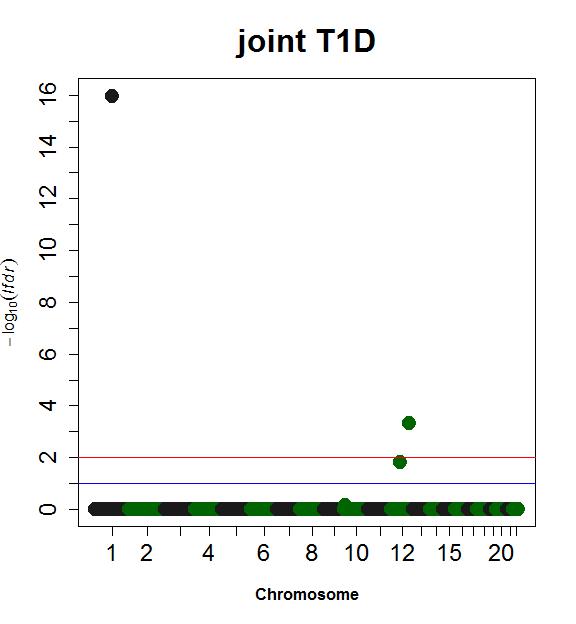}
    \end{subfigure}
    \begin{subfigure}[b]{0.24\textwidth}
        \centering
        \includegraphics[width=\textwidth]{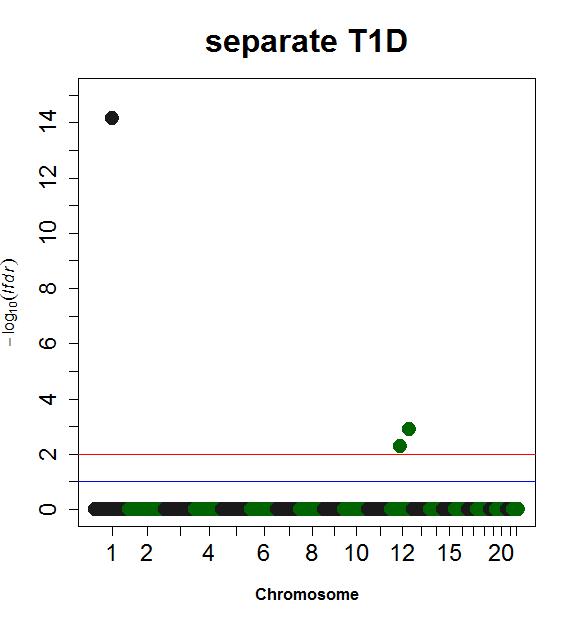}
    \end{subfigure}
   \caption{For the data consisting of 58C controls with T1D and UKBS controls
with RA exclude MHC region SNPs, manhattan plots of separate analysis using BVSR and joint analysis using
FGP.}
\end{figure}

\begin{figure}[H]
  \centering
    \includegraphics[width=0.5\textwidth]{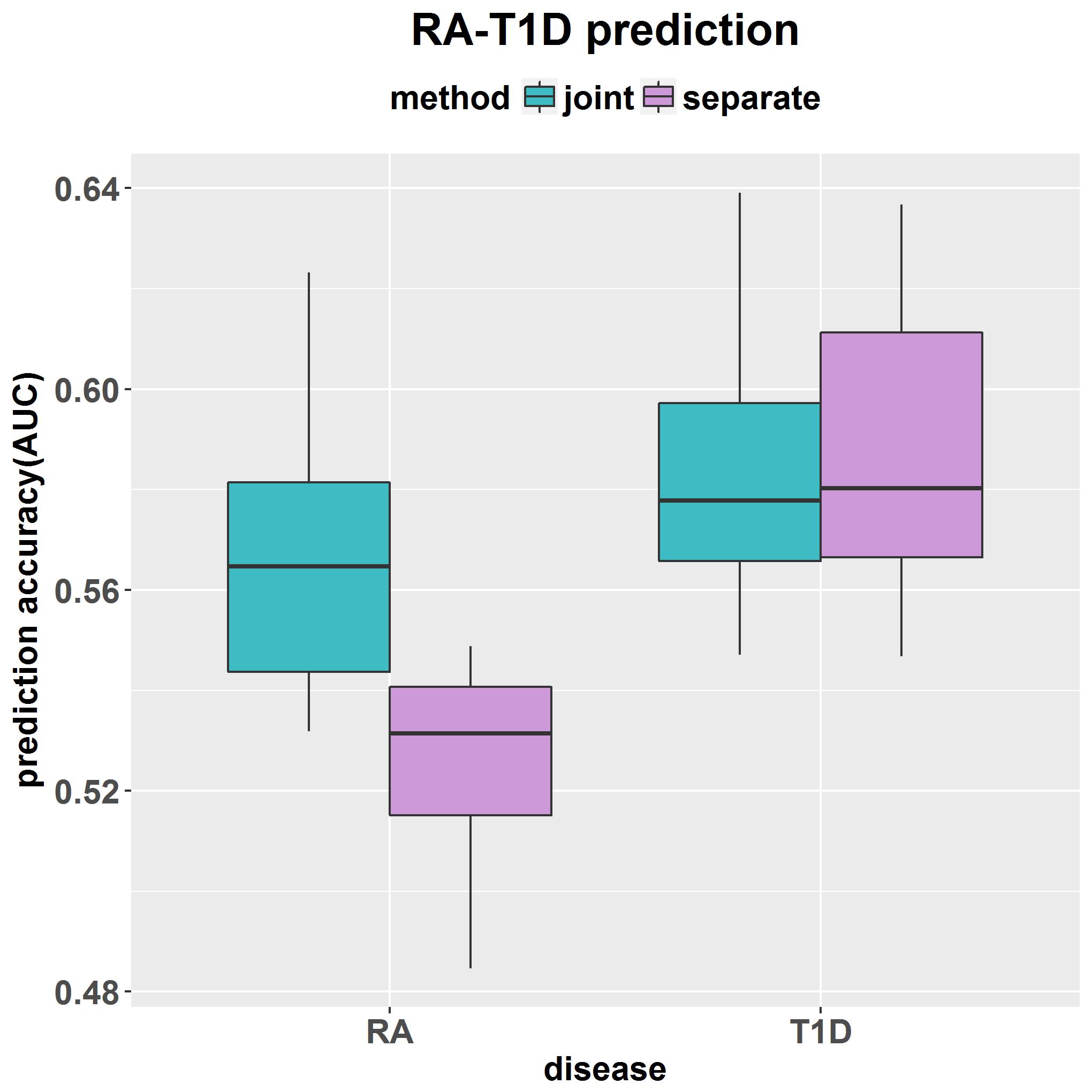}
  \caption{For the data consisting of 58C controls with T1D and UKBS controls
with RA exclude MHC region SNPs, prediction performance of separate analysis using BVSR and joint analysis using
FGP.}
\end{figure}

% latex table generated in R 3.4.1 by xtable 1.8-2 package
% Sun Aug 27 15:14:49 2017
\begin{table}[H]
\centering
\begin{tabular}{rlll}
  \hline
 & Data & number of hits & prediction accuracy(AUC) \\
  \hline
1 & Rheumatoid arthritis(RA)joint & 2 & 56.7\%(2.9\%) \\
  2 & Type 1 diabetes(T1D)joint & 3 & 58.4\%(2.8\%) \\
  3 & Rheumatoid arthritis(RA)separate & 1 & 52.5\%(2.1\%) \\
  4 & Type 1 diabetes(T1D)separate & 3 & 58.8\%(2.9\%) \\
   \hline
\end{tabular}
\caption{For the data consisting of 58C controls with T1D and UKBS controls
with RA exclude MHC region, summary of separate and joint analysis of RA and T1D}
\end{table}

% latex table generated in R 3.4.1 by xtable 1.8-2 package
% Sun Sep 10 14:56:27 2017
\begin{table}[H]
\centering
\begin{tabular}{rlrrllll}
  \hline
 & snp & chr & position & sep RA(fdr) & sep T1D(fdr) & joi RA(fdr) & joi T1D(fdr) \\
  \hline
1 & rs6679677 &   1 & 114303808 & 6.61e-11* & 6.66e-15* & 6.66e-16* & 1.11e-16* \\
  2 & rs11171739 &  12 & 56470625 & 1e+00 & 5.1e-03* & 6.42e-01 & 1.48e-02* \\
  3 & rs17696736 &  12 & 112486818 & 1e+00 & 1.26e-03* & 1.03e-01* & 4.85e-04* \\
   \hline
\end{tabular}
\caption{For the data consisting of 58C controls with T1D and UKBS controls
with RA exclude MHC region, list of SNPs of two modes: separate analysis and joint analysis. * denotes the local
fdr \textless 0.2, sep means separate, joi means joint.}
\end{table}
\clearpage
\subsection{Comparison of FGP and BVSR for the data consisting of 58C controls with RA and UKBS
controls with T1D exclude MHC region}\begin{figure}[H]
    \centering
    \begin{subfigure}[b]{0.24\textwidth}
        \centering
        \includegraphics[width=\textwidth]{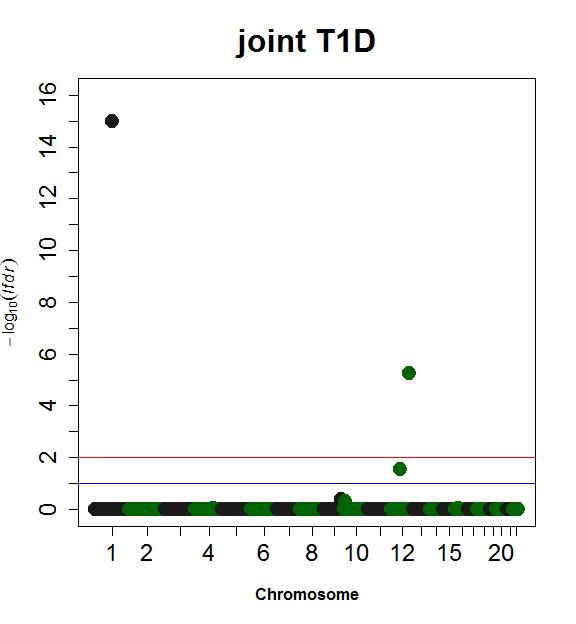}
    \end{subfigure}
    \begin{subfigure}[b]{0.24\textwidth}
        \centering
        \includegraphics[width=\textwidth]{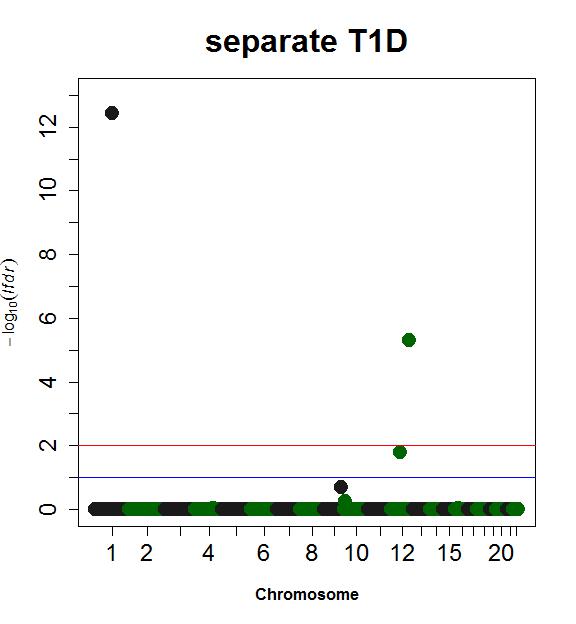}
    \end{subfigure}
    \begin{subfigure}[b]{0.24\textwidth}
        \centering
        \includegraphics[width=\textwidth]{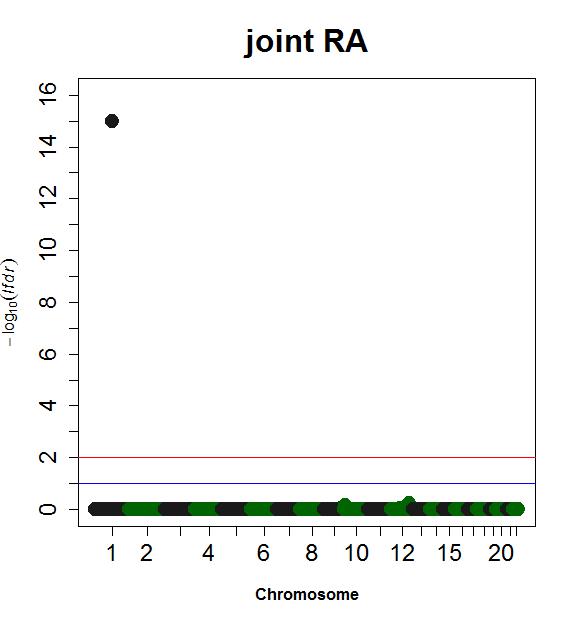}
    \end{subfigure}
    \begin{subfigure}[b]{0.24\textwidth}
        \centering
        \includegraphics[width=\textwidth]{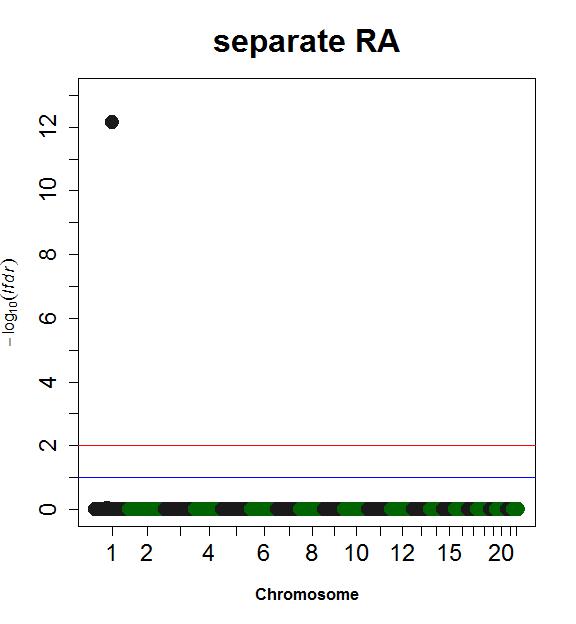}
    \end{subfigure}
   \caption{For the data consisting of 58C controls with RA and UKBS controls
with T1D exclude MHC region SNPs, manhattan plots of separate analysis using BVSR and joint analysis using
FGP.}
\end{figure}

\begin{figure}[H]
  \centering
    \includegraphics[width=0.5\textwidth]{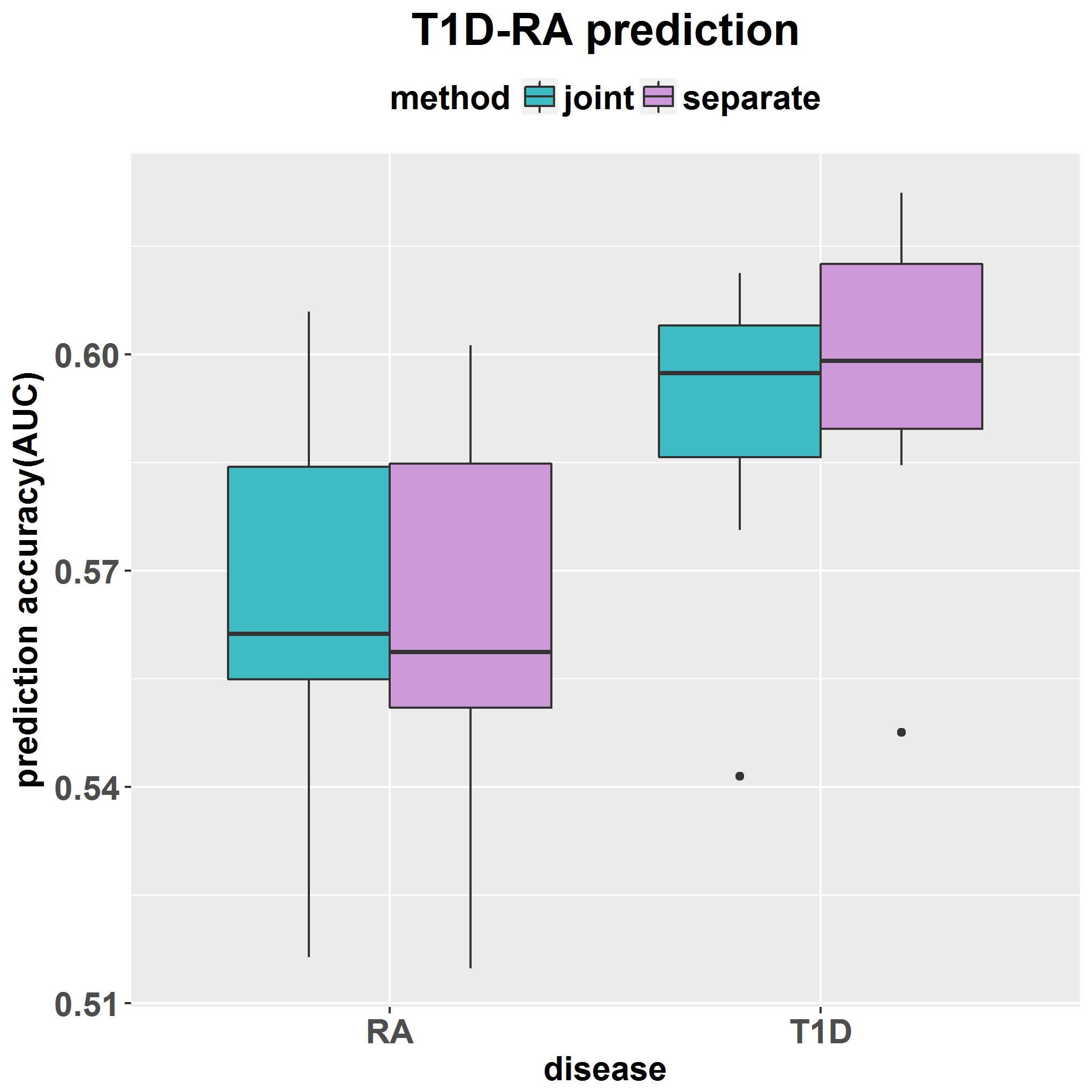}
  \caption{For the data consisting of 58C controls with RA and UKBS controls
with T1D exclude MHC region SNPs, prediction performance of separate analysis using BVSR and joint analysis using
FGP.}
\end{figure}

% latex table generated in R 3.4.1 by xtable 1.8-2 package
% Sun Aug 27 15:14:49 2017
\begin{table}[H]
\centering
\begin{tabular}{rlll}
  \hline
 & Data & number of hits & prediction accuracy(AUC) \\
  \hline
1 & Type 1 diabetes(T1D)joint & 3 & 59.1\%(2.1\%) \\
  2 & Rheumatoid arthritis(RA)joint & 1 & 56.7\%(2.6\%) \\
  3 & Type 1 diabetes(T1D)separate & 3 & 59.8\%(2.2\%) \\
  4 & Rheumatoid arthritis(RA)separate & 1 & 56.4\%(2.7\%) \\
   \hline
\end{tabular}
\caption{For the data consisting of 58C controls with RA and UKBS controls
with T1D exclude MHC region, summary of separate and joint analysis of T1D and RA}
\end{table}

% latex table generated in R 3.4.1 by xtable 1.8-2 package
% Sun Sep 10 14:56:27 2017
\begin{table}[H]
\centering
\begin{tabular}{rlrrllll}
  \hline
 & snp & chr & position & sep T1D(fdr) & sep RA(fdr) & joi T1D(fdr) & joi RA(fdr) \\
  \hline
1 & rs6679677 &   1 & 114303808 & 3.78e-13* & 7.34e-13* & 0e+00* & 0e+00* \\
  2 & rs2292239 &  12 & 56482180 & 1.65e-02* & 1e+00 & 2.82e-02* & 8.81e-01 \\
  3 & rs17696736 &  12 & 112486818 & 4.95e-06* & 1e+00 & 5.5e-06* & 5.96e-01 \\
   \hline
\end{tabular}
\caption{For the data consisting of 58C controls with RA and UKBS controls
with T1D exclude MHC region, list of SNPs of two modes: separate analysis and joint analysis. * denotes the local
fdr \textless 0.2, sep means separate, joi means joint.}
\end{table}
\clearpage
\subsection{Comparison of FGP and BVSR for the data consisting of 58C controls with T1D and UKBS
controls with CD}\begin{figure}[H]
    \centering
    \begin{subfigure}[b]{0.24\textwidth}
        \centering
        \includegraphics[width=\textwidth]{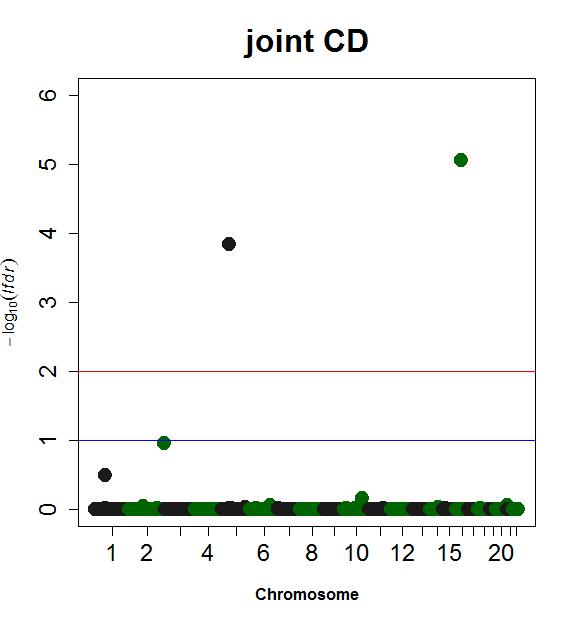}
    \end{subfigure}
    \begin{subfigure}[b]{0.24\textwidth}
        \centering
        \includegraphics[width=\textwidth]{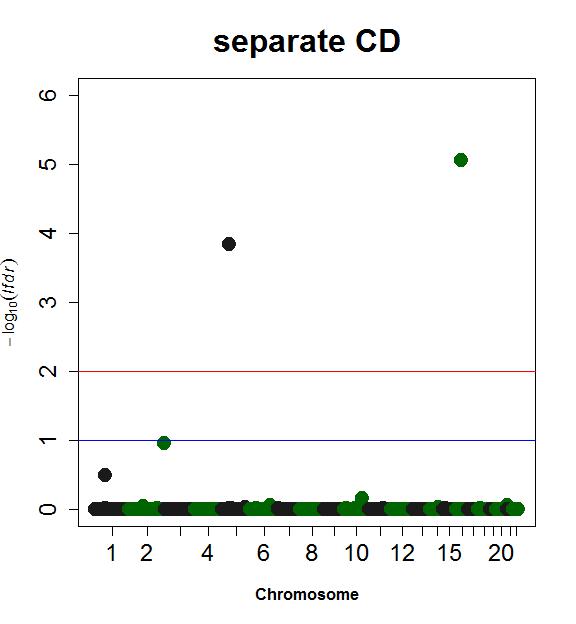}
    \end{subfigure}
    \begin{subfigure}[b]{0.24\textwidth}
        \centering
        \includegraphics[width=\textwidth]{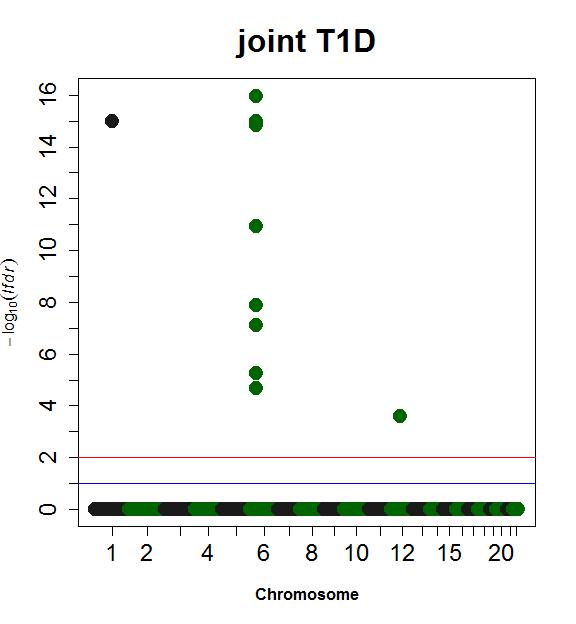}
    \end{subfigure}
    \begin{subfigure}[b]{0.24\textwidth}
        \centering
        \includegraphics[width=\textwidth]{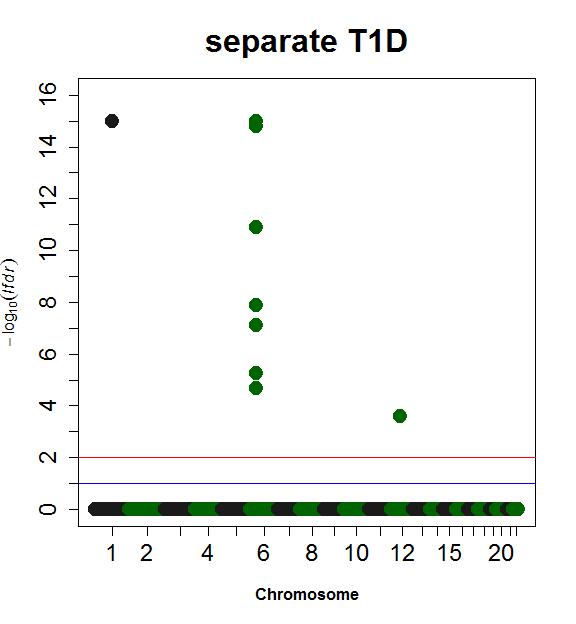}
    \end{subfigure}
   \caption{For the data consisting of 58C controls with T1D and UKBS controls
with CD, manhattan plots of separate analysis using BVSR and joint analysis using
FGP.}
\end{figure}

\begin{figure}[H]
  \centering
    \includegraphics[width=0.5\textwidth]{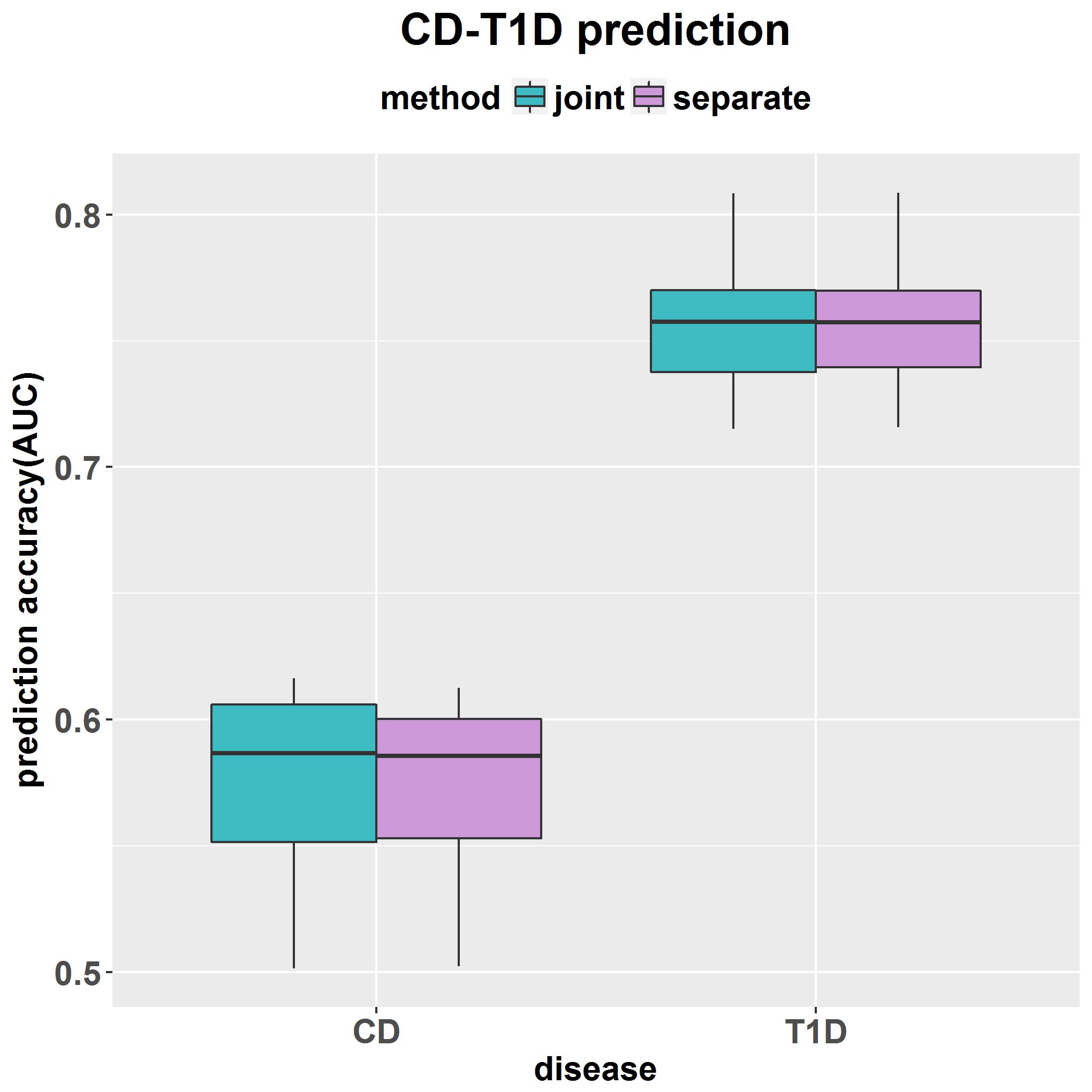}
  \caption{For the data consisting of 58C controls with T1D and UKBS controls
with CD, prediction performance of separate analysis using BVSR and joint analysis using
FGP.}
\end{figure}

% latex table generated in R 3.4.1 by xtable 1.8-2 package
% Sun Aug 27 15:14:49 2017
\begin{table}[H]
\centering
\begin{tabular}{rlll}
  \hline
 & Data & number of hits & prediction accuracy(AUC) \\
  \hline
1 & Crohn's disease(CD)joint & 3 & 57.6\%(3.7\%) \\
  2 & Type 1 diabetes(T1D)joint & 10 & 75.6\%(2.8\%) \\
  3 & Crohn's disease(CD)separate & 3 & 57.4\%(3.6\%) \\
  4 & Type 1 diabetes(T1D)separate & 10 & 75.7\%(2.8\%) \\
   \hline
\end{tabular}
\caption{For the data consisting of 58C controls with T1D and UKBS controls
with CD, summary of separate and joint analysis of CD and T1D}
\end{table}

% latex table generated in R 3.4.1 by xtable 1.8-2 package
% Sun Sep 10 14:56:27 2017
\begin{table}[H]
\centering
\begin{tabular}{rlrrllll}
  \hline
 & snp & chr & position & sep CD(fdr) & sep T1D(fdr) & joi CD(fdr) & joi T1D(fdr) \\
  \hline
1 & rs6679677 &   1 & 114303808 & 1e+00 & 0e+00* & 9.98e-01 & 0e+00* \\
  2 & rs10210302 &   2 & 234158839 & 1.1e-01* & 1e+00 & 1.1e-01* & 9.99e-01 \\
  3 & rs9292777 &   5 & 40437948 & 1.45e-04* & 1e+00 & 1.45e-04* & 1e+00 \\
  4 & rs13200022 &   6 & 31098957 & 1e+00 & 5.58e-06* & 1e+00 & 5.57e-06* \\
  5 & rs3130484 &   6 & 31715882 & 1e+00 & 2.06e-05* & 1e+00 & 2.06e-05* \\
  6 & rs3130287 &   6 & 32050544 & 1e+00 & 0e+00* & 9.98e-01 & 0e+00* \\
  7 & rs9272346 &   6 & 32604372 & 1e+00 & 0e+00* & 9.98e-01 & 1.11e-16* \\
  8 & rs2070121 &   6 & 32781554 & 1e+00 & 1.23e-11* & 1e+00 & 1.21e-11* \\
  9 & rs10484565 &   6 & 32795032 & 1e+00 & 1.55e-15* & 1e+00 & 1.44e-15* \\
  10 & rs241427 &   6 & 32804414 & 1e+00 & 1.29e-08* & 1e+00 & 1.29e-08* \\
  11 & rs12529313 &   6 & 32817130 & 1e+00 & 7.63e-08* & 1e+00 & 7.59e-08* \\
  12 & rs11171739 &  12 & 56470625 & 1e+00 & 2.61e-04* & 1e+00 & 2.61e-04* \\
  13 & rs2066843 &  16 & 50745199 & 8.8e-06* & 1e+00 & 8.81e-06* & 1e+00 \\
   \hline
\end{tabular}
\caption{For the data consisting of 58C controls with T1D and UKBS controls
with CD, list of SNPs of two modes: separate analysis and joint analysis. * denotes the local
fdr \textless 0.2, sep means separate, joi means joint.}
\end{table}
\clearpage
\subsection{Comparison of FGP and BVSR for the data consisting of 58C controls with CD and UKBS
controls with T1D}\begin{figure}[H]
    \centering
    \begin{subfigure}[b]{0.24\textwidth}
        \centering
        \includegraphics[width=\textwidth]{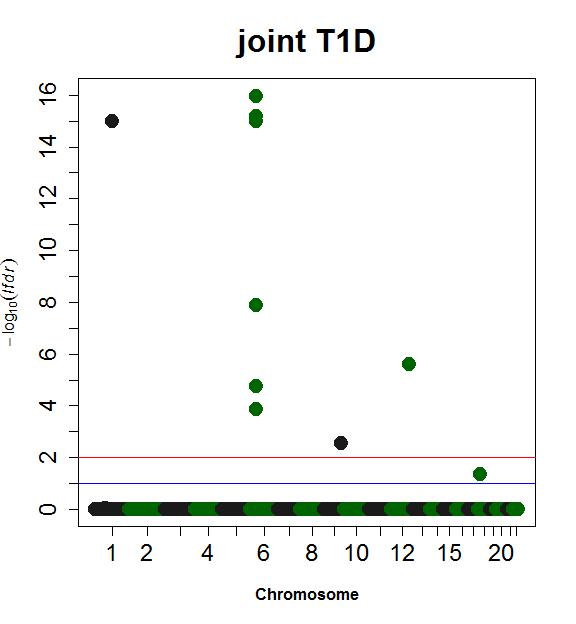}
    \end{subfigure}
    \begin{subfigure}[b]{0.24\textwidth}
        \centering
        \includegraphics[width=\textwidth]{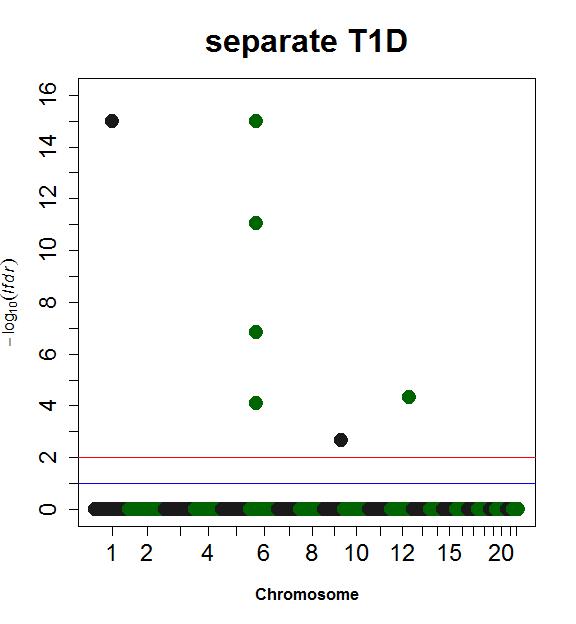}
    \end{subfigure}
    \begin{subfigure}[b]{0.24\textwidth}
        \centering
        \includegraphics[width=\textwidth]{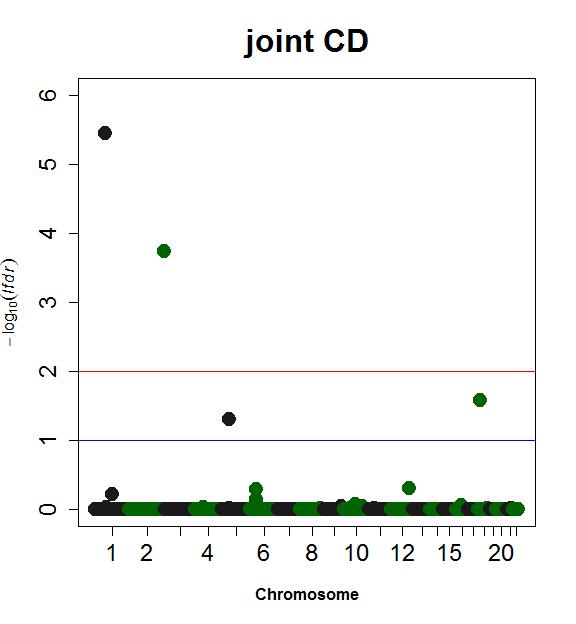}
    \end{subfigure}
    \begin{subfigure}[b]{0.24\textwidth}
        \centering
        \includegraphics[width=\textwidth]{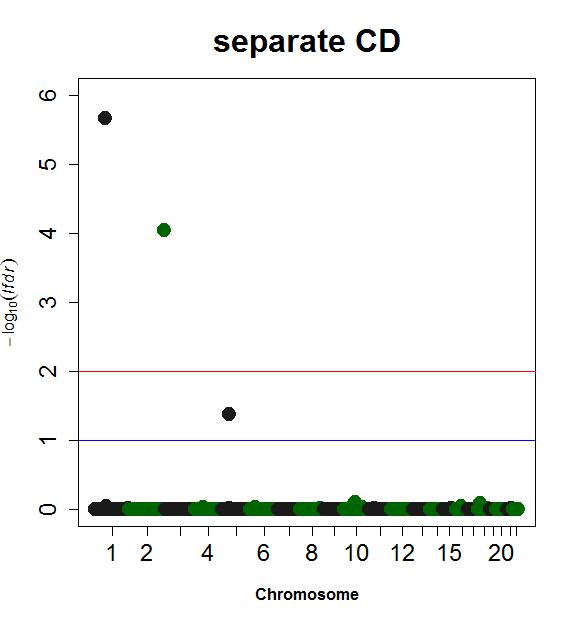}
    \end{subfigure}
   \caption{For the data consisting of 58C controls with CD and UKBS controls
with T1D, manhattan plots of separate analysis using BVSR and joint analysis using
FGP.}
\end{figure}

\begin{figure}[H]
  \centering
    \includegraphics[width=0.5\textwidth]{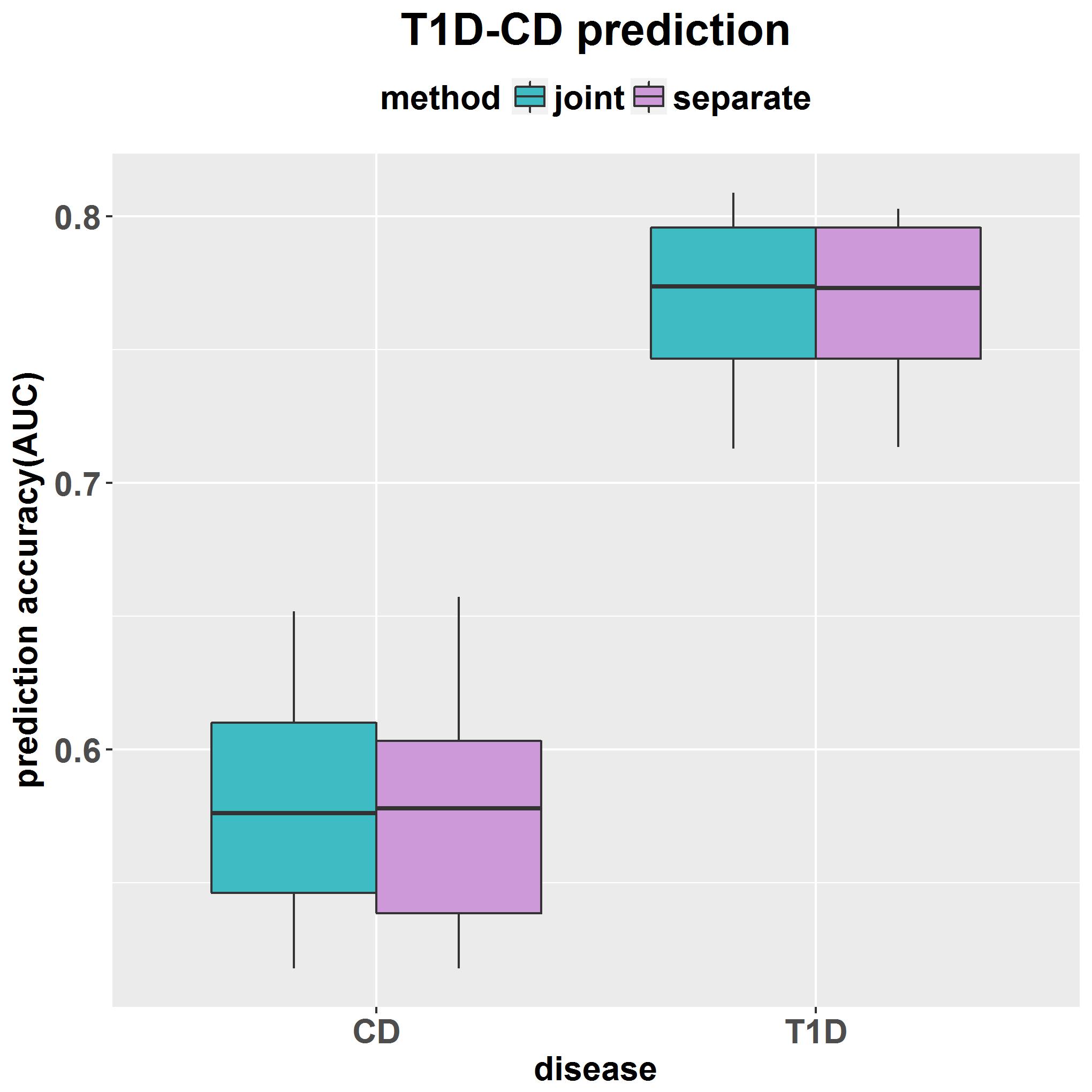}
  \caption{For the data consisting of 58C controls with CD and UKBS controls
with T1D, prediction performance of separate analysis using BVSR and joint analysis using
FGP.}
\end{figure}

% latex table generated in R 3.4.1 by xtable 1.8-2 package
% Sun Aug 27 15:14:49 2017
\begin{table}[H]
\centering
\begin{tabular}{rlll}
  \hline
 & Data & number of hits & prediction accuracy(AUC) \\
  \hline
1 & Type 1 diabetes(T1D)joint & 12 & 76.9\%(3.3\%) \\
  2 & Crohn's disease(CD)joint & 4 & 57.9\%(4.5\%) \\
  3 & Type 1 diabetes(T1D)separate & 10 & 76.8\%(3.2\%) \\
  4 & Crohn's disease(CD)separate & 3 & 57.9\%(4.7\%) \\
   \hline
\end{tabular}
\caption{For the data consisting of 58C controls with CD and UKBS controls
with T1D, summary of separate and joint analysis of T1D and CD}
\end{table}

% latex table generated in R 3.4.1 by xtable 1.8-2 package
% Sun Sep 10 14:56:27 2017
\begin{table}[H]
\centering
\begin{tabular}{rlrrllll}
  \hline
 & snp & chr & position & sep T1D(fdr) & sep CD(fdr) & joi T1D(fdr) & joi CD(fdr) \\
  \hline
1 & rs11805303 &   1 & 67675516 & 1e+00 & 2.2e-06* & 9.45e-01 & 3.58e-06* \\
  2 & rs6679677 &   1 & 114303808 & 0e+00* & 1e+00 & 0e+00* & 6.02e-01 \\
  3 & rs6752107 &   2 & 234161448 & 1e+00 & 9.06e-05* & 9.58e-01 & 1.81e-04* \\
  4 & rs17234657 &   5 & 40401509 & 1e+00 & 4.14e-02* & 9.61e-01 & 4.96e-02* \\
  5 & rs13200022 &   6 & 31098957 & 9.26e-12* & 1e+00 & 0e+00* & 8.11e-01 \\
  6 & rs550513 &   6 & 31920687 & 9.98e-01 & 1e+00 & 1.73e-05* & 8.28e-01 \\
  7 & rs3130287 &   6 & 32050544 & 0e+00* & 1e+00 & -2.22e-16* & 5.14e-01 \\
  8 & rs17421624 &   6 & 32066177 & 9.7e-01 & 1e+00 & 1.27e-08* & 8.66e-01 \\
  9 & rs9272346 &   6 & 32604372 & 0e+00* & 1e+00 & 0e+00* & 8.94e-01 \\
  10 & rs2070121 &   6 & 32781554 & 0e+00* & 1e+00 & 6.66e-16* & 7.21e-01 \\
  11 & rs10484565 &   6 & 32795032 & 0e+00* & 1e+00 & 1.11e-16* & 8.59e-01 \\
  12 & rs241427 &   6 & 32804414 & 1.51e-07* & 1e+00 & 1.34e-04* & 8.93e-01 \\
  13 & rs12529313 &   6 & 32817130 & 8.28e-05* & 1e+00 & 9.84e-01 & 9.98e-01 \\
  14 & rs10759987 &   9 & 121364134 & 2.15e-03* & 1e+00 & 2.84e-03* & 8.92e-01 \\
  15 & rs17696736 &  12 & 112486818 & 4.87e-05* & 1e+00 & 2.45e-06* & 4.93e-01 \\
  16 & rs2542151 &  18 & 12779947 & 1e+00 & 8.09e-01 & 4.59e-02* & 2.61e-02* \\
   \hline
\end{tabular}
\caption{For the data consisting of 58C controls with CD and UKBS controls
with T1D, list of SNPs of two modes: separate analysis and joint analysis. * denotes the local
fdr \textless 0.2, sep means separate, joi means joint.}
\end{table}
\clearpage
\subsection{Comparison of FGP and BVSR for the data consisting of 58C controls with T1D and UKBS
controls with CD exclude MHC region}\begin{figure}[H]
    \centering
    \begin{subfigure}[b]{0.24\textwidth}
        \centering
        \includegraphics[width=\textwidth]{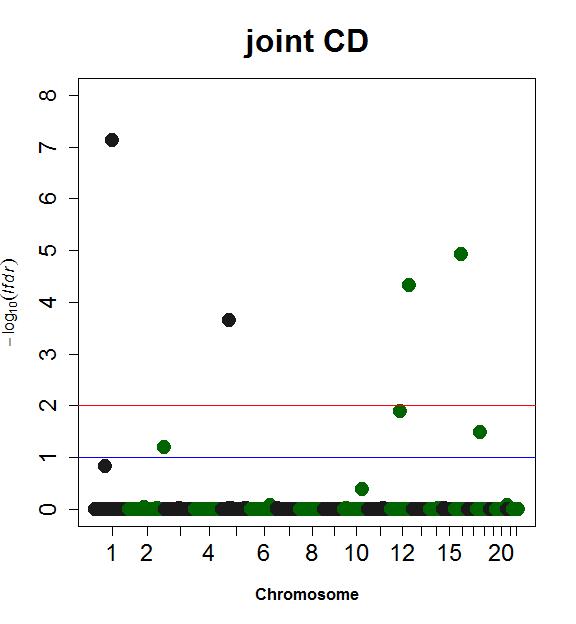}
    \end{subfigure}
    \begin{subfigure}[b]{0.24\textwidth}
        \centering
        \includegraphics[width=\textwidth]{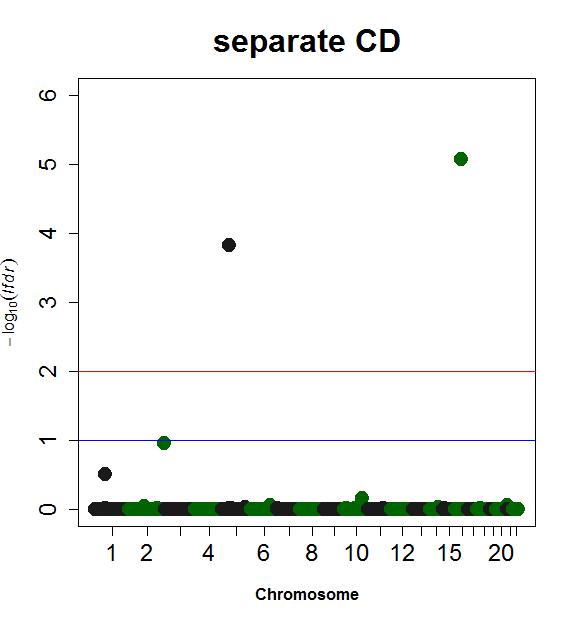}
    \end{subfigure}
    \begin{subfigure}[b]{0.24\textwidth}
        \centering
        \includegraphics[width=\textwidth]{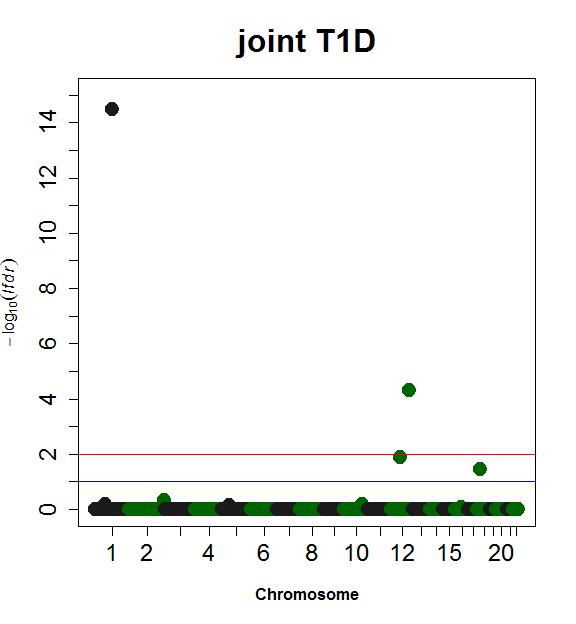}
    \end{subfigure}
    \begin{subfigure}[b]{0.24\textwidth}
        \centering
        \includegraphics[width=\textwidth]{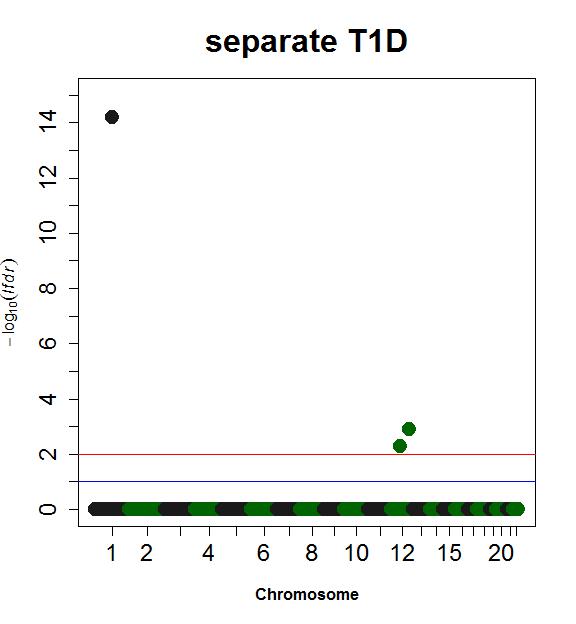}
    \end{subfigure}
   \caption{For the data consisting of 58C controls with T1D and UKBS controls
with CD exclude MHC region SNPs, manhattan plots of separate analysis using BVSR and joint analysis using
FGP.}
\end{figure}

\begin{figure}[H]
  \centering
    \includegraphics[width=0.5\textwidth]{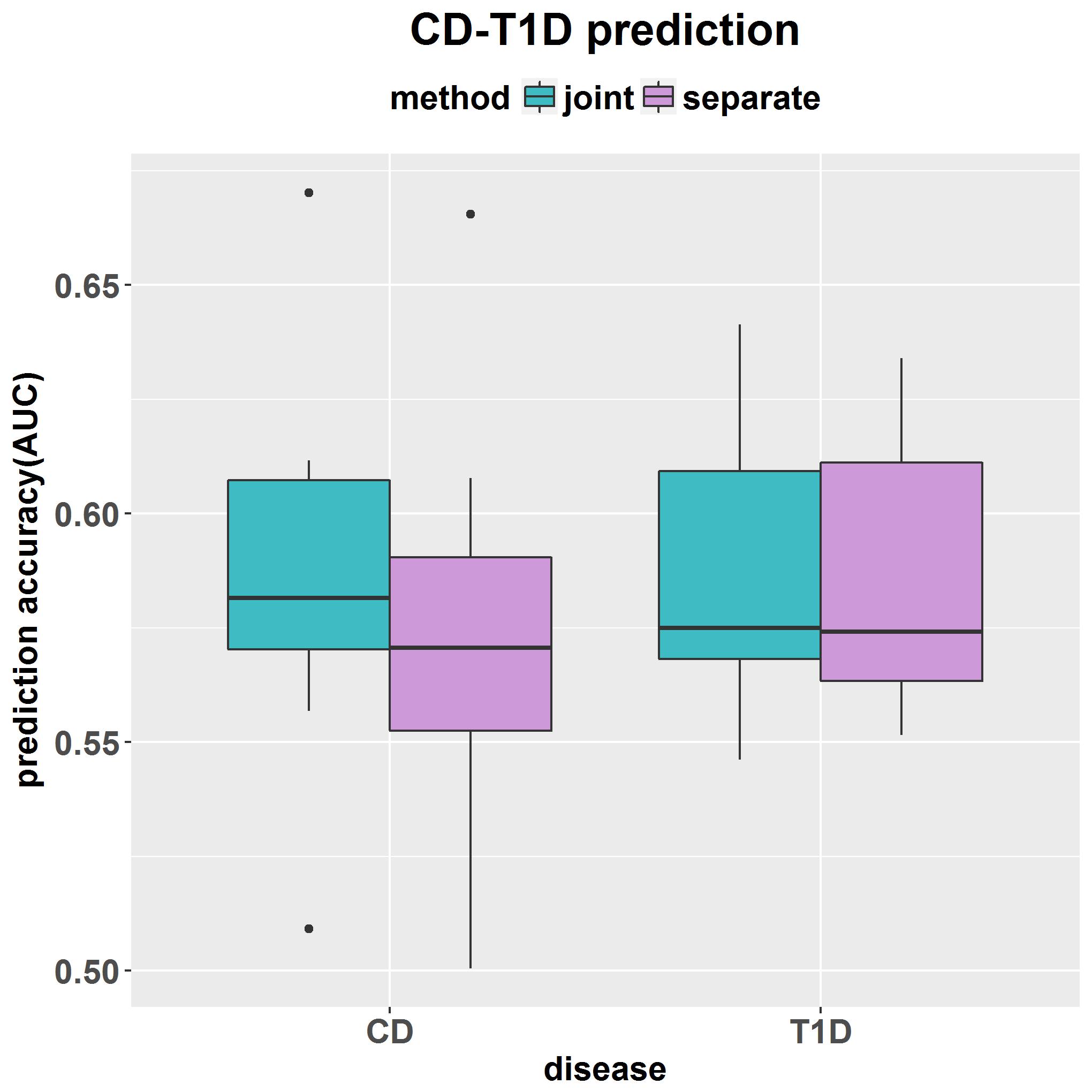}
  \caption{For the data consisting of 58C controls with T1D and UKBS controls
with CD exclude MHC region SNPs, prediction performance of separate analysis using BVSR and joint analysis using
FGP.}
\end{figure}

% latex table generated in R 3.4.1 by xtable 1.8-2 package
% Sun Aug 27 15:14:49 2017
\begin{table}[H]
\centering
\begin{tabular}{rlll}
  \hline
 & Data & number of hits & prediction accuracy(AUC) \\
  \hline
1 & Crohn's disease(CD)joint & 8 & 58.6\%(4.2\%) \\
  2 & Type 1 diabetes(T1D)joint & 4 & 58.6\%(3\%) \\
  3 & Crohn's disease(CD)separate & 3 & 57.4\%(4.4\%) \\
  4 & Type 1 diabetes(T1D)separate & 3 & 58.5\%(2.8\%) \\
   \hline
\end{tabular}
\caption{For the data consisting of 58C controls with T1D and UKBS controls
with CD exclude MHC region, summary of separate and joint analysis of CD and T1D}
\end{table}

% latex table generated in R 3.4.1 by xtable 1.8-2 package
% Sun Sep 10 14:56:27 2017
\begin{table}[H]
\centering
\begin{tabular}{rlrrllll}
  \hline
 & snp & chr & position & sep CD(fdr) & sep T1D(fdr) & joi CD(fdr) & joi T1D(fdr) \\
  \hline
1 & rs4655679 &   1 & 67599657 & 3.11e-01 & 1e+00 & 1.5e-01* & 6.45e-01 \\
  2 & rs6679677 &   1 & 114303808 & 1e+00 & 6.44e-15* & 7.53e-08* & 3.33e-15* \\
  3 & rs10210302 &   2 & 234158839 & 1.09e-01* & 1e+00 & 6.33e-02* & 4.75e-01 \\
  4 & rs9292777 &   5 & 40437948 & 1.49e-04* & 1e+00 & 2.24e-04* & 7.42e-01 \\
  5 & rs11171739 &  12 & 56470625 & 1e+00 & 5.13e-03* & 1.29e-02* & 1.29e-02* \\
  6 & rs17696736 &  12 & 112486818 & 1e+00 & 1.24e-03* & 4.76e-05* & 4.76e-05* \\
  7 & rs2066843 &  16 & 50745199 & 8.5e-06* & 1e+00 & 1.21e-05* & 8.3e-01 \\
  8 & rs2542151 &  18 & 12779947 & 9.69e-01 & 9.97e-01 & 3.28e-02* & 3.65e-02* \\
   \hline
\end{tabular}
\caption{For the data consisting of 58C controls with T1D and UKBS controls
with CD exclude MHC region, list of SNPs of two modes: separate analysis and joint analysis. * denotes the local
fdr \textless 0.2, sep means separate, joi means joint.}
\end{table}
\clearpage
\subsection{Comparison of FGP and BVSR for the data consisting of 58C controls with CD and UKBS
controls with T1D exclude MHC region}
\begin{figure}[H]
    \centering
    \begin{subfigure}[b]{0.24\textwidth}
        \centering
        \includegraphics[width=\textwidth]{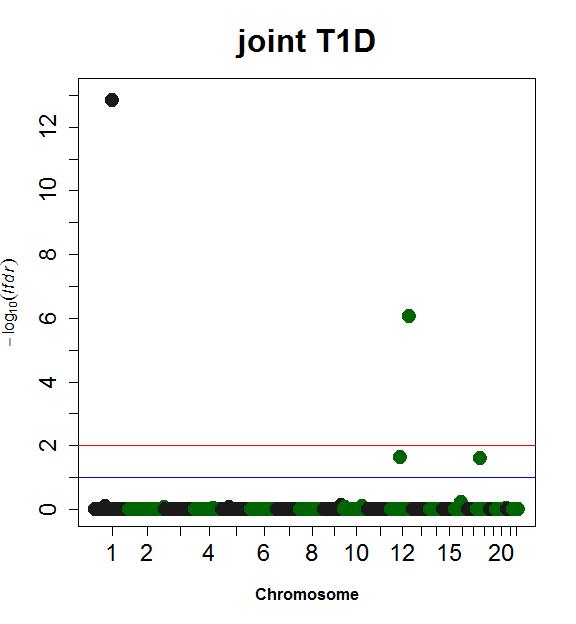}
    \end{subfigure}
    \begin{subfigure}[b]{0.24\textwidth}
        \centering
        \includegraphics[width=\textwidth]{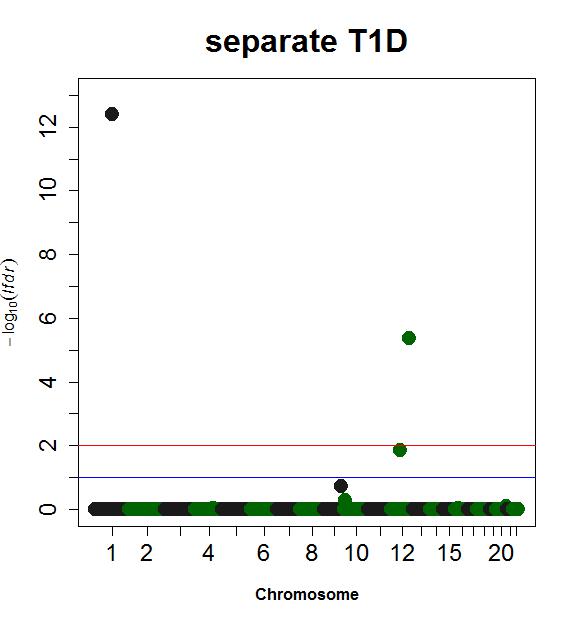}
    \end{subfigure}
    \begin{subfigure}[b]{0.24\textwidth}
        \centering
        \includegraphics[width=\textwidth]{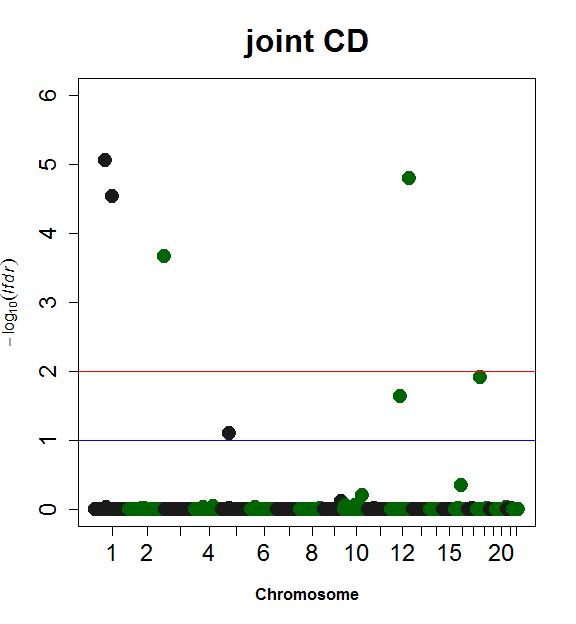}
    \end{subfigure}
    \begin{subfigure}[b]{0.24\textwidth}
        \centering
        \includegraphics[width=\textwidth]{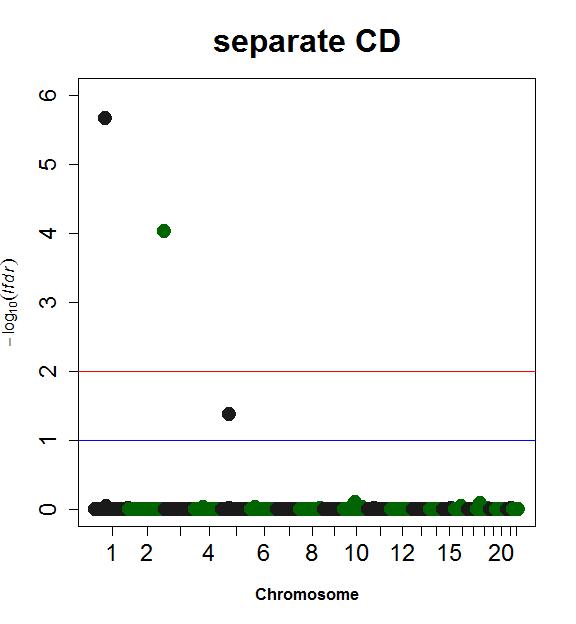}
    \end{subfigure}
   \caption{For the data consisting of 58C controls with CD and UKBS controls
with T1D exclude MHC region SNPs, manhattan plots of separate analysis using BVSR and joint analysis using
FGP.}
\end{figure}

\begin{figure}[H]
  \centering
    \includegraphics[width=0.5\textwidth]{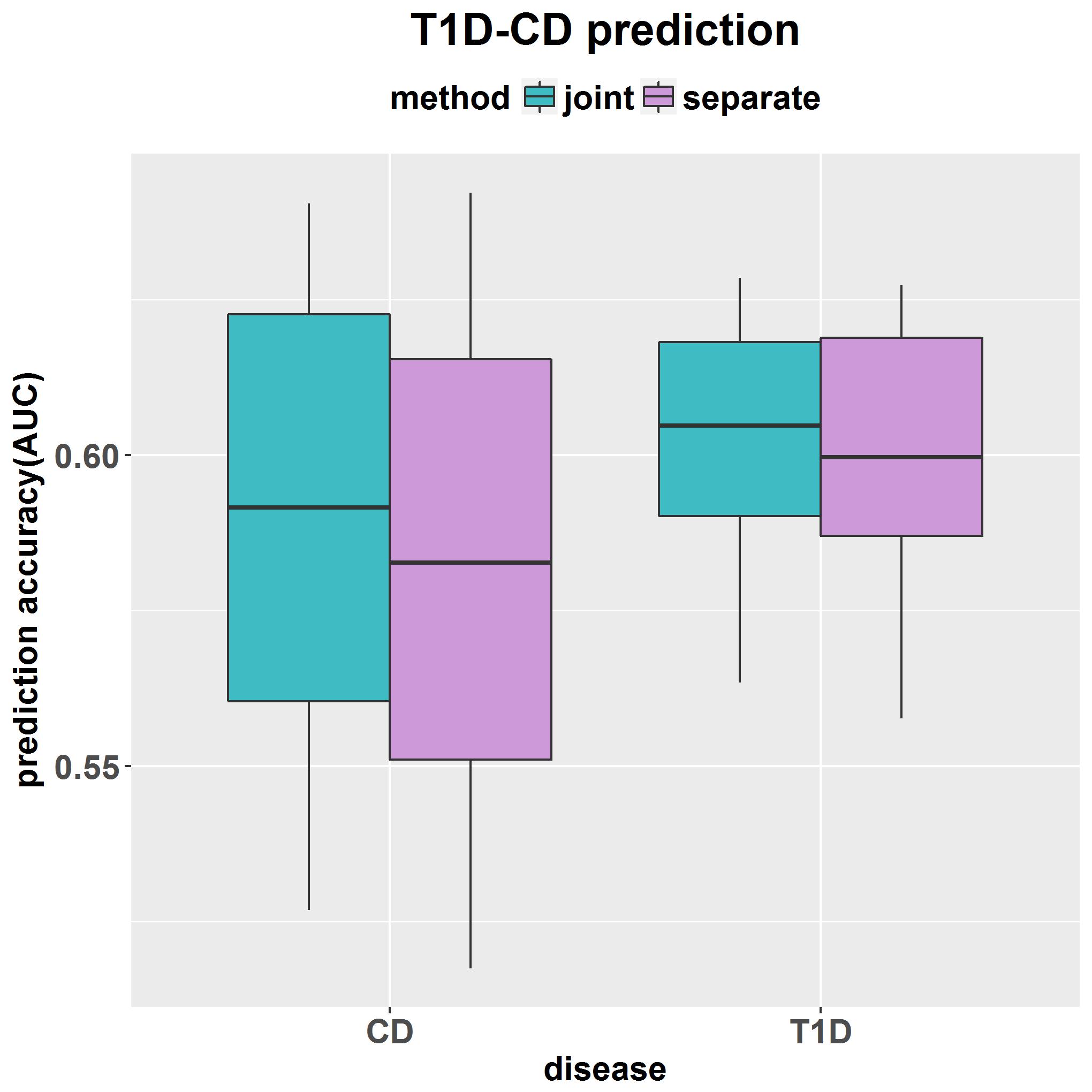}
  \caption{For the data consisting of 58C controls with CD and UKBS controls
with T1D exclude MHC region SNPs, prediction performance of separate analysis using BVSR and joint analysis using
FGP.}
\end{figure}

% latex table generated in R 3.4.1 by xtable 1.8-2 package
% Sun Aug 27 15:14:49 2017
\begin{table}[H]
\centering
\begin{tabular}{rlll}
  \hline
 & Data & number of hits & prediction accuracy(AUC) \\
  \hline
1 & Type 1 diabetes(T1D)joint & 4 & 60.3\%(2\%) \\
  2 & Crohn's disease(CD)joint & 7 & 58.7\%(4.1\%) \\
  3 & Type 1 diabetes(T1D)separate & 4 & 60.1\%(2.3\%) \\
  4 & Crohn's disease(CD)separate & 3 & 58.1\%(4.4\%) \\
   \hline
\end{tabular}
\caption{For the data consisting of 58C controls with CD and UKBS controls
with T1D exclude MHC region, summary of separate and joint analysis of T1D and CD}
\end{table}

% latex table generated in R 3.4.1 by xtable 1.8-2 package
% Sun Sep 10 14:56:27 2017
\begin{table}[H]
\centering
\begin{tabular}{rlrrllll}
  \hline
 & snp & chr & position & sep T1D(fdr) & sep CD(fdr) & joi T1D(fdr) & joi CD(fdr) \\
  \hline
1 & rs11805303 &   1 & 67675516 & 1e+00 & 2.18e-06* & 8.17e-01 & 8.89e-06* \\
  2 & rs6679677 &   1 & 114303808 & 3.98e-13* & 1e+00 & 1.44e-13* & 2.94e-05* \\
  3 & rs6752107 &   2 & 234161448 & 1e+00 & 9.34e-05* & 8.44e-01 & 2.15e-04* \\
  4 & rs17234657 &   5 & 40401509 & 1e+00 & 4.14e-02* & 8.55e-01 & 7.8e-02* \\
  5 & rs10759987 &   9 & 121364134 & 1.88e-01* & 1e+00 & 7.56e-01 & 7.56e-01 \\
  6 & rs2292239 &  12 & 56482180 & 1.41e-02* & 1e+00 & 2.27e-02* & 2.28e-02* \\
  7 & rs17696736 &  12 & 112486818 & 4.17e-06* & 1e+00 & 8.75e-07* & 1.6e-05* \\
  8 & rs2542151 &  18 & 12779947 & 9.99e-01 & 8.11e-01 & 2.49e-02* & 1.22e-02* \\
   \hline
\end{tabular}
\caption{For the data consisting of 58C controls with CD and UKBS controls
with T1D exclude MHC region, list of SNPs of two modes: separate analysis and joint analysis. * denotes the local
fdr \textless 0.2, sep means separate, joi means joint.}
\end{table}

\clearpage
\section{Proof detail for Quantitative trait model}
%\subsection{The definition of lower bound}
%For the ease of presentation, we
%\begin{equation}
%\begin{aligned}
%\log \text{Pr}(&\bfy_1,\bfy_2| \bfX_1,\bfX_2,\bftheta)\\
%=&\log\sum_{\bfgamma_1,\bfgamma_2}\int_{\bfbeta_1,\bfbeta_2}
%\text{Pr}(\bfy_1,\bfy_2,\bfbeta_1,\bfbeta_2,\bfgamma_1,\bfgamma_2|
%\bfX_1,\bfX_2,\bftheta)d\bfbeta_1d\bfbeta_2\\
%\geq& \log\sum_{\bfgamma_1,\bfgamma_2}\int_{\bfbeta_1,\bfbeta_2}
%q(\bfbeta_1,\bfbeta_2,\bfgamma_1,\bfgamma_2)
%\frac{\text{Pr}(\bfy_1,\bfy_2,\bfbeta_1,\bfbeta_2,\bfgamma_1,\bfgamma_2|
%\bfX_1,\bfX_2,\bftheta)}
%{q(\bfbeta_1,\bfbeta_2,\bfgamma_1,\bfgamma_2)}
%d\bfbeta_1d\bfbeta_2\\
%%%%%%%%%%%%%%%%%%%%%%%%%%%%%%%%%%%%%%%%%%%%%%%%%%%%%%%%%%%%%%%%%%%%%%%%%%%%%%%%%%%%%%%%%%%%%%%%%%%%%%%%%%%%%%%%%%%
%=&\mathbb{E}_q[\log\text{Pr}(\bfy_1,\bfy_2,\bfbeta_1,\bfbeta_2,\bfgamma_1,\bfgamma_2|
%\bfX_1,\bfX_2,\bftheta)] - \mathbb{E}_q[\log q(\bfbeta_1,\bfbeta_2,\bfgamma_1,\bfgamma_2)]\\
%:=&\mathcal{L}(q)
%\end{aligned}
%\end{equation}
%where $\mathcal{L}(q)$ denotes the lower bound.
\label{alg_qtm}
To overcome the intractability of marginal likelihood, we derive an efficient algorithm based on variational inference, which makes our model scalable to genome-wide data analysis. The key idea is that we make use of Jensen's inequality to iteratively obtain an adjustable lower bound on the marginal log likelihood \citep{jordan1999introduction}. First, we have a lower bound of the logarithm of the marginal likelihood,
%\begin{eqnarray}
%\label{lob}
%& &\log \pr(\bfy_1,\bfy_2| \bfX_1,\bfX_2;\bftheta) \notag\\ &\geq&\mathbb{E}_q[\log\pr(\bfy_1,\bfy_2,\bfbeta_1,\bfbeta_2,\bfgamma_1,\bfgamma_2| \bfX_1,\bfX_2;\bftheta)] - \mathbb{E}_q[\log q(\bfbeta_1,\bfbeta_2,\bfgamma_1,\bfgamma_2)]
%:=\mathcal{L}(q),
%\end{eqnarray}
\begin{eqnarray}
\label{lob}
& &\log \pr(\bfy_1,\bfy_2| \bfX_1,\bfX_2;\bftheta) = \mathcal{L}(q,\bftheta) + \mathbb{KL}(q||p)
\notag \\
&\ge & \mathbb{E}_q[\log\pr(\bfy_1,\bfy_2,\bfbeta_1,\bfbeta_2,\bfgamma_1,\bfgamma_2| \bfX_1,\bfX_2;\bftheta)] - \mathbb{E}_q[\log q(\bfbeta_1,\bfbeta_2,\bfgamma_1,\bfgamma_2)],
\end{eqnarray}
where we define
\begin{eqnarray}
\label{lb1}
  \mathcal{L}(q,\bftheta) &= &\sum_{\bfbeta_1,\bfbeta_2,\bfgamma_1,\bfgamma_2} q(\bfbeta_1,\bfbeta_2,\bfgamma_1,\bfgamma_2)\log \frac{p(\bfy_1,\bfy_2, \bfbeta_1,\bfbeta_2,\bfgamma_1,\bfgamma_2|\bfX_1,\bfX_2;\bftheta) }{q(\bfbeta_1,\bfbeta_2,\bfgamma_1,\bfgamma_2)}, \notag \\
  \mathbb{KL}(q||p) &=& \sum_{\bfbeta_1,\bfbeta_2,\bfgamma_1,\bfgamma_2} q(\bfbeta_1,\bfbeta_2,\bfgamma_1,\bfgamma_2)\log \frac{q(\bfbeta_1,\bfbeta_2,\bfgamma_1,\bfgamma_2) }{p(\bfbeta_1,\bfbeta_2,\bfgamma_1,\bfgamma_2 | \bfy_1, \bfy_2, \bfX_1, \bfX_2; \bftheta)} .
\end{eqnarray}
%where the inequality follows Jensen's inequality and the equality holds if, and only if, $q(\bfbeta_1,\bfbeta_2,\bfgamma_1,\bfgamma_2)$ is the true posterior $p^*=\pr(\bfbeta_1,\bfbeta_2,\bfgamma_1,\bfgamma_2 | \bfy_1, \bfy_2, \bfX_1, \bfX_2; \bftheta)$. One can easily verify that the difference between left-hand side and right-hand side of the lower bound (\ref{lob}) is the Kullback-Leibler (KL) divergence between the variational posterior probability ($q$) and the true posterior probability ($p^*$),
Note that Kullback-Leibler (KL) divergence satisfies $\mathbb{KL}(q||p)\ge 0$ by using Jensen's inequality, with equality holds if, and only if, that variational posterior probability ($q$) and the true posterior probability ($p$) are equal. Similar to expectation-maximization (EM) algorithm, we can maximize the lower bound $\mathcal{L}(q,\bftheta)$ with respect to the variational distribution $q$, which is equivalent to minimizing the KL divergence~\citep{bishop2006pattern}. To make it computationally efficient to evaluate the lower bound, we use mean-field theory~\citep{opper2001advanced}, and assume that $q(\bfbeta_1,\bfbeta_2,\bfgamma_1,\bfgamma_2)$ can be factorized as %Thus, in each iteration, it is equivalent to minimize the KL divergence with $q$ subject to some factorization. To be specific, we use mean-field method \citep{parisi1988statistical,opper2001advanced}, assuming that $q(\bfbeta_1,\bfbeta_2,\bfgamma_1,\bfgamma_2)$ can be factorized as
\begin{equation}
\label{factorization}
q(\bfbeta_1,\bfbeta_2,\bfgamma_1,\bfgamma_2) =
\prod_{j=1}^p q_j(\bfbeta_{1j},\bfbeta_{2j},\bfgamma_{1j},\bfgamma_{2j}).
\end{equation}
No additional assumption on the posterior distribution is required. This factorization~(\ref{factorization}) is used as a surrogate for the posterior distribution $\pr(\bfbeta_1,\bfbeta_2,\bfgamma_1,\bfgamma_2| \bfy_1,\bfy_2, \bfX_1,\bfX_2;\bftheta)$.

Using the properties of factorized distributions in variational inference \citep{bishop2006pattern}, we can obtain the optimal approximation using the following formula:
\begin{equation}
\log q_j(\beta_{1j},\beta_{2j},\gamma_{1j},\gamma_{2j}) =
\mathbb{E}_{j'\neq j}[\log \pr(\bfy_1,\bfy_2,\bfbeta_1,\bfbeta_2,\bfgamma_1,\bfgamma_2|
\bfX_1,\bfX_2,\bftheta)]
 + \mbox{const}
\end{equation}
where the expectation is taken with respect to all of the other factors
$\{q_{j'}(\beta_{1j'},\beta_{2j'},\gamma_{1j'},\gamma_{2j'})\}$ for $j'\neq j$. After some derivations, we have
\begin{equation}
\label{variational_q}
q(\beta_{1j},\beta_{2j},\gamma_{1j},\gamma_{2j}) = f_{1j}(\beta_{1j})^{\gamma_{1j}}f_0(\beta_{1j})^{1-\gamma_{1j}} f_{2j}(\beta_{2j})^{\gamma_{2j}}f_0(\beta_{2j})^{1-\gamma_{2j}} \prod_{l}\alpha_{lj}^{\textbf{1}{(\gamma_{1k}=l_1,\gamma_{2k}=l_2)}}, \\
%&=&
%[\mathcal{N}(\mu_{1j},s_{1j}^2)]^{\gamma_{1j}}[\mathcal{N}(0,\sigma_{\beta_1}^2)]^{1 - \gamma_{1j}}
%[\mathcal{N}(\mu_{2j},s_{2j}^2)]^{\gamma_{2j}}[\mathcal{N}(0,\sigma_{\beta_2}^2)]^{1 - \gamma_{2j}}
%\prod_{l}\phi_{lj}^{\textbf{1}_{(\gamma_{1k}=l_1,\gamma_{2k}=l_2)}}
\end{equation}
where $\alpha_{lj}$ is the posterior probability of $[\gamma_{1j},\gamma_{2j}] = l$, $f_0(\beta_{kj})$ is the posterior distribution of $\beta_{kj}$ when $\gamma_{kj}=0$, $f_{kj}(\beta_{kj})$ is the posterior distribution of $\beta_{kj}$ under $\gamma_{kj}=1$. With some algebra, it is easy to show that $f_0(\beta_{kj})$ and $f_{kj}(\beta_{kj})$ are the density functions of Gaussian distributions $\mathcal{N}(0,\sigma_{\beta_k}^2)$ and $\mathcal{N}(\mu_{kj},s_{kj}^2)$ with
\begin{equation}
\label{var_u}
\mu_{kj} =  \frac{\bfx_{kj}^\top\bfy_k -
   \sum_{j'\neq j}^p\mathbb{E}_{j'}[\gamma_{kj'}\beta_{kj'}]
                    \bfx_{kj}^\top\bfx_{kj'}}
                 {\bfx_{kj}^\top\bfx_{kj} + \frac{\sigma_{e_k}^2}{\sigma_{\beta_k}^2}},\qquad
s_{kj}^2 = \frac{\sigma_{e_k}^2}
           {\bfx_{kj}^\top\bfx_{kj} + \frac{\sigma_{e_k}^2}{\sigma_{\beta_k}^2}}.\\
\end{equation}
As discussed in \cite{carbonetto2012scalable}, both $s^2_{kj}$ and $\mu_{kj}$ can be interpreted using a single-variable linear model $\bfy_k = \bfx_{kj}\beta_{kj} + \bfepsilon_k$, \ie, update for $s_{kj}^2$ is the posterior variance of effect $\beta_{kj}$ and update for $\mu_{kj}$ is the posterior mean of effect $\beta_{kj}$ by correcting correlations among covariates not included in the single-variable model. With some algebra, we can update log odds of $\alpha_{lj}$ as follows,
\begin{eqnarray}
\label{var_gam}
  A_{00j} & =& \log\alpha_{00}+ \frac{1}{2}\log\sigma_{\beta_1}^2 + \frac{1}{2}\log\sigma_{\beta_2}^2 , \notag\\
  A_{10j} & =& \log\alpha_{10} +\frac{1}{2}\log s_{1j}^2 +\frac{1}{2}\log\sigma_{\beta_2}^2 + \frac{\mu_{1j}^2}{2s_{1j}^2},\notag\\
  A_{01j} & =& \log\alpha_{01} +\frac{1}{2}\log\sigma_{\beta_1}^2+\frac{1}{2}\log s_{2j}^2  + \frac{\mu_{2j}^2}{2s_{2j}^2},\notag\\
  A_{11j} & =& \log\alpha_{11} +\frac{1}{2}\log s_{1j}^2 +\frac{1}{2}\log s_{2j}^2 + \frac{\mu_{1j}^2}{2s_{1j}^2}+ \frac{\mu_{2j}^2}{2s_{2j}^2},\notag\\
  \alpha_{lj} &=& \frac{\exp(A_{lj})}{\sum_{l\in \{00,10,01,11\}} \exp(A_{lj})}.
\end{eqnarray}
Since we take $q(\bfbeta_1,\bfbeta_2,\bfgamma_1,\bfgamma_2)$ as a surrogate to approximate the true posterior probability, the resulting variational probability (\ref{variational_q}) has a nice interpretation. First, the marginal probability $(\alpha_{10j} + \alpha_{11j})$ and $(\alpha_{01j} + \alpha_{11j})$ can be viewed as an approximation of $\pr(\gamma_{1j} = 1|\bfy_1,\bfy_2, \bfX_1, \bfX_2; \bftheta)$ and $\pr(\gamma_{2j} = 1|\bfy_1,\bfy_2, \bfX_1, \bfX_2; \bftheta)$, respectively. Clearly, pair-wise difference of $A_{lj}$ in expression (\ref{var_gam}) is the posterior log odds, \eg,~$A_{10j}- A_{00j} = \log\frac{\alpha_{10}}{\alpha_{00}} + \frac{1}{2} \log\frac{s_{1j}^2}{\sigma_{\beta_1}^2} + \frac{\mu_{1j}^2}{2s_{1j}^2}$ is the posterior log odds of $\beta_{1j}\ne 0$.
%the posterior of $\gamma_{kj} = 1$ is affected by its prior and the evidence
%from both individual-level data sets.
In the case that the $j$-th SNP is irrelevant to the first phenotype $(\gamma_{1j} = 0)$, the approximated posterior distribution of $\beta_{1j}$ remains the same as its prior, \ie, $\beta_{1j}\sim \mathcal{N}(0,\sigma_{\beta_1}^2)$. On the contrary, when $\gamma_{1j} = 1$, the posterior distribution becomes $\mathcal{N}(\mu_{1j},s_{1j}^2)$, where $\mu_{1j}$ is the posterior expectation of $\beta_{1j}$ adjusted for all other variables.

With $q(\bfbeta_1,\bfbeta_2,\bfgamma_1,\bfgamma_2)$ given in (\ref{variational_q}), we can derive the lower bound analytically. Once we have variational lower bound, other parameters can be updated by maximizing the lower bound while keeping variational distribution $q$ fixed:
\begin{equation}
\label{mle_update}
\begin{aligned}
&\sigma_{e_k}^2=\frac{1}{n_k}\Big(\|\bfy_k - \sum_{j=1}^p \sum_{l\in L_k}\alpha_{lj}\mu_{kj}\bfx_{kj}\|^2 +\sum_{j=1}^p
         (\sum_{l\in L_k}\alpha_{lj}(s_{1j}^2 + \mu_{1j}^2) - (\sum_{l\in L_k}\alpha_{lj})^2\mu_{kj}^2)
         \bfx_{kj}^\top\bfx_{kj}\Big),\\
%\sigma_{e_2}^2=&\frac{1}{n_2}\Big(\|\bfy_2 - \sum_{j=1}^p(\alpha_{01j} + \alpha_{11j})\mu_{2j}\bfx_{2j}\|^2\\
%         &+\sum_{j=1}^p
%         ((\alpha_{01j} + \alpha_{11j})(s_{2j}^2 + \mu_{2j}^2) - (\alpha_{01j} + \alpha_{11j})^2\mu_{2j}^2)
%         \bfx_{2j}^\top\bfx_{2j}\Big)\\
&\sigma_{\beta_k}^2 = \frac{\sum_{j=1}^p\sum_{l\in L_k}\alpha_{lj}(\mu_{kj}^2 + s_{kj}^2)}
{\sum_{j=1}^p\sum_{l\in L_k}\alpha_{lj}},  \quad \alpha_l = \frac{\sum_{j=1}^p\alpha_{lj}}{p}, \forall l\in\{00,10,01,11\},
\end{aligned}
\end{equation}
where $L_1=\{10,11\},L_2=\{01,11\}$. Clearly, $\sigma^2_{e_k}$ is equal to the its maximum likelihood estimates adjusted by the variance of $\gamma_{kj}\beta_{kj}$, update for $\sigma^2_{\beta_k}$ is the weighted average of posterior variance, where the weights are the posterior mean of $\gamma_{kj}$, and update for $\alpha_l$ is the average of all posterior mean of $\gamma_{kj}$. Derivation details for parameter updates in Gaussian distribution~(\ref{var_u}), and update equations (\ref{var_gam}) and (\ref{mle_update}) can be found later. The VBEM algorithm (Algorithm S1) performs similarly to coordinate descent algorithm, which comes from the factorization of variational distribution (\ref{variational_q}). Hence, VBEM algorithm developed here is scalable to ultra-high dimensions.

\subsection{The derivation of lower bound}
The lower bound (8) of quantitative FGP model can be written as%As the definition of lower bound is as following, we can derive the analytical form of lower bound.
\begin{equation}
\begin{aligned}
\label{def lowerbound}
\mathcal{L}(q) = &\mathbb{E}_q
[\text{Pr}(\bfy_1,\bfy_2,\bfbeta_1,\bfbeta_2,\bfgamma_1,\bfgamma_2
|\bfX_1,\bfX_2;\bftheta)]
- \mathbb{E}_q
[\log q(\bfbeta_1,\bfbeta_2,\bfgamma_1,\bfgamma_1)].
\end{aligned}
\end{equation}
Algebraically, the first term of lower bound (\ref{def lowerbound}) is
\begin{equation}
\begin{aligned}
\label{firstterm}
\mathbb{E}_q\log[\text{Pr}(&\bfy_1,\bfy_2,\bfbeta_1,\bfbeta_2,\bfgamma_1,\bfgamma_2
|\bfX_1,\bfX_2;\bftheta)]\\
=&\sum_{k=1}^2\Bigg(-\frac{n_k}{2}\log(2\pi \sigma_{e_k}^2)
 - \frac{\bfy_k^T\bfy_k}{2\sigma_{e_k}^2}
  + \frac{\sum_{j=1}^p\mathbb{E}_{qj}[\gamma_{kj}{\beta}_{kj}]\bfx_{kj}^T\bfy_k}{\sigma_{e_k}^2}\\
&-\frac{1}{2\sigma_{e_k}^2}
\sum_{j=1}^p\Big(\mathbb{E}_{qj}[\gamma_{kj}{\beta}_{kj}]^2\bfx_{kj}^T\bfx_{kj}
+\sum_{j^\prime\neq j}^p\mathbb{E}_{qjj'}[\gamma_{kj}{\beta}_{kj}
                                           \gamma_{kj^\prime}{\beta}_{kj^\prime}]
                                            \bfx_{kj}^T\bfx_{kj^\prime}\Big)\\
&-\frac{p}{2}\log(2\pi\sigma_{\beta_k}^2)
-\frac{1}{2\sigma_{\beta_k}^2}\sum_{j=1}^p\mathbb{E}_{qj}{\beta}_{kj}^2\Bigg) +\sum_{j=1}^p\sum_{l}\mathbb{E}_{qj}[\textbf{1}_{([\gamma_{1j},\gamma_{2j}] = l)}]\log \alpha_l
\end{aligned}
\end{equation}
where the variational expectations in (\ref{firstterm}) are listed as below
\begin{equation}
\begin{aligned}
\label{expectation}
&\mathbb{E}_{qj}[\gamma_{kj}{\beta}_{kj}] = \sum_{l\in L_k}\alpha_{lj}\mu_{1j},\\
&\mathbb{E}_{qj}[\gamma_{kj}{\beta}_{kj}]^2 = \sum_{l\in L_k}\alpha_{lj}(\mu_{kj}^2 + s_{kj}^2),\\
&\mathbb{E}_{qjj'}[\gamma_{kj}{\beta}_{kj}\gamma_{kj^\prime}{\beta}_{kj^\prime}]
=\sum_{l\in L_k}\alpha_{lj}\mu_{1j}
\sum_{l\in L_k}\alpha_{lj'}\mu_{1j'},\\
&\mathbb{E}_{qj}{\beta}_{kj}^2=\sum_{l\in L_k}\alpha_{lj}(\mu_{kj}^2 + s_{kj}^2)
                                   +\sum_{l\in L/L_k}\alpha_{lj}\sigma_{\beta_{1}}^2,\\
&\mathbb{E}_{qj}[\textbf{1}_{([\gamma_{1j},\gamma_{2j}] = l)}] = \alpha_{lj}.
\end{aligned}
\end{equation}
The second term of lower bound (\ref{def lowerbound}) is the entropy of posterior distribution.
\begin{equation}
\begin{aligned}
\label{secondterm}
 &- \mathbb{E}_q[\log q(\bfbeta_1,\bfbeta_2,\bfgamma_1,\bfgamma_2)]\\
=& -\sum_{j=1}^p\mathbb{E}_{qj}[\log q({\beta}_{1j},{\beta}_{2j},\gamma_{1j},\gamma_{2j})]\\
=\sum_{k=1}^2\Big(&\frac{p}{2}\log\sigma_{\beta_k}^2 + \frac{1}{2}\sum_{j=1}^p\sum_{l\in L_k}\alpha_{lj}\log\frac{s_{kj}^2}{\sigma_{\beta_k}^2}\Big) - \sum_{j=1}^p\sum_{l\in L}\alpha_{lj}\log{\alpha_{lj}}
\end{aligned}
\end{equation}
Obviously, taking some simplification, we can derive the analytical from of lower bound\\
\begin{equation}
\begin{aligned}
\label{lowerbound}
& \mathcal{L}(q)\\
= &\sum_{k=1}^2\Bigg(-\frac{n_k}{2}\log(2\pi \sigma_{e_k}^2)
- \frac{\|\bfy_k - \sum_{j=1}^p\sum_{l\in L_k}\alpha_{lj}\mu_{kj}\bfx_{kj}\|^2}
         {2\sigma_{e_k}^2}\\
         &-\frac{1}{2\sigma_{e_k}^2}
         \sum_{j=1}^p\mathbb{V}ar[\gamma_{kj}{\beta}_{kj}]\bfx_{kj}^T\bfx_{kj}\Bigg)
-\sum_{j=1}^p\sum_{l}\alpha_{lj}(\log\frac{\alpha_{lj}}{\alpha_l})\\
&+\frac{1}{2}\sum_{k=1}^{2}\sum_{j=1}^p{\sum_{l\in L_k}\alpha_{lj}}\Big(\log\frac{s_{kj}^2}{\sigma_{\beta_k}^2}
- \frac{\mu_{kj}^2 + s_{kj}^2}{\sigma_{\beta_k}^2} + 1\Big) - p\log(2\pi) - p
\end{aligned}
\end{equation}
where
\begin{equation}
\begin{aligned}
\mathbb{V}ar[\gamma_{kj}{\beta}_{kj}] =&
(\sum_{l\in L_k}\alpha_{lj})(\mu_{kj}^2+s_{kj}^2) - (\sum_{l\in L_k}\alpha_{lj})^2\mu_{kj}^2.
\end{aligned}
\end{equation}
\subsection{The derivation of posterior distribution}
We derive the variational posterior distribution by maximizing the lower bound (\ref{lowerbound}). The derivative function with respective to $\mu_{kj}$ and $s_{kj}^2$ are
\begin{equation}
\begin{aligned}
\frac{\partial \mathcal{L}(q)}{\partial \mu_{kj}} = &
\frac{(\bfy_k - \sum_{j=1}^p\sum_{l\in L_k}\alpha_{lj}\mu_{kj}\bfx_{kj})^\top\sum_{l\in L_k}\alpha_{lj}\bfx_{kj}}
{\sigma_{e_k}^2}\\
&- \frac{\sum_{l\in L_k}\alpha_{lj}\mu_{kj} - (\sum_{l\in L_k}\alpha_{lj})^2\mu_{kj}}{\sigma_{e_k}^2}
- \sum_{l\in L_k}\alpha_{lj}\frac{\mu_{kj}}{\sigma_{\beta_k}^2} = 0,\\
%\end{aligned}
%\end{equation}
%\begin{equation}
%\begin{aligned}
\frac{\partial \mathcal{L}(q)}{\partial s_{kj}^2} =&
-\frac{\sum_{l\in L_k}\alpha_{lj}\bfx_{kj}^\top\bfx_{kj}}{2\sigma_{e_k}^2} + \frac{1}{2}\sum_{l\in L_k}\alpha_{lj}(\frac{1}{s_{kj}^2} - \frac{1}{\sigma_{\beta_k}^2}) = 0.
\end{aligned}
\end{equation}
It is obvious to obtain
\begin{equation}
\label{var_u}
\mu_{kj} =  \frac{\bfx_{kj}^\top\bfy_k -
   \sum_{i\neq j}^p\sum_{l\in L_k}\alpha_{li}\mu_{ki}
                    \bfx_{kj}^\top\bfx_{ki}}
                 {\bfx_{kj}^\top\bfx_{kj} + \frac{\sigma_{e_k}^2}{\sigma_{\beta_k}^2}},\qquad
s_{kj}^2 = \frac{\sigma_{e_k}^2}
           {\bfx_{kj}^\top\bfx_{kj} + \frac{\sigma_{e_k}^2}{\sigma_{\beta_k}^2}}.\\
\end{equation}
The posterior distribution is
\begin{equation}
\label{variational_q}
q(\beta_{1j},\beta_{2j},\gamma_{1j},\gamma_{2j}) =
\prod_{k=1}^2\Big(f_{kj}(\beta_{kj})^{\gamma_{kj}}f_0(\beta_{kj})^{1-\gamma_{kj}}\Big) \prod_{l}\alpha_{lj}^{\textbf{1}_{([\gamma_{1j},\gamma_{2j}] = l)}}. \\
\end{equation}
\subsection{The estimation of model parameters}
Then we derive the updating formula of $\alpha_{lj}$ and $\bftheta$ by maximizing the lower bound (\ref{lowerbound}).\\
For the updating formula of posterior $\alpha_{lj}$, we adopt to the lagrange multiplier approach.
\begin{equation}
\begin{aligned}
Lag(q) = \mathcal{L}(q) + \sum_{j=1}^p\lambda_j(1-\sum_l\alpha_{lj})
\end{aligned}
\end{equation}
The derivative with respect to $\alpha_{00j}$ and $\alpha_{01j}$ are
\begin{eqnarray}
\label{meanvariance}
\frac{\partial Lag(q)}{\partial\alpha_{00j}}  &=& -\log\alpha_{00j} - 1 - \lambda_j + \log\alpha_{00} = 0 \notag \\
\frac{\partial Lag(q)}{\partial\alpha_{01j}}
&=&\frac{1}{2}\mu_{2j}\textbf{x}_{2j}^T\textbf{y}_2
-\frac{1}{2}(s_{2j}^2 + \mu_{2j}^2)(\frac{1}{s_{2j}^2}-\frac{1}{\sigma_{\beta_2}^2})
-\mu_{2j}(\frac{\textbf{x}_{2j}^T\textbf{y}_2}{2}- \frac{\mu_{2j}}{s_{2j}^2}) \notag\\
& &-\log\alpha_{01j} - 1 - \lambda_j + \log\alpha_{01}
+\frac{1}{2}\Big(\log\frac{s_{2j}^2}{\sigma_{\beta_2}^2} - \frac{\mu_{2j}^2 + s_{2j}^2}{\sigma^2_{\beta_2}} + 1\Big)
 = 0
\end{eqnarray}

%\begin{equation}
%\begin{aligned}
%\frac{\partial Lag(q)}{\partial\alpha_{01j}}
%=&\frac{1}{2}\mu_{2j}\textbf{x}_{2j}^T\textbf{y}_2
%-\frac{1}{2}(s_{2j}^2 + \mu_{2j}^2)(\frac{1}{s_{2j}^2}-\frac{1}{\sigma_{\beta_2}^2})
%-\mu_{2j}(\frac{\textbf{x}_{2j}^T\textbf{y}_2}{2}- \frac{\mu_{2j}}{s_{2j}^2})\\
%&-\log\alpha_{01j} - 1 - \lambda_j + \log\alpha_{01}
%+\frac{1}{2}\Big(\log\frac{s_{2j}^2}{\sigma_{\beta_2}^2} - \frac{\mu_{2j}^2 + s_{2j}^2}{\sigma^2_{\beta_2}} + 1\Big)
% = 0
%\end{aligned}
%\end{equation}
We can obtain the following equation using last equation (\ref{meanvariance}).
\begin{equation}
\begin{aligned}
\log \frac{\alpha_{01j}}{\alpha_{00j}} =&\log\frac{\alpha_{01}}{\alpha_{00}} + \frac{1}{2}\log\frac{s_{2j}^2}{\sigma_{\beta_2}^2} + \frac{\mu_{2j}^2}{2s_{2j}^2}\\
\end{aligned}
\end{equation}
Similarly, we can get the other three equations. After some algebra, they can expressed in the way of equation 16 in the text.
\begin{eqnarray}
\label{var_gam}
  A_{00j} & =& \log\alpha_{00}+ \frac{1}{2}\log\sigma_{\beta_1}^2 + \frac{1}{2}\log\sigma_{\beta_2}^2 , \notag\\
  A_{10j} & =& \log\alpha_{10} +\frac{1}{2}\log s_{1j}^2 +\frac{1}{2}\log\sigma_{\beta_2}^2 + \frac{\mu_{1j}^2}{2s_{1j}^2},\notag\\
  A_{01j} & =& \log\alpha_{01} +\frac{1}{2}\log\sigma_{\beta_1}^2+\frac{1}{2}\log s_{2j}^2  + \frac{\mu_{2j}^2}{2s_{2j}^2},\notag\\
  A_{11j} & =& \log\alpha_{11} +\frac{1}{2}\log s_{1j}^2 +\frac{1}{2}\log s_{2j}^2 + \frac{\mu_{1j}^2}{2s_{1j}^2}+ \frac{\mu_{2j}^2}{2s_{2j}^2},\notag\\
  \alpha_{lj} &=& \frac{\exp(A_{lj})}{\sum_{l\in \{00,10,01,11\}} \exp(A_{lj})}.
\end{eqnarray}
Next, we maximize the lower bound with respect to $\sigma_{e_k}^2$.
\begin{equation}
\begin{aligned}
\label{up_sigam2e}
\frac{\partial L(q)}
{\partial\sigma_{e_k}^2}
=&-\frac{n_k}{2\sigma_{e_k}^2} +
\frac{\|\bfy - \sum_{j=1}^p\sum_{l\in L_k}\alpha_{lj}\mu_{kj}\textbf{x}_{kj}\|}{2\sigma_{e_k}^4}\\
 &+ \frac{1}{2\sigma_{e_k}^4}\sum_{j=1}^p\mathbb{V}ar[\gamma_{kj}{\beta}_{kj}]\textbf{x}_{kj}^T\textbf{x}_{kj}=0
\end{aligned}
\end{equation}
which yields a maximum at
\begin{equation}
\label{mle_update1}
\begin{aligned}
\sigma_{e_k}^2
=\frac{1}{n_k}\Big(\|\bfy_k - \sum_{j=1}^p \sum_{l\in L_k}\alpha_{lj}\mu_{kj}\bfx_{kj}\|^2 +\sum_{j=1}^p
         \mathbb{V}ar[\gamma_{kj}{\beta}_{kj}]
         \bfx_{kj}^\top\bfx_{kj}\Big)\\
\end{aligned}
\end{equation}
Then, we maximize lower bound with respect to $\sigma_{\beta_k}^2$.
\begin{equation}
\begin{aligned}
\frac{\partial L(q)}{\partial \sigma_{\beta_k}^2}
=&\frac{1}{2}\sum_{j=1}^p\sum_{l\in L_k}\alpha_{lj}\Big(-\frac{1}{\sigma_{\beta_k}^2} + \frac{\mu_{kj}^2 + s_{kj}^2}{\sigma_{\beta_k}^4}\Big)=0
\end{aligned}
\end{equation}
which has a maximum at
\begin{equation}
\sigma_{\beta_k}^2 = \frac{\sum_{j=1}^p\sum_{l\in L_k}\alpha_{lj}(\mu_{kj}^2 + s_{kj}^2)}
{\sum_{j=1}^p\sum_{l\in L_k}\alpha_{lj}}\\
\end{equation}
At last, it is obvious to get updating equation of $\alpha_l$ by lagrange multiplier approach
\begin{equation}
\label{mle_update3}
\begin{aligned}
&\alpha_l = \frac{\sum_{j=1}^p\alpha_{lj}}{p}, \forall l\in\{00,10,01,11\},
\end{aligned}
\end{equation}

\begin{algorithm}[H]
\SetAlgoNoLine
\label{varEM}
\caption{Variational EM algorithm to solve the quantitative FGP model}
%\emph{Parameters}: $\bfTheta = \{\bfx_0,\bfX\}$\\
\emph{Initialization}: $\{\alpha_{kj},\mu_{kj}\}_{j=1,\dots,p},\sigma^2_{\beta_k}, \sigma^2_{e_k}$. Let $\tilde{\bfy}_k=\sum_{j=1}^p\sum_{l\in L_k}\alpha_{lj}\mu_{kj}\bfx_{kj}$ with $L_1=\{10,11\},L_2=\{01,11\}$.\\

\Repeat{Converge}{
      \For{$j=1:p$}{
      \For{$k=1:2$}{
      $\tilde{\bfy}_{k} \leftarrow \tilde{\bfy}_k - \sum_{l\in L_k}\alpha_{lj}\mu_{kj}\bfx_{kj} $ \\
      $s_{kj}^2 \leftarrow \frac{\sigma_{e_k}^2}
           {\bfx_{kj}^\top\bfx_{kj} + \frac{\sigma_{e_k}^2}{\sigma_{\beta_k}^2}}.$ \\
      $\mu_{kj} \leftarrow  \frac{\bfx_{kj}^\top\bfy_k -
   \sum_{i\neq j}^p\mathbb{E}_{i}[\gamma_{ki}\beta_{ki}]
                    \bfx_{kj}^\top\bfx_{ki}}
                 {\bfx_{kj}^\top\bfx_{kj} + \frac{\sigma_{e_k}^2}{\sigma_{\beta_k}^2}}$ \\

      $\alpha_{lj} \leftarrow \frac{\exp(A_{lj})}{\sum_{l\in \{00,10,01,11\}} \exp(A_{lj})}$, where $A_{lj}$ are defined in expression (\ref{var_gam})\\
      $\tilde{\bfy}_{k} \leftarrow \tilde{\bfy}_k + \sum_{l\in L_k}\alpha_{lj}\mu_{kj}\bfx_{kj}$
      }
      }
  $\sigma_{e_k}^2 \leftarrow\frac{1}{n_k}\Big(\|\bfy_k - \sum_{j=1}^p \sum_{l\in L_k}\alpha_{lj}\mu_{kj}\bfx_{kj}\|^2 +\sum_{j=1}^p
         (\sum_{l\in L_k}\alpha_{lj}(s_{kj}^2 + \mu_{kj}^2) - (\sum_{l\in L_k}\alpha_{lj})^2\mu_{kj}^2)
         \bfx_{kj}^\top\bfx_{kj}\Big)$\\
  $\sigma_{\beta_k}^2 \leftarrow \frac{\sum_{j=1}^p\sum_{l\in L_k}\alpha_{lj}(\mu_{kj}^2 + s_{kj}^2)}
{\sum_{j=1}^p\sum_{l\in L_k}\alpha_{lj}}$ \\
  $\alpha_l \leftarrow \frac{\sum_{j=1}^p\alpha_{lj}}{p},\forall l\in\{00,10,01,11\}$
}
\end{algorithm}

\clearpage
\section{Proof detail for the binary-trait model}
\subsection{Accommodating case-control data}
Suppose that we have GWAS datasets $\{\bfy_1,\bfX_1, \bfZ_1\}$ and $\{\bfy_2,\bfX_2, \bfZ_2\}$ for a case-control study with $n_1$ and $n_2$ samples, respectively. All settings remain the same to the quantitative traits except that $\bfy_k \in\mathbb{R}^{n_k\times 1}$ is the vector for disease status taking values -1 and 1 for study $k$ and $\bfZ_k=[\bfz_{k1},\dots,\bfz_{kp_0}]\in\mathbb{R}^{n_k\times p_0}$ is a matrix for $p_0$ covariates. Then conditional on observed genotype $\bfX_k$, hidden status $\bfgamma$, and effects $\bfbeta_k$, we have
\begin{equation}
  \bfy_k|\bfX_k, \bfZ_k, \bfbeta_k, \bfgamma_k,\bfphi_k  \sim \mathrm{Ber}(\bfdelta_k),
\end{equation}
where $\bfdelta_k = [\delta_{k1},\dots,\delta_{kn_k}]^\top$, $\delta_{ki} \left(= \pr(y_{ki}=1|\bfX, \bfbeta_k, \bfgamma_k) = \frac{1}{1+e^{-y_{ki}\eta_{ki}}}\right)$ is the sigmoid function of linear predictor $\eta_{ki}$, $i$ is the index for individuals, $\bfeta_k (= [\eta_{k1},\dots,\eta_{kn_k}]^\top\in\mathbb{R}^{n_k\times 1})$ is the linear predictor of the all individuals in study $k$ such that $\bfeta_{k} = \sum_{j=1}^{p_0}\bfz_{kj}\phi_{kj} + \sum_{j=1}^p \gamma_{kj}\beta_{kj}\bfx_{kj}$. Note that $\bfphi_k$ is a vector of fixed effects including intercept in the model. Here, we include fixed-effect covariates in the binary studies to adjust population stratification in samples. %The rationale for including covariates in the model for binary studies is that most times, we need to adjust the first few principal components from data to correct the population stratification in samples \citep{reed2015guide}.
$\bfbeta$ and $\bfgamma$ are effect sizes and indicator variables as defined in Section 2. Let $\bftheta = \{\sigma_{\beta_1}^2, \sigma_{\beta_2}^2,\bfphi_1, \bfphi_2, \bfalpha\}$ be the collection of model parameters. The probabilistic model can be written as
\begin{equation}
\label{joint_4GPMP_binary}
 \pr(\bfy_1,\bfy_2,\bfbeta_1,\bfbeta_2,\bfgamma_1,\bfgamma_2| \bfX_1,\bfZ_1, \bfX_2, \bfZ_2 ;\bftheta) = \prod_{k=1}^2 \bigg(\pr(\bfy_k|\bfX_k,\bfZ_k, \bfbeta_k,\bfgamma_k;\bftheta) \pr(\bfbeta_k|\bftheta)\bigg) \pr(\bfgamma|\bftheta).
\end{equation}
Note that we take coefficients for covariates ($\bfZ_1$ and $\bfZ_2$) as fixed effects, which are included in parameter space $\bftheta$. Marginalizing over latent variables ($\bfbeta_1,\bfbeta_2,\bfgamma_1,\bfgamma_2$), we can get the marginal likelihood. The primary difficulty for the binary model (\ref{joint_4GPMP_binary}) comes from the evaluation of sigmoid function $\delta_{ki}$. %Similar to the quantitative case, it is tempting make model (\ref{joint_4GPMP_binary}) full Bayes by adding hyper prior on $\bftheta$. In fact, this cannot be done easily as there is no convenient conjugate prior for logistic regression.
%As variational approaches iteratively finds a sequence of increasing lower bounds, it is essential to evaluate lower bounds involving sigmoid functions, which is not trivia at all.
As there is no convenient conjugate prior for sigmoid function, it is not analytically feasible to compute the full posterior over the parameter space. To overcome this limitation,, we use the Bohning bound~\citep{bohning1992multinomial}. %many quadratic lower bounds of sigmoid functions can be applied, \eg,, the Bohning bound~\citep{bohning1992multinomial} and the JJ bound~\citep{jaakkola1997variational}. Here, we adopt the Bohning bound. %because it is computationally efficient than the JJ bound which requires iteratively re-evaluation of the variational parameter for each individual.
Here, we first derive a lower bound of the complete-data likelihood as follows
\begin{equation}
\label{lb_bohn}
\begin{aligned}
&\text{Pr}(\textbf{y}_1,\textbf{y}_2,{\bm{\beta}}_1,{\bm{\beta}}_2,\bm{\gamma}_1,\bm{\gamma}_2|
\textbf{X}_1,\textbf{X}_2,\textbf{Z}_1,\textbf{Z}_2,\bm{\theta})\\
%%%%%%%%%%%%%%%%%%%%%%%%%%%%%%%%%%%%%%%%%%%%%%%%%%%%%%%%%%%%%%%%%%%%%%%%%%%%%%%%%%%%%%%%%%%%%%%%%%%%%%%%%%%%%%%%%%%
\geq& \left( \prod_{k=1}^2 B(\textbf{y}_k|\textbf{X}_k,\textbf{Z}_k,\bm{\beta}_k,\bm{\gamma}_k;\bm{\theta})
\text{Pr}(\bm{\beta}_k;\bm{\theta})\right)
\text{Pr}(\bm{\gamma};\bm{\theta})\\
=&h(\textbf{y}_1,\textbf{y}_2,{\bm{\beta}}_1,{\bm{\beta}}_2,\bm{\gamma}_1,\bm{\gamma}_2
|\textbf{X}_1,\textbf{X}_2,\textbf{Z}_1,\textbf{Z}_2;\tilde{\bm{\theta}}),
\end{aligned}
\end{equation}
where $B(\textbf{y}_k|\textbf{X}_k,\textbf{Z}_k,\bm{\beta}_k,\bm{\gamma}_k;\tilde{\bm{\theta}}) \left(=
\prod_{i=1}^{n_k}\exp(-\frac{1}{2}a\eta_{ki}^2y_{ki}^2+(1+b_{ki})\eta_{ki}y_{ki} - c_{ki})\right)$ denotes the product of lower bound of sigmoid functions with $a=1/4$, $b_{kn} = a\psi_{kn} - (1 + e^{-\psi_{kn}})^{-1}$ and $c_{kn} = \frac{1}{2}a\psi_{kn}^2 - (1 + e^{-\psi_{kn}})^{-1}\psi_{kn} + \log(1 + e^{\psi_{kn}})$, and $\tilde{\bm{\theta}} =\{\sigma_{\beta_1}^2,\sigma_{\beta_2}^2,\bfphi_1,\bfphi_2,\bm{\alpha},\bm{\psi}_1,\bm{\psi}_2\}$ is the new parameter which combines the model parameters $\bm{\theta}$ with variational parameters $\bm{\psi}_1$,$\bm{\psi}_2$. Using Jensen's inequality and the lower bound of complete-data likelihood (\ref{lb_bohn}), we have the following lower bound
\begin{equation}
\begin{aligned}
\label{def_lowerbound_logis}
  &\log \text{Pr}(\textbf{y}_1,\textbf{y}_2| \textbf{X}_1,\textbf{X}_2,\textbf{Z}_1,\textbf{Z}_2,\bm{\theta})\\
=&\log \sum_{\bfbeta_1,\bfbeta_2, \bfgamma_1,\bfgamma_2}  \text{Pr}(\textbf{y}_1,\textbf{y}_2,\bfbeta_1,\bfbeta_2,\bfgamma_1,\bfgamma_2| \textbf{X}_1,\textbf{X}_2,\textbf{Z}_1,\textbf{Z}_2,\bm{\theta})  \\
\geq&\log \sum_{\bfbeta_1,\bfbeta_2, \bfgamma_1,\bfgamma_2}  h(\textbf{y}_1,\textbf{y}_2,\bfbeta_1,\bfbeta_2,\bfgamma_1,\bfgamma_2| \textbf{X}_1,\textbf{X}_2,\textbf{Z}_1,\textbf{Z}_2,\bm{\theta}) \\
\geq&\mathbb{E}_q[\log h(\textbf{y}_1,\textbf{y}_2, \bm{\beta}_1,\bm{\beta}_2,\bm{\gamma}_1,\bm{\gamma}_2
|\textbf{X}_1,\textbf{X}_2,\textbf{Z}_1,\textbf{Z}_2;\tilde{\bm{\theta}})] - \mathbb{E}_q[\log q(\bm{\beta}_1,\bm{\beta}_2,\bm{\gamma}_1,\bm{\gamma}_2)]
:=\mathcal{L}(q),
\end{aligned}ß
\end{equation}
where the first inequality is based on Bohning bound and the second one follows from Jensen's inequality as in lower bound.
By maximizing the lower bound (\ref{def_lowerbound_logis}) with respect to $\mu_{kj}$ and $s_{kj}^2$, we can again obtain the variational distribution in the same fashion as expression (\ref{variational_q}). With some algebra, we have
\begin{equation}
\begin{aligned}
\mu_{kj} =  \frac{\bfx_{kj}^\top\bfy^*_k -a\bfx_{kj}^T\bfZ_k\bfphi_k
   - a\sum_{j'\neq j}^p\sum_{l\in L_k}\alpha_{lj'}\mu_{kj'}
                    \bfx_{kj'}^\top\bfx_{kj}}
                 {a\bfx_{kj}^\top\bfx_{kj} + \frac{1}{\sigma_{\beta_k}^2}},\quad
s_{kj}^2 = \frac{1}
           {a\bfx_{kj}^\top\bfx_{kj} + \frac{1}{\sigma_{\beta_k}^2}},\\
\end{aligned}
\end{equation}
where $\bfy^*_k\left( = ((1 + b_{k1})y_{k1},\cdots,(1 + b_{kn_k})y_{kn_k})^\top\right)$ is the working response and
$\alpha_{lj}$ is the log posterior odds of being $l$ in expression (\ref{var_gam}). The difference only lies in the updating equation for the posterior mean $\mu_{kj}$, which uses working response $\bfy^*_k$ and the re-weighting design matrix $a^{1/2}\bfZ$ and $a^{1/2}\bfX$. As $a$ is a constant, there is no additional computational burden to evaluate the re-weighting design. Once we loop through all variational parameters, other parameters ($\tilde{\bm{\theta}}$) can be updated by maximizing the lower bound while keep variational distribution $q$ fixed, \ie, $\frac{\partial \mathcal{L}}{\partial \tilde{\bm{\theta}}}=0$. %Details for updating equations and corresponding VBEM algorithm are given in the supplementary document.

%\subsection{The derivation of lower bound}
%\begin{equation}
%\begin{aligned}
%\log &\text{Pr}(\bfy_1,\bfy_2| \bfX_1,\bfX_2,\textbf{Z}_1,\textbf{Z}_2,\bftheta)\\ =
%&\log\sum_{\bfgamma_1,\bfgamma_2}\int_{{\bfbeta_1},{\bfbeta_2}}
%\text{Pr}(\bfy_1,\bfy_2,\bfbeta_1,\bfbeta_2,\bfgamma_1,\bfgamma_2|
%\bfX_1,\bfX_2,\textbf{Z}_1,\textbf{Z}_2,\bftheta)d\bfbeta_1d\bfbeta_2\\
%\geq& \log\sum_{\bfgamma_1,\bfgamma_2}\int_{{\bfbeta_1},{\bfbeta_2}}
%q(\bfbeta_1,\bfbeta_2,\bfgamma_1,\bfgamma_2)
%\frac{h(\bfy_1,\bfy_2,\bfbeta_1,\bfbeta_2,\bfgamma_1,\bfgamma_2
%|\bfX_1,\bfX_2,\textbf{Z}_1,\textbf{Z}_2;{\tilde{\bftheta}})}
%{q(\bfbeta_1,\bfbeta_2,\bfgamma_1,\bfgamma_2)}
%d\bfbeta_1d\bfbeta_2\\
%\geq&\mathbb{E}_q[h(\bfy_1,\bfy_2,\bfbeta_1,\bfbeta_2,\bfgamma_1,\bfgamma_2
%|\bfX_1,\bfX_2,\textbf{Z}_1,\textbf{Z}_2;\tbftheta)] - \mathbb{E}_q[\log q(\bfbeta_1,\bfbeta_2,\bfgamma_1,\bfgamma_2)]\\
%:=&\mathcal{L}(q)
%\end{aligned}
%\end{equation}

\subsection{The derivation of lower bound}
As the definition of lower bound for binary trait model is
\begin{equation}
\begin{aligned}
\label{def_lowerbound_logis}
\mathcal{L}(q) = &\mathbb{E}_q[\log h(\bfy_1,\bfy_2,\bfbeta_1,\bfbeta_2,\bfgamma_1,\bfgamma_2|\bfX_1,\bfX_2;{\tbftheta})]
- \mathbb{E}_q[\log q(\bfbeta_1,\bfbeta_2,\bfgamma_1,\bfgamma_2)]
\end{aligned}
\end{equation}
We can derive the analytical form of lower bound by evaluate these two expectation.
The first expectation of lower bound (\ref{def_lowerbound_logis}) is
\begin{equation}
\begin{aligned}
\label{firstterm_logis}
\mathbb{E}_q[\log h(&\bfy_1,\bfy_2,\bfbeta_1,\bfbeta_2,\bfgamma_1,\bfgamma_2|\bfX_1,\bfX_2;{\tbftheta})]\\
= &\sum_{k=1}^2\Bigg(-\sum_{n=1}^{n_k}c_{kn}+\sum_{n=1}^{n_k}(1+b_{kn})y_{kn}\textbf{z}_{kn}^T\bfphi_k
  -\frac{a}{2}\sum_{n=1}^{n_k}(y_{kn}\textbf{z}_{kn}^T\bfphi_k)^2\\
  &+\sum_{j=1}^p\mathbb{E}_j[\gamma_{kj}{\beta}_{kj}]\bfx_{kj}^T\bfy^*_k
   - a\sum_{j=1}^p\mathbb{E}_j[\gamma_{kj}{\beta}_{kj}]\bfx_{kj}^T\textbf{Z}_k\bfphi_k\\
 &-\frac{a}{2}\sum_{j=1}^p\mathbb{E}_j[\gamma_{kj}{\beta}_{kj}]^2\bfx_{kj}^T\bfx_{kj}
 -\frac{a}{2}\sum_{j=1}^p\sum_{j'\neq j}^p
 \mathbb{E}_{qjj'}
 [\gamma_{j}{\beta}_{kj}\gamma_{kj'}{\beta}_{kj'}]\bfx_{kj}^T\bfx_{kj'}\\
 &-0.5a\sum_{j=1}^p\mathbb{V}ar[\gamma_{kj}{\beta}_{kj}]\bfx_{kj}^T\bfx_{kj}\\
&-\frac{p}{2}\log(2\pi\sigma_{\beta_1}^2)
-\frac{1}{2\sigma_{\beta_k}^2}\sum_{j=1}^p\mathbb{E}_{qj}{\beta}_{kj}^2\Bigg)\\
& +\sum_{j=1}^p\sum_{l}\mathbb{E}_{qj}[\textbf{1}_{([\gamma_{1j},\gamma_{2j}] = l)}]\log \alpha_l
 \end{aligned}
\end{equation}
Where $\textbf{z}_{kn}$ denotes the $nth$ row of covariates matrix $\bfZ_k$,
 the outcome of these expectations in the above equation is the same as the expectations in the previous equation (\ref{expectation}). Besides, the analytical form of the second term in the equation (\ref{def_lowerbound_logis}) is also the same as the second term of lower bound (\ref{def lowerbound}). The only difference is that the mean $\mu_{kj}$ and variance $s_{kj}$ are defined in the equation (\ref{var_u}). As we derive the first term and second term of the lower bound, the whole lower bound can be obviously obtained by adding them together.
\begin{equation}
\begin{aligned}
\label{lowerbound_logis}
\mathcal{L}(q)
 =& \sum_{k=1}^2\Bigg(\bfm_k^T\bfX_k^T\bfy^*_k
- a\bfm_k^T\bfX_k^T\textbf{Z}_k\bfphi_k
-0.5a\bfm_k^T\bfX_k^T\bfX_k\bfm_k\\
&-0.5a\sum_{j=1}^p\mathbb{V}ar[\gamma_{kj}{\beta}_{kj}]\bfx_{kj}^T\bfx_{kj}
-\sum_{n=1}^{n_k}c_{kn} + \bfy_k^{*\top}\textbf{Z}_k\bfphi_k - 0.5a\bfphi_k^T\textbf{Z}_k^T\textbf{Z}_k\bfphi_k\Bigg)\\
&-\sum_{j=1}^p\sum_{l}\alpha_{lj}(\log\frac{\alpha_{lj}}{\alpha_l})
+\frac{1}{2}\sum_{k=1}^2\sum_{j=1}^p\sum_{l\in L_k}\alpha_{lj}\Big[\log\frac{s_{kj}^2}{\sigma_{\beta_k}^2}
- \frac{\mu_{kj}^2 + s_{kj}^2}{\sigma_{\beta_k}^2} + 1\Big]\\
& - p\log(2\pi) - p\\
\end{aligned}
\end{equation}
where
\begin{equation}
\begin{aligned}
&\bfm_k = (\sum_{l\in L_k}\alpha_{l1}\mu_{k1},\cdots,\sum_{l\in L_k}\alpha_{lp}\mu_{kp})^T\\
&\mathbb{V}ar[\gamma_{kj}{\beta}_{kj}] =
\sum_{l\in L_k}\alpha_{lj}(\mu_{1j}^2+s_{1j}^2) - (\sum_{l\in L_k}\alpha_{lj})^2\mu_{1j}^2\\
\end{aligned}
\end{equation}
\subsection{The derivation of posterior distribution}
We derive the variational posterior distribution by maximizing the lower bound (\ref{lowerbound_logis}). The derivative function with respective to $\mu_{kj}$ and $s_{kj}^2$ are
\begin{equation}
\begin{aligned}
\frac{\partial \mathcal{L}(q)}{\partial \mu_{kj}} = &
\sum_{l\in L_k}\alpha_{lj}\bfx_{kj}^\top\bfy^*_k -a\sum_{l\in L_k}\alpha_{lj}\bfx_{kj}^T\bfZ_k\bfphi_k
-a\sum_{j=1}^p(\sum_{l\in L_k}\alpha_{lj})^2\mu_{kj}\bfx_{kj}^\top\bfx_{kj}\\
&- a\Big(\sum_{l\in L_k}\alpha_{lj}\mu_{kj} - (\sum_{l\in L_k}\alpha_{lj})^2\mu_{kj}\Big)\bfx_{kj}^\top\bfx_{kj}
- \sum_{l\in L_k}\alpha_{lj}\frac{\mu_{kj}}{\sigma_{\beta_k}^2} = 0
\end{aligned}
\end{equation}
\begin{equation}
\begin{aligned}
\frac{\partial \mathcal{L}(q)}{\partial s_{kj}^2} =
-a\frac{\sum_{l\in L_k}\alpha_{lj}\bfx_{kj}^\top\bfx_{kj}}{2} + \frac{1}{2}\sum_{l\in L_k}\alpha_{lj}(\frac{1}{s_{kj}^2} - \frac{1}{\sigma_{\beta_k}^2}) = 0
\end{aligned}
\end{equation}
It is obvious to obtain
\begin{equation}
\begin{aligned}
\mu_{kj} &=  \frac{\bfx_{kj}^\top\bfy^*_k -a\bfx_{kj}^T\bfZ_k\bfphi_k
   - a\sum_{i\neq j}^p\sum_{l\in L_k}\alpha_{li}\mu_{ki}
                    \bfx_{kj}^\top\bfx_{ki}}
                 {a\bfx_{kj}^\top\bfx_{kj} + \frac{1}{\sigma_{\beta_k}^2}},\quad
s_{kj}^2 &= \frac{1}
           {a\bfx_{kj}^\top\bfx_{kj} + \frac{1}{\sigma_{\beta_k}^2}}\\
\end{aligned}
\end{equation}
The posterior distribution is
\begin{equation}
\label{variational_q}
q(\beta_{1j},\beta_{2j},\gamma_{1j},\gamma_{2j}) =
\prod_{k=1}^2f_{kj}(\beta_{kj})^{\gamma_{kj}}f_0(\beta_{kj})^{1-\gamma_{kj}} \prod_{l}\alpha_{lj}^{\textbf{1}_{([\gamma_{1j},\gamma_{2j}] = l)}} \\
\end{equation}
%%%%%%%%%%%%%%%%%%%%%%%%%%%%%%%%%%%%%%%%%%%%%%%%%%%%%%%%%%%%%%%%%%
\subsection{The estimation of model parameters}
Then we want to derive the updating formula of $\alpha_{lj}$ and $\bftheta$.
The technique to get the updating formula of $\alpha_{lj}$, $\sigma_{\beta_k}^2$ and $\alpha_l$ is similar and the updating equation is almost the same as that in the quantitative trait model.Here we discuss the updating equation of $\bm{\psi}_k$ and $\bfphi_k$. We maximize lower bound with respect to $\bm{\psi}_k$
\begin{equation}
\begin{aligned}
\frac{\mathcal{L}(q)}{\partial \psi_{kn}} = \frac{\partial \bfm_k^T\textbf{X}_k^T\bfy_k^*}{\partial \psi_{kn}} - \frac{\partial c_{kn}}{\partial \psi_{kn}} + \frac{\partial \bfy_k^{*T}\textbf{Z}_k\bfphi_k}{\partial \psi_{kn}} = 0
\end{aligned}
\end{equation}
which gets the maximum at
\begin{equation}
\begin{aligned}
\psi_{kn} &= y_n\sum_{j=1}^p\sum_{l\in L_k}\alpha_{lj}\mu_{kj}x_{knj} + y_n\textbf{z}_{kn}^T\bfphi_k\\
%\psi_{nk} &= \bfm_k^T\textbf{X}_k^T\bfy_k + \bfy_k^T\textbf{Z}_k\bfphi_k\\
\end{aligned}
\end{equation}
Then, we maximize lower bound with respect to $\bfphi_k$
\begin{equation}
\begin{aligned}
\frac{\partial\mathcal{L}(q)}{\partial \bfphi_k} = \textbf{Z}_k^T\bfy^*_k-a\textbf{Z}_k^T\textbf{X}_k\bfm_k - a\textbf{Z}_k^T\textbf{Z}_k\bfphi_k = 0
\end{aligned}
\end{equation}
which obtain the maximum at
\begin{equation}
\begin{aligned}
\bfphi_k &= (\textbf{Z}_k^T\textbf{Z}_k)^{-1}\Big(\frac{1}{a}\textbf{Z}_k^T\bfy^*_k - \textbf{Z}_k^T\textbf{X}_k\bfm_k\Big)\\
\end{aligned}
\end{equation}

\begin{algorithm}[H]
\SetAlgoNoLine
\label{varEM2}
\caption{Variational EM algorithm to solve the FGP model in case-control studies}
%\emph{Parameters}: $\bfTheta = \{\bfx_0,\bfX\}$\\
\emph{Initialization}: $\{\alpha_{kj},\mu_{kj}\}_{j=1,\dots,p},\sigma^2_{\beta_k},\bm{\phi}_k,\bm{\psi}_k$. Let $\tilde{\bfy}_k=\sum_{j=1}^p\sum_{l\in L_k}\alpha_{lj}\mu_{kj}\bfx_{kj}$ with $L_1=\{10,11\},L_2=\{01,11\}$,$a = \frac{1}{4}$.\\

\Repeat{Converge}{
      \For{$n=1:n_k$}{
      \For{$k=1:2$}{
      $b_{kn} \leftarrow a\psi_{kn} - (1 + e^{-\psi_{kn}})^{-1}$\\
      $c_{kn} \leftarrow \frac{1}{2}a\psi_{kn}^2 - (1 + e^{-\psi_{kn}})^{-1}\psi_{kn} + \log(1 + e^{\psi_{kn}})$\\
      }
      }
      \For{$j=1:p$}{
      \For{$k=1:2$}{
      $\tilde{\bfy}_{k} \leftarrow \tilde{\bfy}_k - \sum_{l\in L_k}\alpha_{lj}\mu_{kj}\bfx_{kj} $ \\
      $s_{kj}^2 \leftarrow \frac{1}
           {a\bfx_{kj}^\top\bfx_{kj} + \frac{1}{\sigma_{\beta_k}^2}}.$ \\
      $\mu_{kj} \leftarrow  \frac{\textbf{x}_{kj}^\top\textbf{y}^*_k -a\textbf{x}_j^\top\textbf{Z}_k\bm{\phi}_k
   - a\sum_{i\neq j}^p\mathbb{E}_{i}[\gamma_{ki}\beta_{ki}]
                    \textbf{x}_{kj}^\top\textbf{x}_{ki}}
                 {a\textbf{x}_{kj}^T\textbf{x}_{kj} + \frac{1}{\sigma_{\beta_k}^2}}$ \\

      $\alpha_{lj} \leftarrow \frac{\exp(A_{lj})}{\sum_{l\in \{00,10,01,11\}} \exp(A_{lj})}$, where $A_{lj}$ are defined in expression (\ref{var_gam})\\
      $\tilde{\bfy}_{k} \leftarrow \tilde{\bfy}_k + \sum_{l\in L_k}\alpha_{lj}\mu_{kj}\bfx_{kj}$
      }
      }
  $\sigma_{\beta_k}^2 \leftarrow \frac{\sum_{j=1}^p\sum_{l\in L_k}\alpha_{lj}(\mu_{kj}^2 + s_{kj}^2)}
{\sum_{j=1}^p\sum_{l\in L_k}\alpha_{lj}}$ \\
  $\alpha_l \leftarrow \frac{\sum_{j=1}^p\alpha_{lj}}{p},\forall l\in\{00,10,01,11\}$\\
  $\bfphi_k \leftarrow (\textbf{Z}_k^T\textbf{Z}_k)^{-1}\Big[\frac{1}{a}\textbf{Z}_k^T\textbf{y}^*_k - \textbf{Z}_k^T\tilde{\bfy}_k\Big]$\\
  $\psi_{kn} \leftarrow y_n\sum_{j=1}^p\sum_{l\in L_k}\alpha_{lj}\mu_{kj}x_{knj} + y_n\textbf{z}_{kn}^T\bfphi_k$\\
}
\end{algorithm}

\bibliographystyle{plainnat}
%\bibliography{Document}

\bibliography{4groupsupplementary_ref}